\documentclass[aps,twocolumn,prx,amsmath,amssymb,floatfix,superscriptaddress,footnoteinbib]{revtex4-1}
\usepackage[english]{babel}

\usepackage{graphicx}
\usepackage{float}

\usepackage{amssymb,amsfonts,amsmath}
\usepackage{bm}

\usepackage{times}

\usepackage[colorlinks,bookmarks=false,citecolor=blue,linkcolor=blue,urlcolor=blue]{hyperref}
\usepackage{color}
\usepackage[all]{hypcap} 
\usepackage{ulem} 

\newcommand{\rev}[1]{{\color{black}#1}}
\newcommand{\VEK}[1]{{\color{black}#1}}
\newcommand{\VEKn}[1]{{\color{black}#1}}

\newcommand{\ep}{\varepsilon}

\newcommand{\be}{\begin{equation}}
\newcommand{\ee}{\end{equation}}
\def\ba{\begin{aligned}}
\def\ea{\end{aligned}}
\newcommand{\bea}{\begin{eqnarray}}
\newcommand{\eea}{\end{eqnarray}}

\newcommand{\la}{\left\langle}
\newcommand{\ra}{\right\rangle}
\newcommand{\lb}{\left[}
\newcommand{\rb}{\right]}
\newcommand{\lp}{\left(}
\newcommand{\rp}{\right)}
\renewcommand{\Re}{{\rm \, Re\,}}
\renewcommand{\Im}{{\rm \, Im\,}}

\begin{document}

\title{Correlation-induced localization}
\author{P.~A.~Nosov\footnote[1]{To whom correspondence should be addressed}}
 \address{Department of Physics, St. Petersburg State University, St. Petersburg 198504, Russia}
 \address{NRC Kurchatov Institute, Petersburg Nuclear Physics Institute, Gatchina 188300, Russia}
 \address{Max-Planck-Institut f\"ur Physik komplexer Systeme, N\"othnitzer Stra{\ss}e~38, 01187-Dresden, Germany }

\author{I.~M.~Khaymovich}
 \address{Max-Planck-Institut f\"ur Physik komplexer Systeme, N\"othnitzer Stra{\ss}e~38, 01187-Dresden, Germany }

\author{V.~E.~Kravtsov}
 \address{Abdus Salam International Center for Theoretical Physics - Strada Costiera~11, 34151 Trieste, Italy}
 \address{L. D. Landau Institute for Theoretical Physics - Chernogolovka, Russia}
\address{Kavli Institute for Theoretical Physics, Kohn Hall, University of California at Santa Barbara, Santa Barbara, CA 93106, U.S.A.}

\begin{abstract}
A new paradigm of Anderson localization caused by correlations in the long-range hopping along
with uncorrelated on-site disorder is considered
which requires a more precise formulation of
the basic localization-delocalization principles.
A new class of random Hamiltonians with translation-invariant hopping integrals
 is suggested and
the localization properties of such models are established both in the coordinate and in the momentum
spaces alongside with the corresponding level statistics.
 Duality of translation-invariant models in the momentum and coordinate space is uncovered
and exploited to find a full localization-delocalization phase diagram for such models.
\rev{The crucial role of the spectral properties of hopping matrix is established and
a new matrix inversion trick is suggested to generate a one-parameter family of equivalent localization/delocalization problems.}
Optimization over the free parameter in such a transformation
together with the localization/delocalization principles
allows to establish exact bounds for the localized and ergodic states in long-range hopping models.
When applied to the random matrix models with deterministic power-law hopping this transformation
allows to confirm localization of states at all values of the exponent in power-law hopping
and to prove analytically the symmetry of the exponent in the power-law localized wave functions.
\end{abstract}
\date{\today}

\maketitle
\section{Introduction.}
The standard picture of Anderson localization in a three-dimensional single-particle system
with short-range hopping~\cite{Anderson1958}
is represented by the phase transition between extended ergodic and localized phases at a certain
critical disorder strength or 
energy with a sharp mobility edge separating ergodic and localized states.
Exactly at the Anderson localization transition (AT) non-ergodic (multifractal) extended states
have been proven to appear~\cite{Wegner1980, Roemer}.
It is well-known that in low dimensions $d=1,2$ for any tight-binding (or short-range)
Hamiltonian with uncorrelated disorder all states are localized.

\begin{figure}[h!]
\includegraphics[width=0.9\columnwidth]{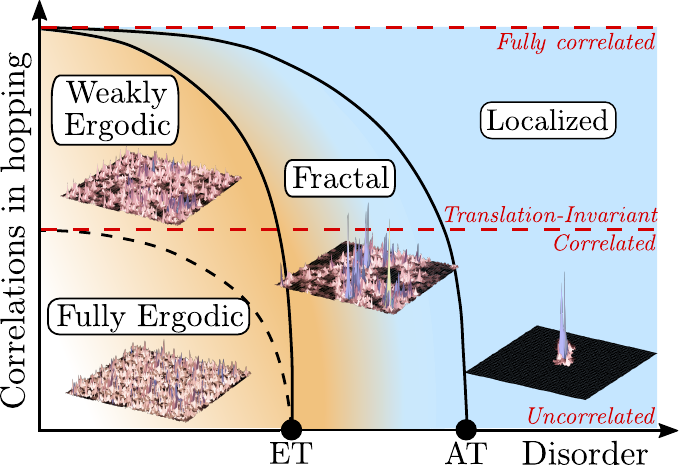}
\caption{{\bf Effect of correlations in long-range hopping on the Anderson
localization.}
The correlations in the long-range hopping results in the sequence~\eqref{sequence} of phase transitions
with degrading ergodicity. The parameter labeled as 'Disorder' is an {\it effective} disorder which determines the ratio of
on-site disorder (fixed in our models) to the hopping integrals at large distances (controlled by the exponents $a$ or $\gamma$).
The phase diagram is shown for the Rosenzweig-Porter (RP) family of ensembles, being an example of the system where all phases
(fully and weakly ergodic, fractal, and localized) are present.
The fractal phase separated by the localization (AT) and the ergodic (ET) transitions from the localized and the ergodic phases
is present in these models even in the absence of correlations.
Increasing the correlations (upwards along the vertical axis) sends the AT and ET to smaller values of disorder and stretches the
critical point of ET into a whole weakly ergodic phase.
Three-dimensional plots show cartoons of spatial distributions of wavefunction intensities in the corresponding localized,
(multi)fractal, and ergodic phases.
For a family of the power-law random banded matrices (PLRBM)
considered in this work (not shown) the fractal phase is
replaced by the weakly ergodic one.
}
\label{Fig:Cartoon}
\end{figure}

However, delocalized states may appear even in one-dimensional systems if the hopping is
long-ranged~\cite{Levitov1989,Levitov1990, MirFyod1996, MirRev}. An archetypical example
of such nominally one-dimensional systems is suggested in Ref.~\cite{MirFyod1996}. In this
\rev{power-law random banded matrix (PLRBM)} model the long-ranged hopping terms are completely
uncorrelated and Gaussian distributed
with a power-law decay of the variance $\langle |H_{nm}|^{2}\rangle\propto (b/ |n-m|)^{2a}$
with the distance $|m-n|$ that saturates $\langle |H_{nm}|^{2}\rangle\sim 1$ at $|m-n|<b$.
The parameter that drives the localization transition in this system is the exponent $a$.
For $a>1$ the states are power-law localized, while at $a<1$ they are extended.
At the critical point $a=1$ multifractal states with variable (depending on the parameter $b$)
strength of multifractality are formed~\cite{MirFyod1996, KravtsovMuttalib1997, MirRev}.

At the first glance this {\it delocalization} at long range hopping is natural and
independent of the uncorrelated nature of the hopping integrals, as at one hop the
particle can reach any point of the system.
Yet, as we show in this paper, localization effects get {\it stronger}
if the long-range hopping integrals are fully or partially correlated (see Fig.~\ref{Fig:Cartoon}).

\begin{figure*}[t]
\includegraphics[width=0.8 \linewidth,angle=0]{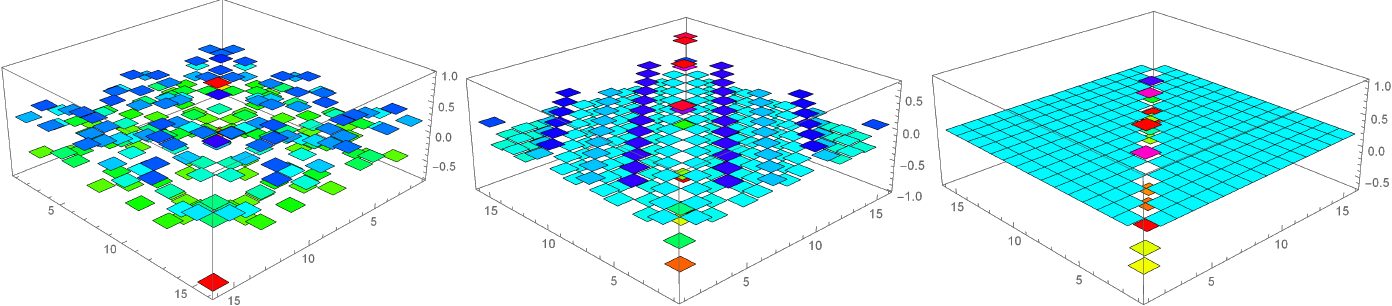}
\caption{(Color online) {\bf Matrix Hamiltonians (in the order of increasing correlations)} with
 fully random, random translation-invariant and deterministic hopping. Squares of different
color and at different heights represent the value of matrix elements for different $m,n=1,...,15$. }
\label{Fig:correlations}
\end{figure*}

It is commonly believed that correlated disorder in the on-site energies ("diagonal disorder") tends
to {\it delocalize} systems. An important example is the Aubry-Andre lattice model~\cite{AA} with an
incommensurate periodic potential that possesses delocalized states and exhibits AT.
This type of {\it quasi-disorder} was widely used in recent experiments on localization of
matter waves of cold atoms~\cite{Cold-loc}.
Our findings show that correlations in the long-range hopping produce an {\it opposite}
effect.
This effect is not restricted to the one-dimensional systems: the tendency towards localization at
correlated long-range hopping is present in higher dimensional
systems thus showing the universality of this phenomenon which we call
{\it 'correlation-induced localization'}.

The physics of long-range interacting systems is now an emerging field. Initially it was
motivated by experiments on trapped cold atoms with dipole moments
(see e.g.~\cite{Cold-dipole1, Cold-dipole2, Cold-dipole3}). However, now the interest
is being shifted towards many-body localization in systems with long-range (e.g. Coulomb) interaction~\cite{Rahul2017}.
Several models with \rev{
such}
interactions have been
suggested in the past~\cite{Burin_Kagan1994} and
recently~\cite{Yao_Lukin2014,BurinPRB2015-1,BurinPRB2015-2,BurinAnnPhys2017,Moessner2017,
halimeh2017dynamical,homrighausen2017anomalous,luitz2018emergent,Botzung_Muller2018,
deTomasi2018,hashizume2018dynamical,tikhonov2018many} in the problem of
entanglement dynamics at many-body localization.
The models with fully-correlated hopping and interaction
terms~\cite{Burin_Kagan1994,Yao_Lukin2014,BurinPRB2015-2,halimeh2017dynamical,
homrighausen2017anomalous,luitz2018emergent,Botzung_Muller2018,hashizume2018dynamical}
show significantly different behavior as compared to the ones with uncorrelated
hopping and interactions~\cite{Yao_Lukin2014,BurinPRB2015-1,BurinAnnPhys2017,tikhonov2018many,deTomasi2018}.
Some of the former works (see, e.g.,~\cite{luitz2018emergent}) even demonstrate explicitly
that many-body properties are formed as superpositions of short-range and power-law decaying
contributions in the complete agreement with the single-particle picture developed in this work.

Moreover in physics of classical dynamical systems long-range interactions also play a significant role leading to
the formation of inhomogeneous spatial temperature distribution anti-correlated with the density profile after a
global spatially homogeneous quench~\cite{casetti2014velocity,teles2015temperature,gupta2016surprises}.
The relaxation times from these emergent inhomogeneous states to the equilibrium are very long and
diverge with the system size~\cite{campa2009statistical,bouchet2010thermodynamics,campa2014physics}.
This physics is relevant, in particular, to the explanation of the heating of the solar corona.
Thus, we believe that the results of this paper are relevant for all above mentioned types of
many-body problems as well.


The correlations in long-range hopping are barely studied.
The earlier study of a single-particle system with {\it deterministic}
(or fully-correlated) power-law decay of long-range hopping~\cite{Burin1989},
 has been nearly unnoticed until recently.
Several recent works
~\cite{Yuzbashyan_NJP2016,Yuzbashyan_JPhysA2009_Exact_solution,Ossipov2013,
Borgonovi_2016,Kravtsov_Shlyapnikov_PRL2018} reported
about localization in such systems with fully-correlated long-range hopping,
confirming (mostly numerically) the conclusion of the renormalization
group (RG) analysis done in Ref.~\cite{Burin1989}.
Neither of the models
~\cite{Yuzbashyan_NJP2016,Yuzbashyan_JPhysA2009_Exact_solution,Ossipov2013,Borgonovi_2016,Kravtsov_Shlyapnikov_PRL2018}
demonstrates a truly delocalized behavior of wave functions in
the bulk of the spectrum for {\it all} strengths of disorder
and {\it all values} of the exponent $a$.

Furthermore, very recently a striking duality $\mu(a)=\mu(2-a)$
in the spatial decay rate $\mu \geq 2$ of the power-law localized
wave functions $|\psi|^{2}\propto |r-r_{0}|^{-\mu}$ was
discovered~\cite{Kravtsov_Shlyapnikov_PRL2018} in the models with
algebraically decaying correlated hopping~\cite{Burin1989,Borgonovi_2016,
Kravtsov_Shlyapnikov_PRL2018}. This implies
enhancement of localization upon making hopping more long-ranged.
In this work we prove this duality and analytically show the absence
of delocalized phase in these models.

Despite all these facts spread in the literature, the systematic
study of correlations at long-range hopping has not been done so far and
importance and generality of the phenomenon of enhancement of
localization by correlations
have not been appreciated. This paper is
aimed to fill in this gap in the theory of Anderson localization.

In all the above mentioned models the long-range hopping integrals are either uncorrelated
or fully correlated (deterministic). The systematic study of the
role of correlations requires a gradual increase of correlations. In this work
we suggest a new class of models that bridge between the
models with uncorrelated hopping and those with fully correlated
hopping (see Fig.~\ref{Fig:correlations}). These are the models
with random long-ranged hopping integrals which are {\it translation-invariant}
(TI). In a given realization the hopping integrals $H_{nm}=H_{|n-m|}$ in TI models
are
fully correlated along a diagonal (see Fig.~\ref{Fig:correlations})
but they are uncorrelated and sign-alternating for different diagonals~\cite{Periodic_bound_conds_footnote}.
Such models emerge naturally, e.g. in the case when hopping is caused
by the RKKY interaction which oscillates with the period incommensurate
with the lattice constant.

In addition to the models with the typical long-range hopping integrals
decreasing algebraically
with the distance which physical realization is more or less obvious,
the models with the typical hopping integrals being distance-independent but
dependent on the system size (as $N^{-\gamma/2}$) have recently come under the spotlight.
The interest to
such models emerged because of the discovery~\cite{Kravtsov_NJP2015} of the
new non-ergodic extended (multi-fractal) phase and the corresponding {\it ergodic transition} in the
generalized Rosenzweig-Porter (RP) model. This model appeared to be relevant for several many-body
problems such as the
Quantum Random Energy Model \cite{faoro2018non} with implications for quantum
computing \cite{smelyanskiy2018non}, as well as for non-ergodic extended states in
the Sachdev-Ye-Kitaev \cite{sachdev1993gapless,Kitaev-talk} ($SYK_{4}+SYK_{2}$)
many-body model \cite{micklitz2019non,Kamenev-talk-1,Kamenev-talk-2}.
The presence of non-ergodic extended phase and of above
mentioned ergodic transition puts on a solid ground the search for
ergodic transition and non-ergodic extended phase on Random Regular
Graphs (RRG) (initiated in Ref.~\cite{Luca2014, RRGAnnals} and discussed
in detail in~\cite{RRGAnnals}) and in real many-body
systems~\cite{Pino-Ioffe-Altshuler2016, Pino-Ioffe-Kr2017}.
 Slow dynamics on RRG~\cite{Bera-Khaym2018,Biro17,Tikhonov_misc}
and in disordered spin chains~\cite{
Gopa2016,Bloch2017a,Bloch2017b} may be a signature of such a phase.
In this work we suggest the translation-invariant extension of the RP model and study the
localization properties of the RP family of models along with the PLRBM family as the correlations in the long-range hopping
increase.

\begin{figure}[t]
\includegraphics[width=1\columnwidth]{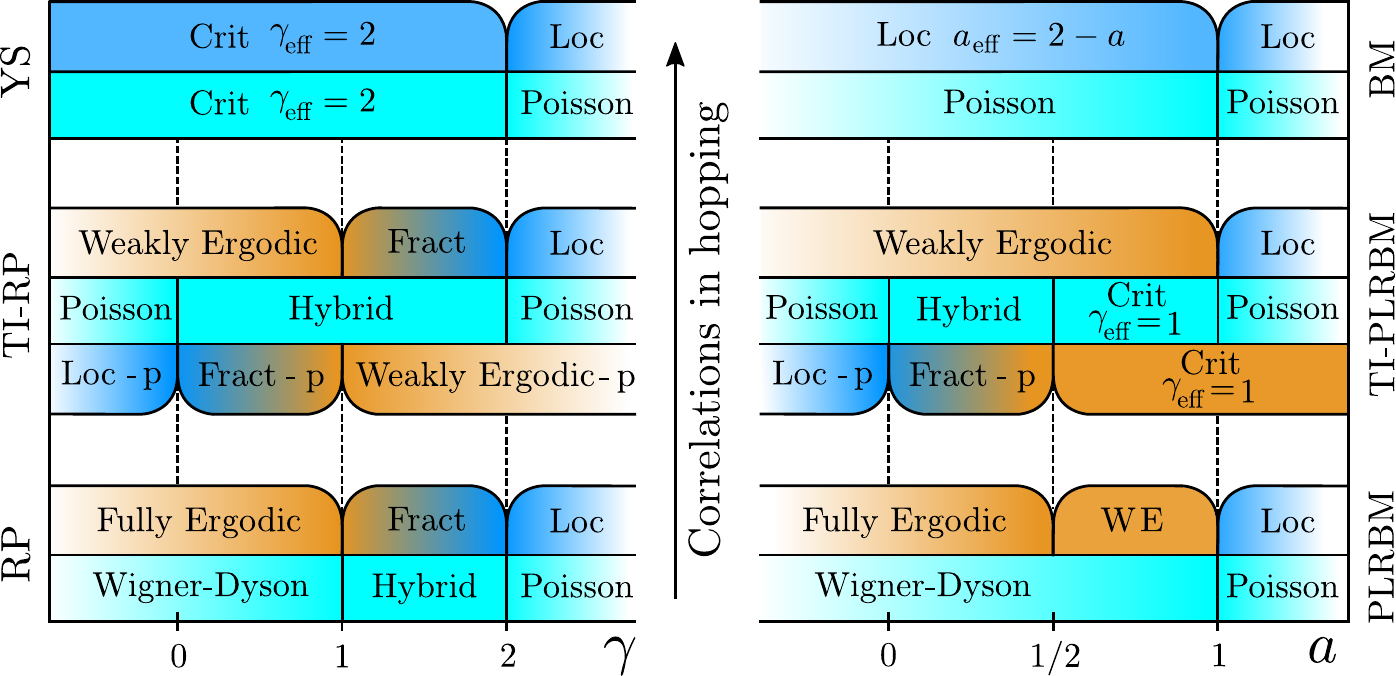}
\caption{{\bf Localization-delocalization phase diagrams for (left)~RP--
and (right)~PLBM--families of ensembles.} 
Additionally to coordinate-space diagrams (above horizontal lines) and level-statistic diagrams (in the middle)
for TI-models the momentum-space diagrams are shown below the lines. The phases
in TI-RP model are symmetric with respect
to duality point $\gamma=1$.
The type of spectral statistics (Wigner-Dyson, Poisson and hybrid) is indicated for each phase. Notice Poisson
level statistics in delocalized phases of TI- models in accordance with general principles formulated in Sec.~\ref{Sec:Principles}.
The increase of correlations in the hopping (from bottom to top) first destroys the fully-ergodic phase in all models,
making TI-systems weakly ergodic (WE), and then (in YS and BM models) localizes wave functions in the coordinate space.
}
\label{Fig:loc-deloc_diagram}
\end{figure}

A remarkable feature of random TI models is the presence of the Poisson
spectral statistics within the delocalized phase
(see Fig.~\ref{Fig:loc-deloc_diagram}). This goes against the
common wisdom that the Poisson statistics signals of localization.
The reason for such a behavior is that the Poisson spectral statistics
emerges in the parameter region where the states in the coordinate space are,
indeed, extended and weakly ergodic~\cite{ergodic_footnote} but those
in the momentum space are localized. The common wisdom assumes by default that
the states in momentum space are always chaotically extended.
The TI models introduced in this paper constitute a class of models where this
assumption fails. We formulated principles to identify the type of basis-invariant
spectral statistics if the statistics of eigenstates in the coordinate and in the
momentum spaces are known (see Fig.~\ref{Fig:loc-deloc_diagram}). One of them reads that the Poisson spectral statistics
emerges each time when the eigenstates are localized either in the coordinate or
in the momentum space~\cite{Special_basis_footnote}. These statement is checked
numerically in the paper.

The results of this paper allow us to formulate a new phase diagram which
is presented in Fig.~\ref{Fig:Cartoon}.
This figure shows a certain hierarchy of phases with respect to
the extent of ergodicity of eigenstates.
The {\it fully ergodic} (FE) phase corresponds to the Porter-Thomas
eigenfunction statistics if it is basis-independent. The corresponding level
statistics is Wigner-Dyson. We denote the states as {\it weakly ergodic} (WE)
if the eigenfunction support set~\cite{RRGAnnals, KravtsovSupportSet}
{\it in a given basis}
scales like the matrix size $N$ but the significant fraction of sites are not
populated. The eigenfunction statistics in the WE phase is
basis-dependent and deviates from the Porter-Thomas one. The non-ergodic extended,
{\it (multi)-fractal} (F) states are characterized by the support sets which
scale as $N^{D}$, where $0<D<1$. Finally the {\it localized} (L) states
correspond to $D=0$. Obviously, the ergodicity of the states decreases in
the following sequence:
\be\label{sequence}
FE\to WE\to F\to L \ .
\ee
The main result of this paper illustrated by Fig.~\ref{Fig:Cartoon} is that
with increasing correlations in the long-range hopping the sequence of
phases at a certain fixed disorder strength is that of Eq.~(\ref{sequence})
where some phases of this sequence may be missing, i.e. with increasing the correlations in
the long-range hopping the ergodicity of eigenstates progressively degrades.
Simultaneously, the lines of localization or ergodic transitions are shifted to
lower disorder.
At fully correlated long-range hopping the delocalized states in the
bulk of the spectrum disappear whatsoever.

It is important that the critical lines of all transitions bend to the left,
i.e. the states which are localized in the absence of correlations remain
 localized when the correlations are present. However the former ergodic
extended states may become weakly ergodic, non-ergodic or even localized in the
presence of correlations in the long-range hopping. This is the essence of
correlation-induced localization.


\section{Localization criteria for models with long-range hopping.}
The most generic \rev{
}free-particle Hamiltonian is defined as follows:
\begin{gather}\label{eq:ham}
H_{nm} =\ep_m \delta_{nm}+j_{nm} \ ,
\end{gather}
where $1\leq m,n\leq N$ are lattice sites,
$\ep_m$ are random on-site energies with zero mean $\la \ep_m\ra=0$ and
the variance $\la\ep_m^2\ra = \Delta^{2}$~\cite{Gaussian_disorder_footnote} represents
uncorrelated {\it diagonal} disorder. The (possibly correlated) hopping integrals
$j_{mn}=j_{nm}^*$ can be deterministic or random
and they are characterized by the averaged value $\langle j_{nm}\rangle$ and the variance
$\langle |j_{nm}|^{2}\rangle$.
Throughout the text we refer to the correlations in the hopping terms $j_{nm}$ simply as correlations.
For simplicity we restrict our consideration to $d=1$, unless stated otherwise.

The basic localization principle originally suggested by Mott~\cite{Mott1966} states that the
wave functions are localized (extended) when the disorder strength $\Delta$ is larger (smaller)
than the bandwidth $\Delta_{p}$ in the absence of diagonal disorder.
The results of this paper and other recent works
~\cite{Burin1989,Yuzbashyan_JPhysA2009_Exact_solution,Ossipov2013,Kravtsov_NJP2015,
Yuzbashyan_NJP2016,Borgonovi_2016,Kravtsov_Shlyapnikov_PRL2018,BogomolnyPLRBM2018,RP_R(t)_2018},
however, show that this principle should be reformulated.

Let us first consider the case when the spectrum of the off-diagonal part of the Hamiltonian
Eq.~(\ref{eq:ham}) $\rev{\hat{j}}=\hat{H}(\ep_{n}=0)$
is bounded both from above and from below in the limit $N\rightarrow\infty$, and we are
concerned with the
statistics of eigenstates in the bulk of the spectrum.
We claim that the Mott's criterion:
\be\label{eq:Mott_condition_in_r}
 \Delta\lesssim\Delta_p,\;\;\;\Rightarrow\;\; \text{weak ergodicity} \ ,
\ee
is the {\it sufficient condition}
for (at least weakly) {\it ergodic} {\it delocalization} when $\Delta$ is
smaller than
the bandwidth $\Delta_p$ of $\rev{\hat{j}}$. We will be using this criterion
in a weak sense as the condition:
\be\label{eq:Mott-weak}
\lim_{N\rightarrow\infty}\frac{\Delta}{\Delta_{p}}=0,\;\;\;\Rightarrow\;\; \text{weak ergodicity} \ ,
\ee
In the absence of correlations in $\rev{\hat{j}}$ the bandwidth is given by:
\be\label{width}
\Delta_{p}^{2} = \rev{\frac1N}\sum_{ n,m,\\ m\neq n}^{N}\langle |H_{nm}|^{2}\rangle.
\ee
For this particular case the criterion equivalent to
Eq.~(\ref{eq:Mott-weak}) was mentioned in Ref.~\cite{BogomolnyPLRBM2018}.

For correlated long-range hopping, specifically for the translation-invariant hopping described in Section~\ref{Sec:TI-models},
the spectrum of $\rev{\hat{j}}$ is often non-compact, with an infinite support
set in the energy space. In this case the bandwidth $\Delta_{p}$ should be defined as the width of the
energy domain where mean level spacing $\delta(N)$ takes a typical
value. The observation energy $E$ should be also chosen from inside of this domain.

As was explained in Introduction, weak ergodicity~\cite{Luca2014, RRGAnnals} defined in a given (e.g. coordinate) basis
does not
imply invariance of wave function statistics under basis rotation.
In some models (see, e.g.,~\cite{BogomolnyPLRBM2018}) weak ergodicity may survive beyond the
condition~\eqref{eq:Mott_condition_in_r},
showing that~\eqref{eq:Mott_condition_in_r} is only the sufficient but not the necessary condition of weak ergodicity
and, thus, $\Delta = \Delta_p$ is the lower bound for the ergodic transition between
the weakly ergodic extended phase and the non-ergodic phases (localized or extended).

The criterion of {\it localization} suggested for systems with long-range hopping by
Levitov~\cite{Levitov1989,Levitov1990} following Anderson's ideas of locator expansion, reads:
\be\label{eq:Anderson_principle_in_r}
\delta_R > |j_R|,\,\,\,\Rightarrow\;\; {\rm localization}.
\ee
The key point of~\cite{Levitov1989,Levitov1990} is that one should compare the mean level spacing $\delta_R \sim \Delta/R^d$
of a $d$-dimensional system at a certain length scale $1\lesssim |m-n|\sim R\lesssim N$ with the width of a resonance governed by the average absolute value of hopping integrals
$j_{R}$ within the same scale.
Then most eigenstates (except measure zero) are localized if~\eqref{eq:Anderson_principle_in_r}
holds for \rev{
almost all} $R$.
Indeed, in order to find the eigenstates one can use the perturbation theory in the small
parameter $j_R/\delta_R$.
The inequality~\eqref{eq:Anderson_principle_in_r} means convergence of the perturbation
series and thus localization. \VEKn{A more strict condition
\be\label{convergecond}
\frac{|j_R|}{\delta_{R}}< R^{-\epsilon},\quad \epsilon>0
\ee
 as $R\rightarrow\infty$ implies convergence
of the series $\sum_{R}|j_R|$.
For random matrices the corresponding criterion of
convergence reads as follows:
\bea\label{converge}
&&\lim_{N\rightarrow\infty}S/\Delta <\infty,\;\;\;\Rightarrow {\rm localization},
\\ \nonumber
&&S=  \frac{1}{N}\sum_{n,m, n\neq m}\langle|j_{nm}
|\rangle .
\eea
}

If the criterion~(\ref{eq:Anderson_principle_in_r}) is violated, both a multifractal
~\cite{Kravtsov_NJP2015} and a weakly ergodic~\cite{BogomolnyPLRBM2018} extended phases
may emerge. More surprisingly, violation of~\eqref{eq:Anderson_principle_in_r} does not
exclude localization either, provided that the hopping integrals are correlated.
Indeed, the presence of correlations cannot destroy localization if the condition~\eqref{eq:Anderson_principle_in_r} is fulfilled and the perturbation series is
convergent. Under this condition the main contribution to the eigenfunction amplitude
away from the localization center comes from the first-order perturbation theory which
knows nothing about correlations in the hopping matrix elements.
The situation changes completely when the perturbation theory diverges.
In this case all orders in perturbation theory contribute to the eigenfunction amplitude
on an equal footing and correlations come into play. As recent examples~\cite{Yuzbashyan_NJP2016,Yuzbashyan_JPhysA2009_Exact_solution,Ossipov2013,
Borgonovi_2016,
Kravtsov_Shlyapnikov_PRL2018} show, the effect of correlations when~\eqref{eq:Anderson_principle_in_r}
is violated may be localization of states which were extended in the absence of correlations.
These examples, in which hopping is deterministic, prove that~\eqref{eq:Anderson_principle_in_r}
is a {\it sufficient} but not a necessary condition of localization.

The Anderson and Mott criteria may be made {\it sufficient and necessary} criterion of localization by means of the
{\it matrix inversion trick} described in Section~\ref{Sec:Matrix_inv}. This trick converts the initial Shr\"odinger problem into a
{\it family} of equivalent problems with modified Hamiltonians $\hat{H}_{{\rm eq}}(E_{0})$ parameterized by a
continuous parameter $E_{0}\propto N^{\beta}$. The effective disorder strength \rev{
saturating
Eq.~(\ref{eq:Anderson_principle_in_r}) to an equality} is a function of this parameter $\beta$. The true border of the localized
phase then corresponds to the optimal $\beta$ that minimizes \rev{this} effective disorder.

The domains of validity of~\eqref{eq:Mott_condition_in_r} and~\eqref{eq:Anderson_principle_in_r}
are in general non-complimentary.
This is the reason why the non-ergodic extended phase may exist~\cite{Kravtsov_NJP2015} in the parameter region where
neither~\eqref{eq:Mott_condition_in_r} nor~\eqref{eq:Anderson_principle_in_r} holds true.

\section{Translation-invariant (Toeplitz) models.}\label{Sec:TI-models}
An important sub-class of the models~\eqref{eq:ham} is a family of translation-invariant (TI) models
with the hopping term $j_{nm}=j_{n-m}$~\cite{j_0_sum_e_p_footnote} depending only on the
directed distance $m-n$ between coupled sites~\cite{Periodic_bound_conds_footnote}
(Toeplitz random matrix models).
For such models a special role is played by the momentum basis.
An equivalent dual form of the Hamiltonian $H_{nm} = \ep_m \delta_{nm}+j_{n-m}$ in the momentum basis is
$H_{pq} = \tilde E_p \delta_{p,q}+\tilde J_{p-q}$, with
new on-site energies
\be\label{Ep}
\tilde E_p =\tilde E_p^* = \sum_{m=0}^{N-1}j_{m}e^{-2\pi i \frac{p m}N},
\ee
and new hopping integrals
\be\label{Jp}
\tilde J_p = \tilde J_{-p}^* = \frac{1}{N}\sum_{m=0}^{N-1}\ep_{m}e^{-2\pi i \frac{p m}N},
\ee
{\it exchanging their roles} after Fourier transforming (FT).
It allows one to generalize the Levitov's localization principle~\eqref{eq:Anderson_principle_in_r}
to the momentum space, with $|p-q|\sim P$
\be\label{eq:Anderson_principle_in_p}
\tilde\delta_P > \tilde J_P, \; \tilde\delta_P =
\frac{\big\langle |\tilde E_p - \tilde E_q|\big\rangle_{P}}{\rev{P}}, \; \tilde J_{P} = \big\langle |\tilde J_{p-q}|\big\rangle_{P} \ .
\ee

The dual counterpart of the weak ergodicity criterion~\rev{\eqref{eq:Mott-weak}}
for TI-models differs from~\rev{\eqref{eq:Mott-weak}} only by the opposite sign of the inequality:
\be\label{eq:ALT_Delta_p}
\rev{
\lim_{N\rightarrow\infty}\frac{\Delta_p}{\Delta}=0,}
\;\;\;\Delta_p = \Big<\max_{p,p'} |\tilde E_p - \tilde E_{p'}|\Big>.
\ee

\rev{Equations~~(\ref{eq:Mott-weak}}, \ref{eq:Anderson_principle_in_r},
\ref{eq:Anderson_principle_in_p}, \ref{eq:ALT_Delta_p}) form the basis of
our qualitative analysis throughout the manuscript~\cite{Special_basis_footnote}.
Below we consider two families of random matrix models as examples and
show the effect of correlations on their localization properties
applying two dual pairs of these localization-delocalization principles.

\section{Principles of level statistics.}\label{Sec:Principles}
It is frequently taken for granted that the level and eigenfunction statistics
are in one-to-one correspondence: delocalized states correspond to the
Wigner-Dyson level statistics and the localized ones correspond to
the Poisson level statistics. However, there is a big problem with this
statement: the level statistics is basis-invariant while the eigenfunction
statistics is generically basis-dependent. We will show in this paper that
in the random TI ensembles eigenfunctions may be extended in the coordinate
space and localized in the momentum one.

The key phenomenological principles to identify the level statistics in such a
situation are the following:
\begin{enumerate}
 \item[(i)] if there is a basis 
 in which the states are localized and uncorrelated with the
	corresponding eigenvalues then the level statistics is Poisson,
 \item[(ii)] the Wigner-Dyson (WD) level statistics hold if and
	only if the eigenfunction statistics is fully ergodic and basis-invariant, and
 \item[(iii)] when neither (i) or (ii) holds, the level statistics is of the
	hybrid nature that shares the features of both WD and the Poisson statistics.
\end{enumerate}

It follows from these principles that a coexistence of Poisson levels statistics and the delocalized (but not fully ergodic!) character of eigenstates is possible in a certain (e.g., coordinate) basis.
Indeed, according to Eqs.~(\ref{Ep}~-~\ref{eq:Anderson_principle_in_p}) in TI models at small enough disorder $\Delta$ there exist states localized in the momentum space (p-localized states). At the same time in the coordinate space these states must be extended due to the dual criterion~(\ref{eq:Mott-weak}). Then using the principles (i) and (ii) we come to the conclusion that the level statistics in TI models at small disorder must be Poisson, despite the states are delocalized in the coordinate space.

Below we consider the level statistics in TI models in more detail.

\section{Matrix inversion trick.}\label{Sec:Matrix_inv}
In this section we describe a useful trick that allows to reduce the problem
with the spectrum of the hopping matrix
$\rev{\hat j}$ which is unbounded from above or from below to the equivalent problem
with the bounded spectrum.

The initial problem is given by the Schr\"odinger equation:
\be\label{Schroed}
(E-\ep_{n})\,\psi_{E}(n)=\sum_{m=1\atop m\neq n}^{N}j_{n-m}\,\psi_{E}(m).
\ee
Let us introduce the matrix $M_{n-m}$:
\be\label{J}
\hat{M}=(\hat{1}+\hat{j}/E_{0})^{-1}=
\sum_{p}\frac{|p\rangle\,\langle p|}{1+\frac{\tilde{E}_{p}}{E_{0}}},
\ee
where \rev{$|p\rangle$ is the momentum-space basis vector and }
$E_{0}\propto N^{\beta}$ is a certain energy, such that \rev{
$(-E_0)$}
lies
outside the
spectrum of $\tilde{E}(p)$ of $\rev{\hat j}$ or inside the gaps in this spectrum.

\VEK{Singling out the diagonal term $M_{nn}=M_{0}$ and symmetrizing the hopping matrix,
 one arrives at the
equivalent eigenvalue/eigenfunction problem:
\be\label{equiv-psi}
 (\tilde{E} -\ep_{n})\,(E+E_{0}-\ep_{n})\,\psi_{E}(n)=
\sum_{m=1 \atop m\neq n}^{N} J_{nm}\,\psi_{E}(m),
\ee
where $\tilde{E}=E+E_{0}\,(1- M_{0}^{-1})$, and
\be\label{J_sym}
J_{nm}=-(E+E_{0}-\ep_{n})\frac{M_{n-m}}{M_{0}}\,(E+E_{0}-\ep_{m}).
\ee
}

The spectrum of the \rev{
}matrix $M_{n-m}$ ($n-m\neq 0$) is given
by Eq.~(\ref{J}), and it is bounded in the limit $N\rightarrow\infty$ even
if $\tilde{E}_{p}$ is
unbounded.
The same is true for the constant $M_{0}$.
For an unbounded $\tilde{E}_{p}\gg E_{0}$ one obtains at large $|n-m|$:
\be\label{expan-J}
M_{n-m}\approx \rev{\frac{E_0}N} \sum_{p} \frac{ e^{2\pi i(n-m)\,p/N}}{\tilde{E}_{p}},
\ee
in contrast to
\be\label{j-p}
j_{n-m}=\rev{\frac1N}\sum_{p}\tilde{E}_{p}\,e^{2\pi i(n-m)\,p/N}.
\ee

Eq.~(\ref{expan-J},~\ref{j-p}) will be useful to prove the
duality discovered in Ref.~\cite{Kravtsov_Shlyapnikov_PRL2018}.
However, the main idea of introducing the matrix inversion trick is applying the
localization/delocalization criteria Eqs.~(\ref{eq:Mott-weak},~\ref{converge}) to
a new hopping matrix \VEK{$J_{nm}(E_{0})$, Eq.~(\ref{J_sym}), and to find the true borders of the
localized and ergodic
 phases by optimization over $E_{0}$}.

In the next Section\rev{s} we show how does it work for
\rev{RP-family, Sec.~\ref{Sec:RP}, and PLRBM-family, Sec.~\ref{Sec:PLBM}, of matrix ensembles.
}

\section{Rosenzweig-Porter family}\label{Sec:RP}
\subsection{Yuzbashyan-Shastry model.}
The simplest model with long-range {\it fully-correlated} hopping is the model with the constant
$j_{nm}=N^{-\gamma /2}$
\cite{Ossipov2013,Yuzbashyan_NJP2016,Yuzbashyan_JPhysA2009_Exact_solution}
which we will
refer to as
the Yuzbashyan-Shastry (YS) model.
It is a particular case of a wider
class of exactly
solvable random matrix ensembles with the rank-1 hopping matrix
$j_{nm}=g_{n}\,g^{*}_{m}$ suggested in Ref.
\cite{Yuzbashyan_NJP2016,Yuzbashyan_JPhysA2009_Exact_solution}.
We use this exactly solvable model to illustrate the
general method of identifying the eigenfunction and spectral
statistics developed in this paper.

In the absence of (diagonal) disorder in this model
the single zero-momentum ($p=0$)
level $\tilde E_0$ is decoupled from the degenerate set
of the rest of states in the momentum space:
\be\label{YS-spectr}
\tilde E_0 = N^{1-\gamma /2}, \;\;\tilde E_{p\ne0}=0 \ .
\ee
Thus for $\gamma<2$ the spectrum of $\rev{\hat j}$ is unbounded from above.
Then applying the matrix inversion trick Eq.~\rev{(\ref{J})} and taking
into account that $|0\rangle\langle 0| = N^{-1}$ and
$\sum_{p}|p\rangle\langle p|=\hat{1}$ one obtains for the
 matrix $\hat{M}$:
\be
\hat{M}=\hat{1}-\frac{N^{-\gamma/2}}{E_{0}+N^{1-\gamma/2}},
\ee
where $\hat{1}$ is the unit matrix and $E_{0}\propto N^{\beta}$.
Then the matrix elements
$M_{n\neq m}$ are
independent of $n,m$ and scale with $N$ like:
\be\label{scaling-M}
M_{n\neq m}\propto\left\{\begin{matrix} N^{-\gamma/2 -\beta}, & {\rm if} \;\;
\gamma>2(1-\beta) \cr N^{-1}, & {\rm if}\;\; \gamma<2(1-\beta). \end{matrix} \right.
\ee
\VEKn{The ratio of  the hopping matrix   $J_{nm}(E_{0})$
to effective diagonal disorder
\be\label{Delta_E0}
\Delta(E_{0})\sim N^{{\rm max}(0,\beta)}
\ee
of the equivalent problem, Eq.~(\ref{equiv-psi}),
 scales as:
\be\label{equiv-hopping}
\frac{\langle |J_{nm}(E_{0})|\rangle}{\Delta(E_{0})} \propto \left\{\begin{matrix}
N^{-\gamma/2-{\rm min}(0,\beta)},&
{\rm if}\;\;\gamma>2(1-\beta)\cr N^{-(1-{\rm max}(0,\beta))}, & {\rm if}\;\;
\gamma<2(1-\beta).
\end{matrix} \right.
\ee
}
As the result, the border line for \VEKn{Eq.~(\ref{convergecond})} with \VEK{$j_{nm}\Rightarrow
J_{nm}(E_{0})$}
is (see Fig.~\ref{Fig:loc-domains}(a)):
\be\label{gamma-beta-border}
\gamma(\beta)=\left\{\begin{matrix} 2(1-\beta), & {\rm if \;\;\beta\leq 0}
\cr 2, & {\rm if}\;\;\beta>0\end{matrix} \right..
\ee
The minimal value $\gamma_{{\rm min}}=2$ of $\gamma(\beta)$
is reached at
the optimal value of $\beta=\beta_{{\rm opt}}=0$. Thus we conclude that the
true border
 of localization for YS model is $\gamma=2$.

At $\beta=\beta_{{\rm opt}}=0$, Eq.~(\ref{equiv-hopping}) gives:
\VEKn{
\be\label{J-M-opt}
\left(\frac{\langle |J_{nm}|\rangle}{\Delta}\right)_{{\rm opt}} \propto \left\{\begin{matrix}
N^{-\gamma/2},&
{\rm if}\;\;\gamma>2\cr N^{-1}, & {\rm if}\;\;
\gamma\leq 2.
\end{matrix} \right.
\ee
}
Eq.~(\ref{J-M-opt}) implies that \VEKn{
  $(S/\Delta)_{{\rm opt}}\sim N^{0}$} for {\it all} $\gamma\leq 2$. It corresponds to the critical state
similar to the one in the point of Anderson transition on the Bethe lattice
\cite{Luca2014}. In many respects this state may be considered as the limiting
localized state which we refer to as the 'critically localized' state.

The absence of truly extended states in YS model can be further confirmed by
the Mott's criterion Eq.~(\ref{eq:Mott-weak}). Indeed, the spectrum of
$\rev{\hat{j}}$ given by Eq.~(\ref{YS-spectr}) consists of the $(N-1)$-fold
degenerate band and a single level. Thus the {\it typical} level spacing of
$\rev{\hat{j}}$ is exactly zero, the same as the corresponding bandwidth
$\Delta_{p}$. This means that the Mott's criterion of delocalization
Eq.~(\ref{eq:Mott-weak}) is never fulfilled.

We come to the conclusion that for YS model the
delocalized phase in the coordinate space
is absent, \rev{in agreement with the results in the literature}
\cite{Ossipov2013,Yuzbashyan_NJP2016,Yuzbashyan_JPhysA2009_Exact_solution},
despite infinitely long-ranged hopping integrals. This is the most spectacular
effect of destructive interference of long-range hopping trajectories
on Anderson localization.

\subsection{Rosenzweig-Porter ensemble.}
The destructive interference in long-range hopping is drastically sensitive to
correlations
in the hopping integrals. The best studied relative of YS model is the
Rosenzweig-Porter (RP) ensemble~\cite{RP,Kravtsov_NJP2015,
Pandey,BrezHik,Guhr,AltlandShapiro,ShapiroKunz,Shukla2000,Shukla2005,
Biroli_RP,Ossipov_EPL2016_H+V,Amini2017,vonSoosten2017phase,vonSoosten2017non,
Monthus,BogomolnyRP2018,RP_R(t)_2018}.

The Hamiltonian of the RP-ensemble takes the form~\eqref{eq:ham} with
{\it totally uncorrelated} hopping matrix elements $j_{nm}$ with zero mean and
the variance
$\la |j_{nm}|^2\ra = \Delta^2 N^{-\gamma}$
scaling with the matrix size $N$ in the same way as $|j_{nm}|^{2}$ in YS model.
The diagonal elements are characterized by $\la\ep_m^2\ra = \Delta^{2}$.

In contrast to YS model, there are three phases in RP model
\cite{Kravtsov_NJP2015}: fully ergodic (FE)
for $\gamma<1$,
fractal(F) for
$1<\gamma<2$ and localized (L) for $\gamma>2$, of which two (FE and F) are extended.
 They are separated by two phase transitions: the Anderson localization
transition (AT) at $\gamma=2$ and
the ergodic \rev{transition (ET)} at $\gamma=1$. At $\gamma=2$ the eigenfunctions are
critically localized like in the corresponding point of YS model, while at
$\gamma=1$ a different type of critical statistics emerges.

The level statistics of
RP-ensemble
~\cite{Kravtsov_NJP2015,Biroli_RP,Pandey,BrezHik,
Guhr,AltlandShapiro,ShapiroKunz,RP_R(t)_2018} is
of Wigner-Dyson form for $\gamma<1$ and Poisson for $\gamma>2$.
For $1<\gamma<2$ it shows the Wigner-Dyson-like level repulsion at
small 
level spacings $s_k = E_{n+k}-E_n<k\delta $
and the Poisson statistics at 
$s_k\gg k\,\delta $~\cite{RP_R(t)_2018}.
 Further on we refer to this level statistics as the {\it hybrid} one.

Low-energy level repulsion is well-represented by a so-called ratio-
or $r$-statistics, see Fig.~\ref{Fig:r-statistics}:
\be
r = \la\min\left(r_n,\frac{1}{r_n}\right)\ra \ , \quad r_n=\frac{E_{n}-E_{n-1}}{E_{n+1}-E_{n}}\ ,
\ee
showing the value $r\approx0.5307$ 
for Gaussian orthogonal ensemble (GOE), $r\approx0.5996$ 
for Gaussian unitary ensemble (GUE), and $r=2\ln2-1\simeq 0.3863$ for Poisson
level statistics~\cite{Bogomol2013}.

We would like to stress once again that
despite the $r$-statistics is widely used to locate the localization transition,
it is not capable to distinguish between the WD level statistics of fully
ergodic phases and the hybrid statistics. In order to distinguish between
them one should study the spectral statistics at energy scale much larger
than the mean level spacing $\delta$. An example of such statistics is
the level number variance
$\langle n^{2}\rangle-\langle n\rangle^2 $ at a large average
number $\langle n\rangle\gtrsim \Gamma/\delta\sim N^{2-\gamma}$ of levels
in the studied energy interval (here $\Gamma\sim N^{1-\gamma}$
and $\delta\sim N^{-1}$)~\cite{Kravtsov_NJP2015} which for
the hybrid statistics should show the quasi-Poisson behavior
$\langle n^{2}\rangle-\langle n\rangle^2 =
\chi\, \langle n\rangle$ ($0<\chi<1$). Another possibility is to study
the probability distributions of
several consecutive level spacings
$s_k = E_{n+k}-E_{n}$~\cite{Mehta2004random,RP_R(t)_2018}.

A relevant measure of eigenfunction statistics is the distribution of amplitudes $P(|\psi_E(n)|^2)$ encoded in the {\it spectrum of fractal dimensions}~\cite{MirRev}
\be
f(\alpha) = 1-\alpha+\lim_{N\to\infty} \ln[P(|\psi_E(n)|^2=N^{-\alpha})]/\ln N \ .
\ee
As shown in~\cite{Kravtsov_NJP2015} for RP $f(\alpha)$ takes a simple linear form
\be\label{eq:RP_f(a)}
f(\alpha) = \left\{\begin{array}{lc}
1+(\alpha-\gamma)/2, & \max(0,2-\gamma)<\alpha<\gamma \\
-\infty, & \text{otherwise } \\
\end{array}
\right. \ ,
\ee
for $\gamma\geq1$ with an additional point $f(0)=0$ for $\gamma>2$.
The $f(\alpha)$ in the ergodic phase, $\gamma<1$, coincides with the
one at $\gamma=1$ and is represented by a single point $f(1)=1$,
see Fig.~\ref{Fig:RP_f(a)}.

Simple arguments based on the Mott's and Anderson's criteria,
Eq.~(\ref{eq:Mott-weak}), (\ref{eq:Anderson_principle_in_r}) allow
to locate the localized and ergodic phase without going into cumbersome mathematics.
Indeed, the Anderson's criterion $\delta\sim N^{-1}\gtrsim \langle|j_{nm}|\rangle
\sim N^{-\gamma/2}$ predicts localization for $\gamma>2$. At the same time, the
Mott's criterion $\Delta \lesssim \Delta_{p}$ predicts ergodic
delocalized states for $\gamma<1$, as $\Delta\sim 1$, and
$\Delta_{p}\sim N^{(1-\gamma)/2}$ according to Eq.~(\ref{width}).

\rev{
Note that using the spectral properties of the hopping term of the RP-model in its eigenbasis and the optimization
procedure for Eqs.~(\ref{eq:Mott-weak}) and
(\ref{eq:Anderson_principle_in_r}) one may show that the latter
are not only sufficient but also
necessary conditions for weak ergodicity and localization, respectively.
The corresponding analysis in the translation-invariant model is given in the next subsection.
}

\subsection{TI-RP ensemble and the coordinate-momentum
space duality}\label{Sec:TI-RP}
We extend the {\it Rosenzweig-Porter family} of random matrix ensembles by
introducing a translation-invariant RP ensemble (TI-RP). It is
described by \rev{
the Hamiltonian}
\be\label{Ham-TI-RP}
H_{nm} = \ep_m \delta_{nm}+j_{n-m},\;\;\;\;\la |j_{n-m}|^2\ra =
\Delta^2 N^{-\gamma} \ ,
\ee
with independent identically distributed (i.i.d.) Gaussian random (GR)
hopping integrals $j_{n-m}$ with zero mean and the variance independent
of $m$ and $n$.

Because of translation invariance $j_{nm}=j_{n-m}$, the TI-RP model possesses
the duality of properties in the coordinate and the momentum spaces
~\footnote{This duality is similar to the one in the Aubry-Andr{\'e}
model~\cite{AA}.}.
Indeed, FT of i.i.d. real $\{\ep_n\}$ or complex $\{j_n = j_{-n}^*\}$ GR numbers
are i.i.d. complex $\{\tilde J_p = \tilde J_{-p}^*\}$ or real $\{\tilde E_p\}$
GR numbers with the {\it dual} variances~\cite{SM}. Then from Eqs.~(\ref{Ep}),
(\ref{Jp}) one obtains:
\be\label{EP}
\big\langle \tilde E_p^2\big\rangle \simeq N\la |j_n|^2\ra \propto N^{1-\gamma},
\ee
\be
\big\langle |\tilde J_p|^2\big\rangle \simeq N^{-1}\, \la \ep_n^2\ra \ .
\ee
To avoid complications related to the correlations (degeneracy)
$\{\tilde E_p = \tilde E_{-p}\}$ of FT of real symmetric GR
$\{j_n = j_{-n}^* = j_n^*\}$
here and further we consider the class of Gaussian unitary ensembles.
For discussion of orthogonal class of ensembles see~\cite{SM}.

Thus the ratio $\big\langle |\tilde J_p|^2\big\rangle/\big\langle
\tilde E_p^2\big\rangle\propto N^{-\gamma_{p}}$ determines a parameter
$\gamma_p$
dual to $\gamma$ in the momentum space
\be\label{eq:TIRP_symmetry}
\gamma_p = 2-\gamma \ .
\ee
Eq.~(\ref{eq:TIRP_symmetry}) implies that in TI-RP model the phases in
the coordinate and
momentum spaces are symmetric with respect to the point $\gamma=1$
(see Fig.~\ref{Fig:loc-deloc_diagram}).

The Mott's criterion Eqs.~(\ref{eq:Mott_condition_in_r},~\ref{eq:Mott-weak}) ensures
existence of {\it weakly ergodic} phase for $\gamma<1$,
since according to Eq.~(\ref{EP}) $\Delta_{p}\sim N^{(1-\gamma)/2}$ and $\Delta \sim N^0$~\footnote{Here the width $\Delta_p$ of the energy domain where mean level spacing $\delta(N)$ takes a typical value coincides with the total hopping bandwidth.}.
This result is corroborated by numerics (see Fig.~\ref{Fig:RP_f(a)}(a)).

In order to use efficiently the Anderson localization criterion we first apply
the matrix-inversion trick. Consider first the case when $E_{0}\propto N^{\beta}
\gg \Delta_{p}\sim N^{(1-\gamma)/2}$. Then expanding Eq.~(\ref{J}) in $\hat{j}/E_{0}$ one obtains:
\be
\hat{M}=\hat{1}-\hat{j}/E_{0}.
\ee
The new hopping matrix \VEK {$J_{nm}(E_{0})$, Eq.~(\ref{J}),}
is estimated as:\VEKn{
\be\label{j-E0-TI-RP}
\frac{\langle |J_{nm}(E_{0})|\rangle}{\Delta(E_{0})} \sim \left\{\begin{matrix} |j_{nm}|= N^{-\gamma/2},&
{\rm for}\;\; \beta>0 \cr |j_{nm}|/E_{0}\sim N^{-\gamma/2 -\beta}, & {\rm for}\;\;\beta\leq 0.
\end{matrix} \right.
\ee}
Then the border line $\gamma(\beta)$ for the Anderson localization criterion
Eq.~(\ref{converge}) takes the form (see Fig.~\ref{Fig:loc-domains}) identical to Eq.~(\ref{gamma-beta-border})~\footnote{One can check that even at $\beta<0$ the assumption $E_{0}\sim
N^{\beta}\gg \Delta_{p}(\beta)\sim N^{(1-\gamma(\beta))}$ still holds true.}.

Thus we find the same optimal $\beta_{{\rm opt}}=0$ as for the YS model.
However, being substituted into Eq.~(\ref{j-E0-TI-RP}), this optimal $\beta$
results in a {\it different} optimal \VEK{ $\langle |J_{nm}^{({\rm opt})}|\rangle =
\langle |J_{nm}(E_{0}\sim N^{\beta_{{\rm opt}}})|\rangle$ (cf. Eq.~(\ref{J-M-opt})):
\be\label{J-opt-TI-RP}
\langle |J_{nm}^{({\rm opt})}|\rangle\rev{/\Delta(E_{0})} \sim N^{-\gamma/2}.
\ee}
With the optimal \VEK{ $\hat{J}^{({\rm opt})}$} Eq.~(\ref{converge}) becomes
the {\it necessary and sufficient} criterion of localization.
Thus we conclude that in the TI-RP model the localized phase
in the coordinate space
corresponds \rev{only} to $\gamma>2$. Numerics fully confirms this conclusion (see Fig.~\ref{Fig:RP_f(a)}(c)).
Due to the duality Eq.~(\ref{eq:TIRP_symmetry}) the
localized phase in the momentum space (which corresponds to the ballistic
propagation) is realized for $\gamma<0$.

\VEK{
To establish the character of wave function statistics in the remaining interval
$1<\gamma<2$ in the coordinate space and in the dual interval $0<\gamma<1$
in the momentum space we apply the Mott's criterion to the equivalent
problem Eq.~(\ref{equiv-psi}). The bandwidth
$\Delta_{p}(\beta)$
for this problem determined by Eqs.~(\ref{width}), (\ref{j-E0-TI-RP}) is given by:
\be\label{deltaP}
\Delta_{p}(\beta)\sim \left\{\begin{matrix}N^{(1-\gamma)/2 -\beta},&{\rm if}\;\;
(1-\gamma)/2 <\beta\leq 0\cr N^{(1-\gamma)/2},&{\rm if}\;\;\beta>0 \end{matrix}
\right.
\ee

The bandwidth is small $\Delta_{p}(\beta)\ll 1$ in the entire
region $\gamma>1$ and $\beta>0$, and thus the borderline for the Mott's
criterion is $\gamma(\beta)=1$ for all $\beta>0$. For $\beta\leq 0$
the domain of validity of the Mott's criterion seems to be wider, as the
bandwidth increases with decreasing $\beta$. This is, however, not true.
The reason is that the l.h.s. of Eq.~(\ref{equiv-psi}) is sign definite
$(E-\ep_{n})^{2}\geq 0$ at $E_{0}=N^{\beta}\ll 1$. For the usual
Schr\"odinger equation this corresponds to the energy $E$ outside the band
of on-site energies or on the border of it. If, in addition, the
hopping matrix bandwidth is small, the true eigenstates will be either
absent (as in band insulator) or localized as in the Lifshitz tail.
We conclude that the special structure of l.h.s. of Eq.~(\ref{equiv-psi}) prohibits
extended states in the region $\beta\leq 0$. Thus optimization over $\beta$
helps to establish a true border line $\gamma=1$ of the WE states in
the TI-RP model. In the region $1<\gamma<2$ in the coordinate space and the dual region $0<\gamma<1$ in the momentum space wave functions are neither
ergodic nor localized, i.e. they are {\it non-ergodic extended}. Numerics
confirms (see Fig.5b) that they are {\it fractal}, like in the RP model.}

A non-trivial property that follows from the above analysis which is also
confirmed by numerics, see
Figs.~\ref{Fig:loc-deloc_diagram} and~\ref{Fig:RP_f(a)})
is that the sequence of phases in the {\it coordinate space}
of RP and TI-RP ensembles \rev{and the positions of phase transitions} are {\it the same} \rev{ 
along with the spectra of fractal dimensions (see Fig.~\ref{Fig:RP_f(a)}).}
The only difference is that {\it fully} ergodic phase is not realized
for the TI-RP ensembles, as due to duality Eq.~(\ref{eq:TIRP_symmetry})
the phases in the coordinate and in the momentum space
never coincide at the same value of the disorder parameter $\gamma$.



\begin{figure}[t]
\includegraphics[width=0.9\columnwidth]{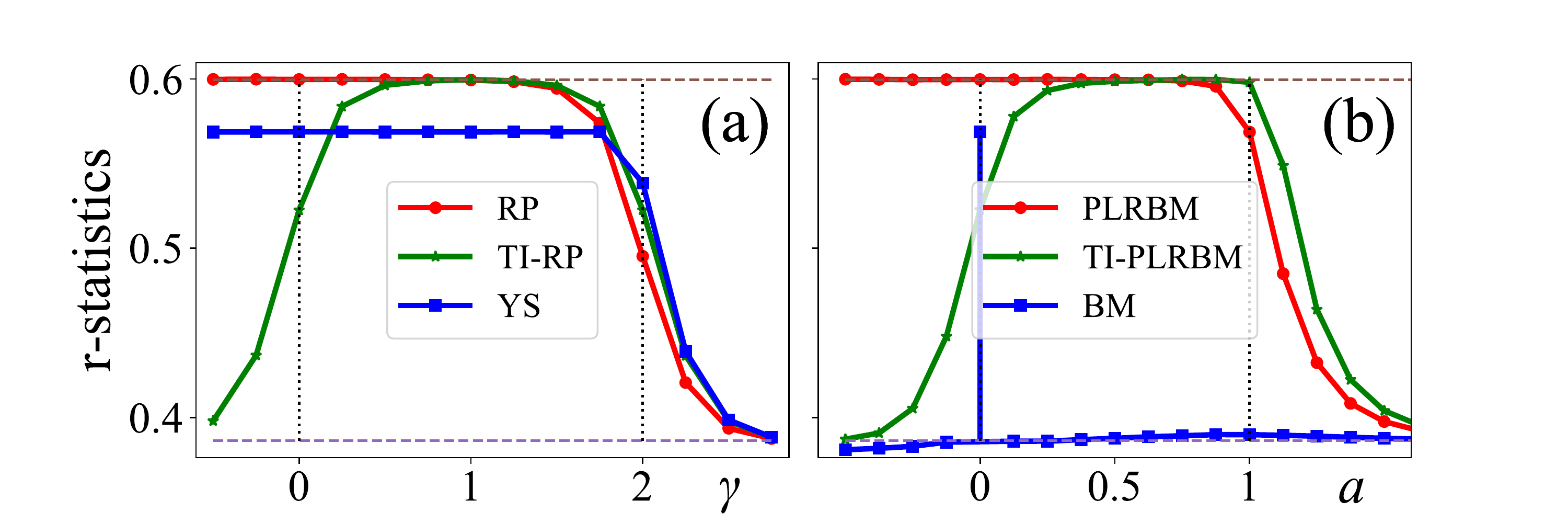}
\caption{{\bf $r$-statistics (average level-spacing ratio) for (a)~RP--
and (b)~PLBM--families of models} numerically calculated for Random Matrix
Ensembles of {\it unitary} symmetry for the system size $N=2^{14}$
and $N_r = 10^3$ disorder realizations.
In both cases deterministic models (YS and BM) show only the localized
or the critical behavior,
while TI-models (TI-RP and TI-PLRBM) demonstrate delocalized behavior in a
finite range of parameters turning to
Poisson statistics both at small and large hopping integrals,
corresponding to localization in real and momentum space.
Upper (lower) horizontal line shows the $r$-values for
Wigner-Dyson (Poisson) statistics.
Right (left) vertical line shows the Anderson localization
transition in real (momentum) space for TI-models.
}
\label{Fig:r-statistics}
\end{figure}

\begin{figure*}[t]
\centering
\includegraphics[width=0.75\textwidth]{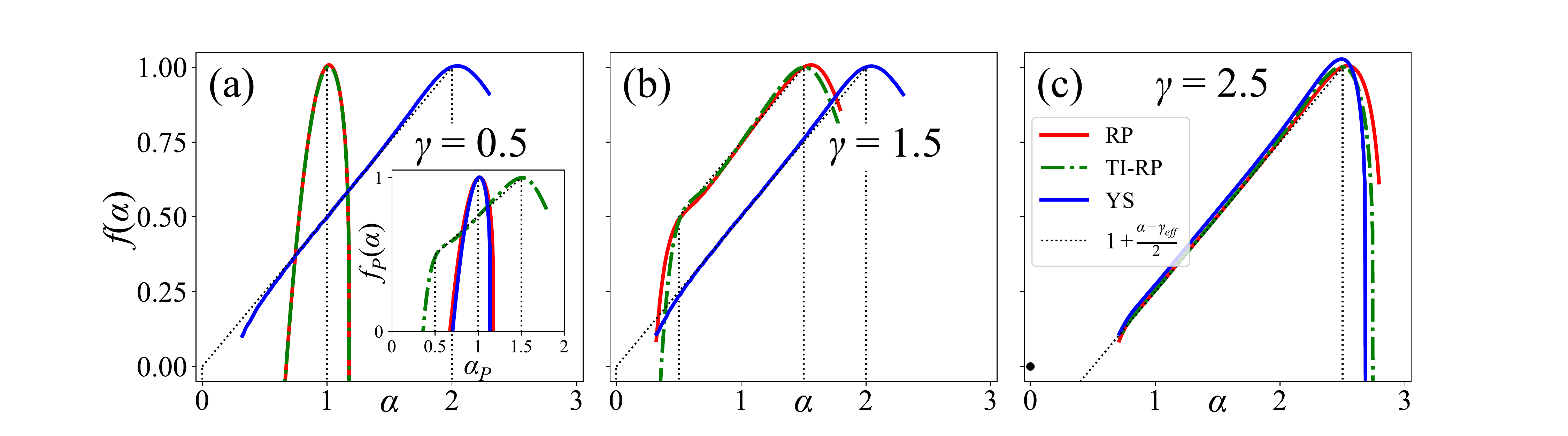}
\caption{{\bf Spectrum of fractal dimensions $f(\alpha)$ for the Rosenzweig-Porter family of models} (RP, TI-RP, YS) for (a)~$\gamma=0.5$, (b)~$1.5$, (c)~$2.5$ numerically extrapolated from system sizes $N=2^{9}\ldots 2^{14}$ with $N_r = 10^3$ disorder realizations in each.
Dashed lines show analytical predictions~\eqref{eq:RP_f(a)} for $f(\alpha)$.
(inset)~Spectrum of fractal dimensions in the momentum space $f_p(\alpha_p)$ with analytical predictions~\eqref{eq:RP_f(a)} (dashed lines) and $\gamma_p=2-\gamma$ for TI-RP, demonstrating the difference between RP and TI-RP ensembles in their delocalized phases.
}
\label{Fig:RP_f(a)}
\end{figure*}

In agreement with the principles formulated in Section~\ref{Sec:Principles}, the level statistics of
TI-RP is {\it symmetric} with respect to the dual point
$\gamma=\gamma_p=1$, see Fig.~\ref{Fig:r-statistics}. It
shows the hybrid behavior (the same as for RP in the interval $1<\gamma<2$)
in the entire interval $0<\gamma<2$ and the Poisson behavior outside of it.

Note that for $\gamma<0$ the Poisson level statistics coexists
(because of localization in the momentum space) with the
weakly ergodic delocalized wave function statistics in the coordinate space.
This is fully confirmed by numerics presented in Fig.~\ref{Fig:r-statistics}.
In contrast to RP model, the Wigner-Dyson level statistics in TI-RP model
do not occur, as the eigenfunction statistics in the coordinate and in the
momentum spaces never coincide.

\begin{figure*}[t]
\includegraphics[width=0.75\textwidth]{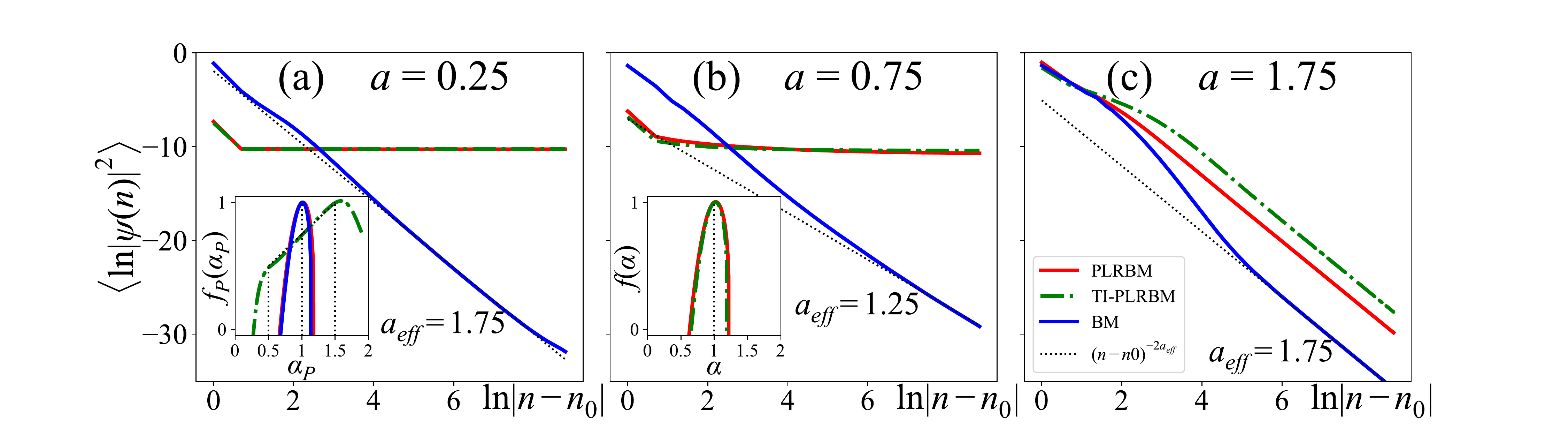}
\caption{{\bf Average $\langle \ln(|\psi_E(n)|^{2}\rangle=
\ln|\psi_{E}(n)|^{2}_{{\rm typ}}$
for power-law banded matrix family} (PLRBM, TI-PLRBM, BM) for different
exponents in the power-law decay of hopping (a)~$a=0.25$, (b)~$0.75$, (c)~$1.75$
numerically
calculated for the system size $N=2^{14}$ and $N_r = 10^3$ disorder realizations.
All models are power-law localized for $a>1$, while for $a<1$ only
BM shows
localization with effective exponent $a_{\rm eff}=2-a$.
Dashed lines show analytical prediction~\eqref{eq:PLBM_WF_space_scaling} of this
power-law decay.
(inset)~(a) spectrum of fractal dimensions in the momentum space $f_p(\alpha_p)$
with
analytical predictions~\eqref{eq:RP_f(a)} (dashed lines)
demonstrating the difference between PLRBM and TI-PLRBM ensembles in their
delocalized phases $a<1$; (b) spectrum of fractal dimensions in the
coordinate space for TI-PLBRM coinciding with that of PLBRM for $1/2<a<1$.
}
\label{Fig:PLBM_WF_space_scaling}
\end{figure*}

\section{Power-law banded matrix family.}\label{Sec:PLBM}
The next family we consider is the one of the power-law random banded matrices
(PLRBM)~\cite{MirFyod1996,MirRev}. The Hamiltonian of the conventional
(fully random) PLBRM is of the form of~\eqref{eq:ham},
with $\la j_{nm}\ra=0$ and $\la |j_{nm}|^2\ra = [1+(|n-m|/b)^{2a}]^{-1}$.
Its fully correlated counterpart, to which we refer further as the
Burin-Maksimov (BM) model~\cite{Burin1989}, is characterized by the {\it deterministic} sign-fixed
power-law decaying hopping integrals~\cite{Burin1989,Malyshev2000,
Malyshev_PRL2003,Malyshev2004,Malyshev2005,Borgonovi_2016,Kravtsov_Shlyapnikov_PRL2018}
$j_{nm} = j_0(1-\delta_{nm})/|n-m|^a$.

PLRBM shows ALT at $a=1$, with ergodic states for $a<1$ and
localized states for $a>1$.
The parameter $b$ matters only at the transition point $a=1$ and determines
the strength of multifractality~\cite{MirRev,MirFyod1996}.

By contrast, the BM-model demonstrates the {\it power-law localization} for most of the states
(except measure zero) not only at $a>1$,
but also at $a<1$~\cite{Kravtsov_Shlyapnikov_PRL2018} with an
intriguing symmetry of the exponents in the power-law decay of wave functions.

The level statistics of PLRBM is of the Wigner-Dyson form
 at least for $a<1/2$ and Poisson for $a>1$~\cite{MirFyod1996,MirRev}.
 Recently it has been shown~\cite{BogomolnyPLRBM2018} that in PLRBM the
wavefunction statistics is not Porter-Thomas for $1/2<a<1$ \rev{(see also Fig.~\ref{Fig:PLBRM-distr}(a-d))} implying the presence of weakly
ergodic phase in this interval.
In contrast, for the BM-model the level statistics is always Poisson,
except for an integrable point $a=0$ coinciding with the YS-model with $\gamma=0$,
where the statistics is critical, see right panel in Fig.~\ref{Fig:r-statistics}.

Both power-law models have a built-in spatial structure. Therefore the
eigenstate statistics allows more detailed characterization than in RP-family.
Indeed, considering the typical decay of the wave function intensity
$|\psi_E(n)|^2$ with the distance $|n-n_0|$~\cite{Periodic_bound_conds_footnote}
from its maximal value $|\psi_E(n_0)|^2$, one finds at large distances:
\be\label{eq:PLBM_WF_space_scaling}
|\psi_E(n)|^2_{typ}\equiv \exp\lb\la \ln|\psi_E(n)|^2\ra\rb \sim |n-n_0|^{-\mu} \ ,
\ee
with $\mu=2a$ for $a>1$ both in PLRBM and in the BM-model by the
perturbation theory. At $a<1$
the fully random model shows $\mu=0$, while the deterministic one
gives $\mu=2a_{{\rm eff}}=2\,(2-a) $, as shown numerically
in~\cite{Kravtsov_Shlyapnikov_PRL2018}, see also Fig.~\ref{Fig:PLBM_WF_space_scaling}.

A random TI relative of the PLRBM model, namely TI-PLRBM, is described
by $H_{nm} = \ep_m \delta_{nm}+j_{n-m}$,
with i.i.d. GR hopping integrals with zero mean and the variance:
\be
\la |j_{n-m}|^2\ra = (1-\delta_{nm})/|n-m|^{2a}.
\ee

In the momentum space both BM and TI-PLRBM ensemble are characterized by
i.i.d. GR hopping integrals which can be found from Eq.~(\ref{Jp}):
$\tilde J_{p-q}$ with
\be\label{JP-PLBRM}
\big\langle |\tilde J_p|^2\big\rangle \simeq \Delta^2/N.
\ee
The momentum-space on-site energies $\tilde{E}_{p}$ (which coincide with the spectrum
of the corresponding $\rev{\hat{j}}$) are given by Eq.~(\ref{Ep})
and depend
crucially on correlations in the hopping matrix elements.

For TI-PLRBM the spectrum of $\tilde{E}_{p}$ is random with zero mean and
the variance:
\be\label{EP-PLBRM}
\langle |\tilde{E}_{p}|^{2}\rangle=\sum_{m}\langle |j_{n-m}|^{2}\rangle \sim \left\{\begin{matrix} N^{1-2a},
&{\rm if}\;\;a<1/2 \cr
\ln N,& {\rm if}\;\;a=1/2 \cr
1,&{\rm if}\;\;a>1/2. \end{matrix} \right..
\ee
In contrast, for BM model with fully correlated hopping $j_{0}\,|n-m|^{-a}$
($a\neq 0$) one obtains:
\be \label{eq:h_p_a-1}
\tilde E_{p}/(2 j_0) \simeq \zeta_a + A_a
\left(\frac{N}{|p|}\right)^{1-a},\; \text{ for } |p|\ll N \ ,
\ee
\be \label{eq:h_p_a-1_q}
\tilde E_{p}/(2 j_0) \simeq \tilde E_{\min} + B_a
\left(\frac{2q}{N}\right)^{2}, \;\text{ for } |N/2-p|\ll N. \
\ee
$\zeta_a$ is the Riemann zeta-function $\zeta_a$, and dimensionless
constants $A_a$, $B_a$, and $\tilde E_{\min}$ given in~\cite{SM}.

One can see that the spectrum $\tilde{E}_{p}$ for TI-PLBRM is either bounded
($a>1/2$)
or unbounded from both sides ($a\leq 1/2$). In contrast, for BM model the spectrum,
while also bounded for $a>1$, is unbounded {\it only from one side}
for all $a<1$.

This difference appears to have crucial consequences for the
eigenfunctions statistics.

\subsection{Wave function statistics in BM model}
In this section we consider the wave function statistics of BM model
in the coordinate space.
Before employing the Mott and Anderson localization/delocalization
criteria to BM model at $a<1$ we have to define the effective bandwidth of a
highly stretched spectrum $\tilde{E}_{p}$ in this case. Eqs.~(\ref{eq:h_p_a-1}),
(\ref{eq:h_p_a-1_q}) show that the {\it typical} level spacing $\delta(N)=
d\tilde{E}_{p}/dp \sim N^{-1}$ corresponds to $|p|\sim N\rev{\sout{/2}}$ and $|N/2-p|\sim \rev{N}$ .
The corresponding $\tilde{E}_{p} \sim 1$ \rev{
gives}
the right estimation of the
effective bandwidth:
\be\label{width-BM}
\Delta_{p}^{({\rm eff})}\sim 1,
\ee
for {\it typical states} of BM problem with $a<1$. The remaining part of the spectrum
of $\tilde{E}_{p}$ has an increasing mean level spacing up to the maximal level spacing
of the order of $\delta_{{\rm max}}\sim N^{1-a}$ at $\tilde{E}_{p}\sim N^{1-a}$.
This part of spectrum, as well as the properties of the separate state
 in the YS model, requires a special study \cite{Malyshev2000,
Malyshev_PRL2003,Malyshev2004,Malyshev2005}. In this paper we limit ourselves by
the case when the energy $E\sim 1$ lies inside of the band of typical states.

For $a>1$, the spectrum is bounded with the bandwidth of order 1, so that
Eq.~(\ref{width-BM}) is valid for all $a$.

Eq.~(\ref{width-BM}) implies that the Mott's delocalization criterion is
never fulfilled in the sense of Eq.~(\ref{eq:Mott-weak}) and thus ergodic
delocalization is nowhere guaranteed.

To apply the Anderson localization criterion Eq.~(\ref{converge}) we first
compute the ``inverted'' hopping matrix $\hat{M}(E_{0})$ given by Eq.~(\ref{J}) with
$E_{0}\sim N^{\beta}$.
We start by the case $a>1-\beta$, where $|\tilde{E}_{p}|/E_{0} \ll 1$, and
the analysis may be carried out similar to the case of TI-RP model. One obtains:
\be\label{Jeff-BM_1}
\VEKn{\frac{\langle |J_{nm}(E_{0})|\rangle}{\Delta(E_{0})}}\sim \left\{\begin{matrix} j_R,& {\rm if}
\;\;\beta>0\cr N^{-\beta}\,j_R,&{\rm if} \;\; 1-a<\beta<0\end{matrix} \right.,
\ee
where $j_R=R^{-a}$.

For $a>1$, the sum $S(E_0)\rev{=N^{-1}\sum_{n,m\atop n\neq m}\langle|j_{nm}
|\rangle}$ in Eq.~(\ref{converge}) converges and one obtains:
\be\label{S1}
\VEKn{\frac{S(E_{0})}{\Delta(E_{0})}}=\left\{\begin{matrix}N^{0},& {\rm if}\;\;\beta>0\cr
N^{-\beta},& {\rm if}\;\;1-a<\beta<0 \end{matrix} \right.
\ee
For $1-\beta<a<1$ and $\beta>0$ one obtains
\VEKn{$S(E_{0})/\Delta(E_{0})\sim N^{1-a}$}.

Now consider the case $a<1-\beta$, where $\tilde{E}_{p}/E_{0}\gg 1$ and
\rev{the matrix inversion trick, }Eq.~(\ref{expan-J}), applies. Then we obtain:
\be\label{eq:M_n_res}
\frac{M_{n-m}}{E_{0}} = C_1 e^{-\varkappa |n-m|}+C_2\frac{(1-\delta_{nm})}
{j_{0}\,|n-m|^{2-a}}.
\ee
where dimensionless constants $C_1$, $C_2$, and $\varkappa$ can be
found in~\cite{SM}.

Notice that due to inverted position of $\tilde{E}_{p}$ in Eq.~(\ref{expan-J})
compared to Eq.~(\ref{j-p}) a new exponent:
\be\label{a_eff}
a_{{\rm eff}}=2-a
\ee
emerges in the place of $a$.

With this modification, Eq.~(\ref{Jeff-BM_1}), takes the form:
\be\label{Jeff-BM_2}
\VEKn{\frac{\langle |J_{nm}(E_{0})|\rangle}{\Delta(E_{0})}} \sim \left\{\begin{matrix} N^{2\beta}\,
R^{-a_{{\rm eff}}},&
{\rm if}
\;\;1-a>\beta>0\cr N^{\beta}\,R^{-a_{{\rm eff}}} ,&{\rm if} \;\;
\beta<0\end{matrix} \right..
\ee
The corresponding expression for \VEKn{ $S(E_{0})/\Delta(E_{0})$ } reads as follows:
\be
\VEKn{\frac{S(E_{0})}{\Delta(E_{0})}}=\left\{\begin{matrix} N^{2\beta},& {\rm if}\;\;a<1;\,1-a>\beta>0
\cr N^{\beta},&{\rm if}\;\;a<1;\,\beta<0 \cr
N^{a-1+\beta},&{\rm if}\;\;a>1;\,\beta<1-a \end{matrix} \right..
\ee
\begin{figure}[t]
\includegraphics[width=1\columnwidth]{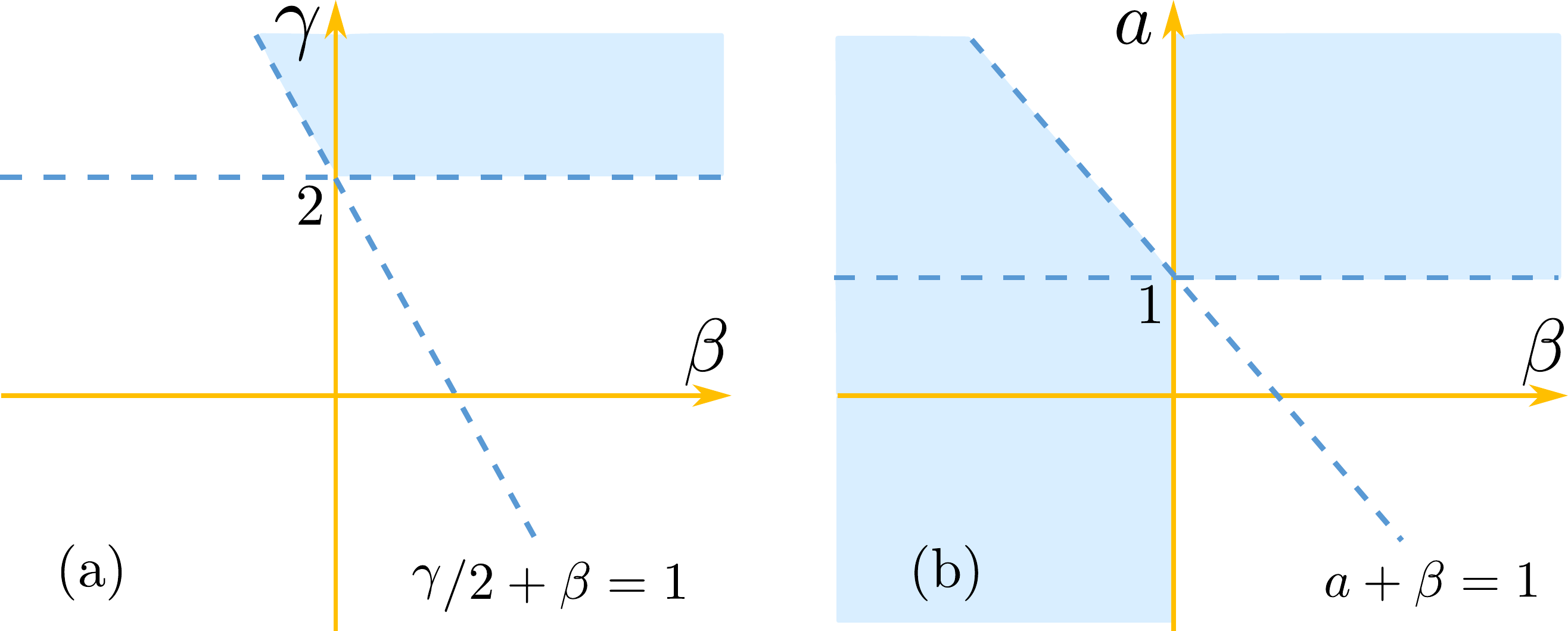}
\caption{{\bf Optimized Anderson localization criterion.}
Domains of validity of Eq.~(\ref{converge}): (a) for the YS and TI-RP models
and (b) for BM model. The true domain of the localized states corresponds to
the optimal $\beta_{{\rm opt}}$ at which the domain of validity of
Eq.~(\ref{converge}) is the
widest possible. In both cases $\beta_{{\rm opt}}=0$ but for the BM model
the typical states are localized
at all values of the exponent $a$, while for YS and TI-RP models the truly
localized
states exist only at $\gamma>2$.}
\label{Fig:loc-domains}
\end{figure}
As the result of this analysis we obtain a diagram which shows the domains
on the plane $(\beta,a)$ where the {\it sufficient} condition for localization,
Eq.~(\ref{converge}), is fulfilled in BM ensemble (see Fig.~\ref{Fig:loc-domains}(b)). The optimal $\beta$ corresponds
to the widest domain of validity of Eq.~(\ref{converge}) which
is the {\it true} domain of
the localized phase. In Fig.~\ref{Fig:loc-domains} such domains are shown in blue
for the BM model (Fig.~\ref{Fig:loc-domains}(b)) and for the YS and TI-RP models
(Fig.~\ref{Fig:loc-domains}(a)). It is seen that for BM model at the optimal
$\beta_{{\rm opt}}=0$ the states inside the band Eq.~(\ref{width-BM})
are localized at {\it all} values of the parameter $a$. The corresponding spectral
statistics is therefore Poisson.

Note that the above analysis corresponding to the energy $E\sim 1$ does
not apply to the states outside the effective band Eq.~(\ref{width-BM}) (i.e.
in the stretched part of
the spectrum) though the method itself is applicable everywhere.

\subsection{Duality of the exponent $\mu$ in BM model.}
The fact that the typical states in BM ensemble are localized at all values of $a$
can be traced back to the divergence of the spectrum $\tilde{E}_{p}$ and as a consequence
to a possibility to use the matrix inversion trick and Eq.~\rev{
\eqref{J}} to derive
Eq.~(\ref{eq:M_n_res}) for $a<1$ and define $a_{{\rm eff}}$ as in Eq.~(\ref{a_eff}).
The same Eq.~(\ref{eq:M_n_res}) helps to prove the duality of the exponents $\mu(a)$,
Eq.~(\ref{eq:PLBM_WF_space_scaling}), of the power-law localization:
\be\label{dual_mu}
\mu(a)=\mu(2-a),
\ee
suggested recently in Ref. \cite{Kravtsov_Shlyapnikov_PRL2018}.

At $a>1$ the conventional representation of the eigenproblem
\be\label{eq:BM_eigenproblem}
E \psi_E(n) = \ep_n \psi_E(n) + j_0 \sum_{m\ne n} \psi_E(m)/|m-n|^a
\ee
gives the standard solution from the locator expansion method~\cite{Anderson1958}.
It converges to the power-law decaying large-distance asymptotics of the eigenstate
\be\label{eq:BM_psi_asymp}
|\psi_E(n)|_{{\rm typ}} \sim 1/|n-n_0|^{a} \ , \quad |n-n_0|\gg 1 \ ,(a>1),
\ee
with the decay exponent coinciding with the matrix element exponent $a$
due to the convergence of the sum in the r.h.s. of Eq.~\eqref{eq:BM_eigenproblem}.
Note that this method applies to all PLBRM-model at $a>1$ irrespectively
to their hopping correlations.

At $a<1$ the usual locator expansion fails to converge. However,
the locator expansion can be applied to the equivalent eigenproblem
Eq.~(\ref{equiv-psi}) with the ``inverted'' hopping matrix given by
Eq.~(\ref{eq:M_n_res}). The latter contains the power-law decaying part
characterized by the exponent $a_{{\rm eff}}=2-a$.

Thus by the same token as Eq.~(\ref{eq:BM_psi_asymp}) we obtain a similar
expression
for $|\psi_{E}(n)|_{{\rm typ}}$ at $a<1$ but with $a_{{\rm eff}}=(2-a)$ instead of
$a$. Thus we conclude that:
\be
\mu=\left\{ \begin{matrix} 2a,& {\rm if}\;\;a>1 \cr
2\,(2-a),&{\rm if}\;\;a<1 \end{matrix}\right. ,
\ee
which proves the duality Eq.~(\ref{dual_mu}).

Note that the duality concerns only the exponents in the
power-law tail of the localized wave
functions and not to the amplitude of this tail and the length scale at which the
power-law asymptotics sets in (see Fig.~\ref{Fig:PLBM_WF_space_scaling}).
\subsection{TI-PLBRM ensemble}
 \begin{figure*}[t]
 \centering{
 \includegraphics[width=0.75\textwidth]{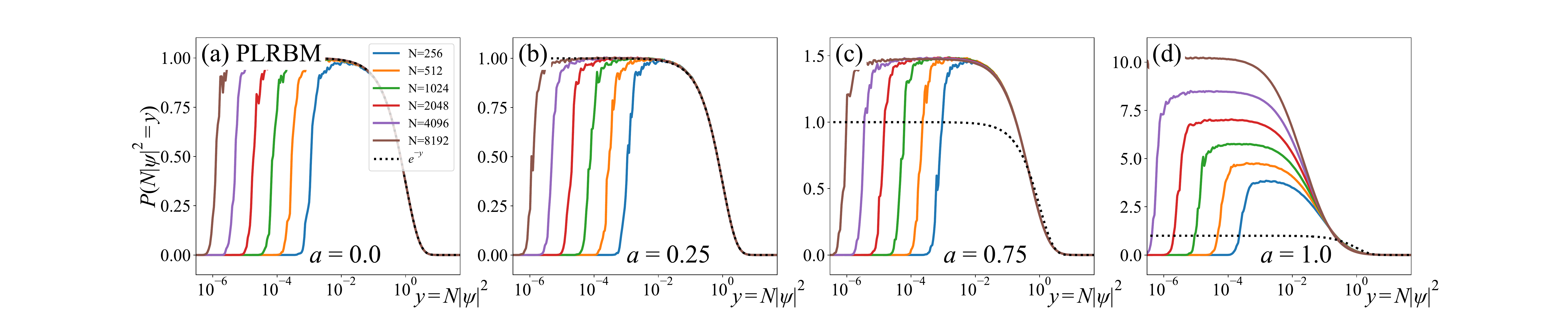}
 \includegraphics[width=0.75\textwidth]
{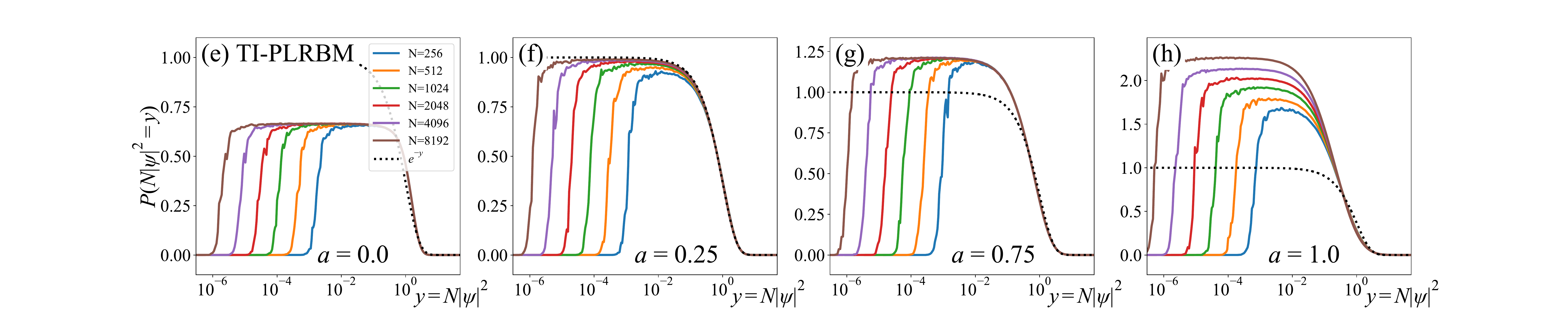}
 }
 \caption{{\bf Comparison of eigenstate probability distributions}
$P(N|\psi|^2=y)$ in (a-d) PLRBM and
(e-h) TI-PLRBM models (solid lines for different system sizes $N$)
with the GUE Porter-Thomas distribution $P(y) = e^{-y}$ (black dashed line).}
 \label{Fig:PLBRM-distr}
\end{figure*}

Finally, we turn to statistics of eigen-data for the translation-invariant PLBRM.
We start by the statistics of wave functions in the {\it momentum} space.
Using Eqs.~(\ref{JP-PLBRM},~\ref{EP-PLBRM}) one finds:
\be
\frac{\langle |J_{p}|^{2}\rangle}{\langle |\tilde{E}_{p}|^{2}\rangle}\propto
N^{-\gamma_{p}^{({\rm eff})}},
\ee
where
\be\label{gammaP}
\gamma_{p}^{({\rm eff})}=\left\{\begin{matrix} 2(1-a),& {\rm if}\;\;a<1/2\cr
1,& {\rm if}\;\;a>1/2 \end{matrix} \right.
\ee
Now the problem of wave function statistics of TI-PLBRM ensemble
in the momentum space is reduced to the
one for \rev{TI-}RP ensemble in the coordinate space with the replacement
$\gamma\rightarrow\gamma_{p}^{({\rm eff})}$. The result is presented on
Fig.~\ref{Fig:loc-deloc_diagram} where we denote by $Loc-p$, $Frac-p$ and
$Crit\,\gamma_{{\rm eff}}=1$ the localized, fractal and critical phase at the
point of ergodic transition, respectively.

As for the phases in the coordinate space, one can easily see from the Mott's
criterion Eq.~(\ref{eq:Mott-weak}) with
$\Delta_{p}\sim \sqrt{\langle |\tilde{E}_{p}|^{2}\rangle}$ that for $a<1/2$ (when
$\Delta_{p}\propto N^{1/2 -a}$ according to Eq.~(\ref{EP-PLBRM})) there is a weakly
ergodic (WE) extended state. The Anderson localization criterion
Eq.~(\ref{converge}) ensures existence of the localized phase for $a>1$.

\VEK{The most difficult is the characterization of phase in the interval $1/2 <a<1$.
The matrix inversion also does not help to establish a true border for
the localized phase. The reason is that for the case $a>1/2$ we are concerned, the
bandwidth $\Delta_{p}\sim \tilde{E}_{p}\sim 1$ and thus $E_{0}$
could be chosen only to be $E_{0}\gtrsim 1$ which according to Eq.~(\ref{j-E0-TI-RP})
leaves
the scaling of effective hopping matrix unchanged.

In this situation we can rely only on numerics presented in
Fig.~\ref{Fig:PLBM_WF_space_scaling}(b) and Fig.~\ref{Fig:PLBRM-distr}.
Indeed, Fig.~\ref{Fig:PLBM_WF_space_scaling}(b)
demonstrates a narrow $f(\alpha)$ in the coordinate space of TI-PLBRM at $a=0.75$
which
is typical for weakly ergodic states and identical
to the one of non-TI PLBRM for the same value of $a$. }

Additionally, Fig.~\ref{Fig:PLBRM-distr}(g)
shows much smaller deviation from the Porter-Thomas distribution
of the distribution function of $|\psi_{E}|^{2}$ for
TI-PLBRM at $a=0.75$ than that for the known multifractal case of $a=1$ of
PLBRM on
Fig.~\ref{Fig:PLBRM-distr}(d).
This makes us to conclude that in the interval $1/2<a<1$ of TI-PLBRM a weakly ergodic
phase is realized, as well as for the non-TI PLBRM.

\VEK{ We note also that in contrast to the TI-RP case,
the phases in the TI-PLBRM are {\it not} symmetric with respect to the
point $a=1/2$. The reason is that the typical off-diagonal matrix elements
have a power-law decay in the coordinate space of TI-PLBRM ensemble, while in the momentum space they have no structure, similar to the coordinate space of \rev{TI-RP} ensemble. In contrast, for TI-RP ensemble the typical off-diagonal
elements are similar in a sense that they do not have structure
both in the coordinate and in the momentum space. This
allows to apply the duality relation Eq.~(\ref{eq:TIRP_symmetry}) and
establish the symmetry of phases with respect to $\gamma=1$. }

The level statistics in TI-PLRBM (see Fig.~\ref{Fig:loc-deloc_diagram})
can be easily identified using the three principles formulated Sec.~\ref{Sec:Principles}
and checked numerically, see Fig.~\ref{Fig:r-statistics}. It is
Poisson at $a<0$ and $a>1$, a hybrid one at $0<a<1/2$ and an
ergodically-critical, like in
the point $\gamma=1$ of RP ensemble,
at $1/2<a<1$. As mentioned above, the latter interval in TI-PLRBM
corresponds to $\gamma_{p}^{({\rm eff})}=1$. Therefore the behavior
of $\langle n^{2} \rangle-\langle n\rangle^2 = \chi \,\langle n\rangle$
(with level compressibility $0<\chi\leq 1$) should be quasi-Poisson,
as in the point $\gamma=1$ of ergodic transition in RP
ensemble~\cite{Kravtsov_NJP2015}. In the interval $0<a<1/2$
the hybrid character of level statistics follows from the lack
of basis invariance of the eigenfunction statistics
(see Fig.~\ref{Fig:loc-deloc_diagram}).

\section{Conclusion and discussions.}

The main result of this paper is the picture of
{\it correlation-induced localization} which is presented in Fig.~\ref{Fig:Cartoon} and
Fig.~\ref{Fig:loc-deloc_diagram}.
We demonstrate that
the correlations in long-range hopping may change drastically
the localization-delocalization phase diagram of many models
turning extended phase into the (multi)fractal or even localized one.

We show that the well-known localization principles~(\ref{eq:Mott_condition_in_r})
and~(\ref{eq:Anderson_principle_in_r}) are not complimentary and are in fact the
sufficient (but not the necessary) conditions for weakly ergodic delocalization
and localization, respectively. Thus they are not able to determine exact
bounds for localization/delocalization and may leave room for
non-ergodic delocalized phases. However, by applying the matrix inversion
trick and the optimization procedure
suggested in this paper one \rev{can make Eqs.~\eqref{eq:Mott-weak} and~\eqref{eq:Anderson_principle_in_r} also necessary conditions for ergodicity and localization and} may in certain cases determine exact phase diagram.
We believe that
the same arguments apply to the models with sign-alternating non-random
hopping integrals~\cite{Garttner2015disorder,Cantin2018}
 as the general applicability of the matrix inversion method is
related with the presence of the finite gaps or edges of the spectrum,
but the consideration of these models is out of scope of the current paper.

We suggest a natural extension of the class of models with correlated
long-range hopping integrals by introducing the translation-invariant (TI)
random matrix models, where hopping integrals are fully correlated {\it along}
the diagonals but the correlations {\it between} the diagonals are absent. We
identify phases with different character of localization/delocalization in
these models both in the coordinate and in the momentum spaces together with
the spectral statistics. The results are summarized in
Fig.~\ref{Fig:loc-deloc_diagram}.
It is shown that at moderately weak disorder the delocalized
phases in TI-models are never fully ergodic, as
the eigenfunction statistics are different in the momentum and the coordinate
spaces.

We formulated the principles to identify the level statistics in the considered
models as belonging to the Wigner-Dyson, Poisson or the new hybrid class.
In particular, the spectral statistics is Poisson if the eigenfunction statistics
shows localization in either coordinate or in the momentum basis. This implies that
in TI random matrix ensembles the spectral statistics
may be Poisson despite the states are extended in the coordinate basis (but localized
in the momentum one). This statement is confirmed by numerics.

The considered models with fully-correlated, TI-correlated, and uncorrelated
hopping can be easily generalized to a whole class of matrix models with
the continuous parameter correlations in the hopping integrals, see
Fig.~\ref{Fig:Cartoon}.
Indeed, in TI-models hopping integrals are fully correlated {\it along} the
diagonals, while in uncorrelated models they are statistically independent.
In between one can consider, e.g., the models with hopping terms in each
diagonal to be correlated in such a way that $M_1$ elements in each diagonal
are equal, where $M_1$ changes from $1$ for uncorrelated models to $N$ for
TI-models.
In the similar way one can consider the continuous correlation parameter from
TI- to fully-correlated models.
Indeed, as in TI-models the correlations {\it between} the diagonals are absent
one can partially add them by considering blocks of $M_2$ diagonals to
be equal, where $M_2$ changes from $1$ for TI-models to $N$ for fully-correlated
models.
The overall number of independent hopping terms in the matrix that scales as
$N^2/(M_1 M_2)$ can be considered as a continuous hopping correlation parameter.
Of course this is not a unique way to include hopping correlations in
uncorrelated models, but this kind of correlations is natural as it emerges
in physical models such as the RKKY where hopping integrals deterministically
oscillate as a function of $|n-m|$ with the period incommensurate with the
lattice constant.

Within the same method, for the random matrix models with deterministic
power-law decaying hopping integrals $j_{n-m}\sim |n-m|^{-a}$
we confirm that both for $a>1$ (see~\cite{Malyshev2000,Malyshev_PRL2003,
Malyshev2004,Malyshev2005}) and $a<1$~\cite{Kravtsov_Shlyapnikov_PRL2018}
the typical states are localized with the power-law tails $\psi_{E_n}(m)
\sim |n-m|^{-a_{{\rm eff}}}$ at $a\ne 0$ and analytically prove the duality
$a_{{\rm eff}} = \max(a,2-a)$.

It is also worth noticing that our arguments are not restricted only to the
one-dimensional case, $d=1$. Recent work~\cite{Cantin2018} has shown the
presence of localized states for isotropic deterministic power-law hopping
with $a<d=3$ in three-dimensional cubic lattices. This problem might be
understood within our formalism.

Another intriguing direction of research is the interplay between correlations
in the hopping integrals and in the on-site energies. As recently shown the
correlated on-site ``disorder'' (quasi-periodic potential~\cite{AA}) may
destroy localization and produce a whole bunch of (multi)fractal phases
depending on the power $a$~\cite{Deng_AA2018} in the BM-model with deterministic
power-law hopping integrals.

 Finally, the most challenging problem motivated by our paper is the
effect of correlations on Many Body Localization in the long-range
interacting models (see, e.g., \cite{Burin_Kagan1994,Yao_Lukin2014,
BurinPRB2015-1,BurinPRB2015-2,BurinAnnPhys2017,Moessner2017,
halimeh2017dynamical,homrighausen2017anomalous,luitz2018emergent,
Botzung_Muller2018,deTomasi2018,hashizume2018dynamical,tikhonov2018many}).

\begin{acknowledgments}
We are grateful to D.N.~Aristov, A.L.~Burin, R.~Moessner, and A.S.~Ovchinnikov
for stimulating discussions.
P. N. appreciates warm hospitality of the Max-Planck Institute for the Physics
of Complex Systems, Dresden, Germany, extended to him during his visits when
this work was done.
P.~N. acknowledges funding by the RFBR, Grant No. 17-52-50013, and the Foundation
for the Advancement to Theoretical Physics and Mathematics BASIS
Grant No. 17-11-107.
I.~M.~K. acknowledges the support of German Research Foundation (DFG)
Grant No. KH~425/1-1 and the Russian Foundation for Basic Research under Grant No. 17-52-12044.
V.E.K. is grateful to KITP of University of California at Santa Barbara
where the final part of this research was done. The support by the National
Science Foundation under Grant No. NSF PHY-1748958 is greatly appreciated.
\end{acknowledgments}

\bibliography{Lib}

\begin{thebibliography}{98}%
\makeatletter
\providecommand \@ifxundefined [1]{%
 \@ifx{#1\undefined}
}%
\providecommand \@ifnum [1]{%
 \ifnum #1\expandafter \@firstoftwo
 \else \expandafter \@secondoftwo
 \fi
}%
\providecommand \@ifx [1]{%
 \ifx #1\expandafter \@firstoftwo
 \else \expandafter \@secondoftwo
 \fi
}%
\providecommand \natexlab [1]{#1}%
\providecommand \enquote  [1]{``#1''}%
\providecommand \bibnamefont  [1]{#1}%
\providecommand \bibfnamefont [1]{#1}%
\providecommand \citenamefont [1]{#1}%
\providecommand \href@noop [0]{\@secondoftwo}%
\providecommand \href [0]{\begingroup \@sanitize@url \@href}%
\providecommand \@href[1]{\@@startlink{#1}\@@href}%
\providecommand \@@href[1]{\endgroup#1\@@endlink}%
\providecommand \@sanitize@url [0]{\catcode `\\12\catcode `\$12\catcode
  `\&12\catcode `\#12\catcode `\^12\catcode `\_12\catcode `\%12\relax}%
\providecommand \@@startlink[1]{}%
\providecommand \@@endlink[0]{}%
\providecommand \url  [0]{\begingroup\@sanitize@url \@url }%
\providecommand \@url [1]{\endgroup\@href {#1}{\urlprefix }}%
\providecommand \urlprefix  [0]{URL }%
\providecommand \Eprint [0]{\href }%
\providecommand \doibase [0]{http://dx.doi.org/}%
\providecommand \selectlanguage [0]{\@gobble}%
\providecommand \bibinfo  [0]{\@secondoftwo}%
\providecommand \bibfield  [0]{\@secondoftwo}%
\providecommand \translation [1]{[#1]}%
\providecommand \BibitemOpen [0]{}%
\providecommand \bibitemStop [0]{}%
\providecommand \bibitemNoStop [0]{.\EOS\space}%
\providecommand \EOS [0]{\spacefactor3000\relax}%
\providecommand \BibitemShut  [1]{\csname bibitem#1\endcsname}%
\let\auto@bib@innerbib\@empty
\bibitem [{\citenamefont {Anderson}(1958)}]{Anderson1958}%
  \BibitemOpen
  \bibfield  {author} {\bibinfo {author} {\bibfnamefont {P.~W.}\ \bibnamefont
  {Anderson}},\ }\href@noop {} {\bibfield  {journal} {\bibinfo  {journal}
  {Phys. Rev.}\ }\textbf {\bibinfo {volume} {109}},\ \bibinfo {pages} {1492}
  (\bibinfo {year} {1958})}\BibitemShut {NoStop}%
\bibitem [{\citenamefont {Wegner}(1980)}]{Wegner1980}%
  \BibitemOpen
  \bibfield  {author} {\bibinfo {author} {\bibfnamefont {F.}~\bibnamefont
  {Wegner}},\ }\href@noop {} {\bibfield  {journal} {\bibinfo  {journal}
  {Z.Phys.B}\ }\textbf {\bibinfo {volume} {36}},\ \bibinfo {pages} {209}
  (\bibinfo {year} {1980})}\BibitemShut {NoStop}%
\bibitem [{\citenamefont {A.Rodriguez}\ \emph {et~al.}(2011)\citenamefont
  {A.Rodriguez}, \citenamefont {L.J.Vasquez}, \citenamefont {K.Slevin},\ and\
  \citenamefont {R.A.Romer}}]{Roemer}%
  \BibitemOpen
  \bibfield  {author} {\bibinfo {author} {\bibnamefont {A.Rodriguez}}, \bibinfo
  {author} {\bibnamefont {L.J.Vasquez}}, \bibinfo {author} {\bibnamefont
  {K.Slevin}}, \ and\ \bibinfo {author} {\bibnamefont {R.A.Romer}},\
  }\href@noop {} {\bibfield  {journal} {\bibinfo  {journal} {Phys. Rev. B}\
  }\textbf {\bibinfo {volume} {84}},\ \bibinfo {pages} {134209} (\bibinfo
  {year} {2011})}\BibitemShut {NoStop}%
\bibitem [{\citenamefont {Levitov}(1989)}]{Levitov1989}%
  \BibitemOpen
  \bibfield  {author} {\bibinfo {author} {\bibfnamefont {L.~S.}\ \bibnamefont
  {Levitov}},\ }\href@noop {} {\bibfield  {journal} {\bibinfo  {journal}
  {Europhysics. Lett.}\ }\textbf {\bibinfo {volume} {9}},\ \bibinfo {pages}
  {83} (\bibinfo {year} {1989})}\BibitemShut {NoStop}%
\bibitem [{\citenamefont {Levitov}(1990)}]{Levitov1990}%
  \BibitemOpen
  \bibfield  {author} {\bibinfo {author} {\bibfnamefont {L.~S.}\ \bibnamefont
  {Levitov}},\ }\href@noop {} {\bibfield  {journal} {\bibinfo  {journal} {Phys.
  Rev. Lett.}\ }\textbf {\bibinfo {volume} {64}},\ \bibinfo {pages} {547}
  (\bibinfo {year} {1990})}\BibitemShut {NoStop}%
\bibitem [{\citenamefont {Mirlin}\ \emph {et~al.}(1996)\citenamefont {Mirlin},
  \citenamefont {Fyodorov}, \citenamefont {Dittes}, \citenamefont {Quezada},\
  and\ \citenamefont {Seligman}}]{MirFyod1996}%
  \BibitemOpen
  \bibfield  {author} {\bibinfo {author} {\bibfnamefont {A.~D.}\ \bibnamefont
  {Mirlin}}, \bibinfo {author} {\bibfnamefont {Y.~V.}\ \bibnamefont
  {Fyodorov}}, \bibinfo {author} {\bibfnamefont {F.-M.}\ \bibnamefont
  {Dittes}}, \bibinfo {author} {\bibfnamefont {J.}~\bibnamefont {Quezada}}, \
  and\ \bibinfo {author} {\bibfnamefont {T.~H.}\ \bibnamefont {Seligman}},\
  }\href@noop {} {\bibfield  {journal} {\bibinfo  {journal} {Phys. Rev. E}\
  }\textbf {\bibinfo {volume} {54}},\ \bibinfo {pages} {3221} (\bibinfo {year}
  {1996})}\BibitemShut {NoStop}%
\bibitem [{\citenamefont {Evers}\ and\ \citenamefont {Mirlin}(2008)}]{MirRev}%
  \BibitemOpen
  \bibfield  {author} {\bibinfo {author} {\bibfnamefont {F.}~\bibnamefont
  {Evers}}\ and\ \bibinfo {author} {\bibfnamefont {A.~D.}\ \bibnamefont
  {Mirlin}},\ }\href@noop {} {\bibfield  {journal} {\bibinfo  {journal} {Rev.
  Mod. Phys}\ }\textbf {\bibinfo {volume} {80}},\ \bibinfo {pages} {1355}
  (\bibinfo {year} {2008})}\BibitemShut {NoStop}%
\bibitem [{\citenamefont {Kravtsov}\ and\ \citenamefont
  {Muttalib}(1997)}]{KravtsovMuttalib1997}%
  \BibitemOpen
  \bibfield  {author} {\bibinfo {author} {\bibfnamefont {V.~E.}\ \bibnamefont
  {Kravtsov}}\ and\ \bibinfo {author} {\bibfnamefont {K.~A.}\ \bibnamefont
  {Muttalib}},\ }\href@noop {} {\bibfield  {journal} {\bibinfo  {journal}
  {Phys. Rev. Lett.}\ }\textbf {\bibinfo {volume} {79}},\ \bibinfo {pages}
  {1913} (\bibinfo {year} {1997})}\BibitemShut {NoStop}%
\bibitem [{\citenamefont {Aubry}\ and\ \citenamefont {Andr{\'e}}(1980)}]{AA}%
  \BibitemOpen
  \bibfield  {author} {\bibinfo {author} {\bibfnamefont {S.}~\bibnamefont
  {Aubry}}\ and\ \bibinfo {author} {\bibfnamefont {G.}~\bibnamefont
  {Andr{\'e}}},\ }\href@noop {} {\bibfield  {journal} {\bibinfo  {journal}
  {Ann. Israel Phys. Soc.}\ }\textbf {\bibinfo {volume} {3}},\ \bibinfo {pages}
  {18} (\bibinfo {year} {1980})}\BibitemShut {NoStop}%
\bibitem [{\citenamefont {Roati}\ \emph {et~al.}(2008)\citenamefont {Roati},
  \citenamefont {D'Errico}, \citenamefont {Fallani}, \citenamefont {Fattori},
  \citenamefont {Fort}, \citenamefont {Zaccanti}, \citenamefont {Modugno},
  \citenamefont {Modugno},\ and\ \citenamefont {Inguscio}}]{Cold-loc}%
  \BibitemOpen
  \bibfield  {author} {\bibinfo {author} {\bibfnamefont {G.}~\bibnamefont
  {Roati}}, \bibinfo {author} {\bibfnamefont {C.}~\bibnamefont {D'Errico}},
  \bibinfo {author} {\bibfnamefont {L.}~\bibnamefont {Fallani}}, \bibinfo
  {author} {\bibfnamefont {M.}~\bibnamefont {Fattori}}, \bibinfo {author}
  {\bibfnamefont {C.}~\bibnamefont {Fort}}, \bibinfo {author} {\bibfnamefont
  {M.}~\bibnamefont {Zaccanti}}, \bibinfo {author} {\bibfnamefont
  {G.}~\bibnamefont {Modugno}}, \bibinfo {author} {\bibfnamefont
  {M.}~\bibnamefont {Modugno}}, \ and\ \bibinfo {author} {\bibfnamefont
  {M.}~\bibnamefont {Inguscio}},\ }\href@noop {} {\bibfield  {journal}
  {\bibinfo  {journal} {Nature}\ }\textbf {\bibinfo {volume} {453}},\ \bibinfo
  {pages} {895} (\bibinfo {year} {2008})}\BibitemShut {NoStop}%
\bibitem [{\citenamefont {Richerme}\ \emph {et~al.}(2014)\citenamefont
  {Richerme}, \citenamefont {Gong}, \citenamefont {Lee}, \citenamefont {Senko},
  \citenamefont {Smith}, \citenamefont {Foss-Feig}, \citenamefont {Michalakis},
  \citenamefont {Gorshkov},\ and\ \citenamefont {Monroe}}]{Cold-dipole1}%
  \BibitemOpen
  \bibfield  {author} {\bibinfo {author} {\bibfnamefont {P.}~\bibnamefont
  {Richerme}}, \bibinfo {author} {\bibfnamefont {Z.-X.}\ \bibnamefont {Gong}},
  \bibinfo {author} {\bibfnamefont {A.}~\bibnamefont {Lee}}, \bibinfo {author}
  {\bibfnamefont {C.}~\bibnamefont {Senko}}, \bibinfo {author} {\bibfnamefont
  {J.}~\bibnamefont {Smith}}, \bibinfo {author} {\bibfnamefont
  {M.}~\bibnamefont {Foss-Feig}}, \bibinfo {author} {\bibfnamefont
  {S.}~\bibnamefont {Michalakis}}, \bibinfo {author} {\bibfnamefont {A.~V.}\
  \bibnamefont {Gorshkov}}, \ and\ \bibinfo {author} {\bibfnamefont
  {C.}~\bibnamefont {Monroe}},\ }\href@noop {} {\bibfield  {journal} {\bibinfo
  {journal} {Nature}\ }\textbf {\bibinfo {volume} {511}},\ \bibinfo {pages}
  {198} (\bibinfo {year} {2014})}\BibitemShut {NoStop}%
\bibitem [{\citenamefont {Jurcevic}\ \emph {et~al.}(2014)\citenamefont
  {Jurcevic}, \citenamefont {Lanyon}, \citenamefont {Hauke}, \citenamefont
  {Hempel}, \citenamefont {Zoller}, \citenamefont {Blatt},\ and\ \citenamefont
  {Roos}}]{Cold-dipole2}%
  \BibitemOpen
  \bibfield  {author} {\bibinfo {author} {\bibfnamefont {P.}~\bibnamefont
  {Jurcevic}}, \bibinfo {author} {\bibfnamefont {B.~P.}\ \bibnamefont
  {Lanyon}}, \bibinfo {author} {\bibfnamefont {P.}~\bibnamefont {Hauke}},
  \bibinfo {author} {\bibfnamefont {C.}~\bibnamefont {Hempel}}, \bibinfo
  {author} {\bibfnamefont {P.}~\bibnamefont {Zoller}}, \bibinfo {author}
  {\bibfnamefont {R.}~\bibnamefont {Blatt}}, \ and\ \bibinfo {author}
  {\bibfnamefont {C.~F.}\ \bibnamefont {Roos}},\ }\href@noop {} {\bibfield
  {journal} {\bibinfo  {journal} {Nature}\ }\textbf {\bibinfo {volume} {511}},\
  \bibinfo {pages} {202} (\bibinfo {year} {2014})}\BibitemShut {NoStop}%
\bibitem [{\citenamefont {Hung}\ \emph {et~al.}(2016)\citenamefont {Hung},
  \citenamefont {Gonzalez-Tudelac}, \citenamefont {Cirac},\ and\ \citenamefont
  {Kimble}}]{Cold-dipole3}%
  \BibitemOpen
  \bibfield  {author} {\bibinfo {author} {\bibfnamefont {C.-L.}\ \bibnamefont
  {Hung}}, \bibinfo {author} {\bibfnamefont {A.}~\bibnamefont
  {Gonzalez-Tudelac}}, \bibinfo {author} {\bibfnamefont {J.~I.}\ \bibnamefont
  {Cirac}}, \ and\ \bibinfo {author} {\bibfnamefont {H.~J.}\ \bibnamefont
  {Kimble}},\ }\href@noop {} {\bibfield  {journal} {\bibinfo  {journal} {PNAS}\
  }\textbf {\bibinfo {volume} {113}},\ \bibinfo {pages} {E4946} (\bibinfo
  {year} {2016})}\BibitemShut {NoStop}%
\bibitem [{\citenamefont {Nandkishore}\ and\ \citenamefont
  {Sondhi}(2017)}]{Rahul2017}%
  \BibitemOpen
  \bibfield  {author} {\bibinfo {author} {\bibfnamefont {R.~M.}\ \bibnamefont
  {Nandkishore}}\ and\ \bibinfo {author} {\bibfnamefont {S.~L.}\ \bibnamefont
  {Sondhi}},\ }\href@noop {} {\bibfield  {journal} {\bibinfo  {journal} {Phys.
  Rev. X}\ }\textbf {\bibinfo {volume} {7}},\ \bibinfo {pages} {041021}
  (\bibinfo {year} {2017})}\BibitemShut {NoStop}%
\bibitem [{\citenamefont {Burin}\ and\ \citenamefont
  {Kagan}(1994)}]{Burin_Kagan1994}%
  \BibitemOpen
  \bibfield  {author} {\bibinfo {author} {\bibfnamefont {A.~L.}\ \bibnamefont
  {Burin}}\ and\ \bibinfo {author} {\bibfnamefont {Y.}~\bibnamefont {Kagan}},\
  }\href@noop {} {\bibfield  {journal} {\bibinfo  {journal} {Zh. Eksp. Teor.
  Fiz.}\ }\textbf {\bibinfo {volume} {106}},\ \bibinfo {pages} {633} (\bibinfo
  {year} {1994})}\BibitemShut {NoStop}%
\bibitem [{\citenamefont {Yao}\ \emph {et~al.}(2014)\citenamefont {Yao},
  \citenamefont {Laumann}, \citenamefont {Gopalakrishnan}, \citenamefont
  {Knap}, \citenamefont {M\"uller}, \citenamefont {Demler},\ and\ \citenamefont
  {Lukin}}]{Yao_Lukin2014}%
  \BibitemOpen
  \bibfield  {author} {\bibinfo {author} {\bibfnamefont {N.~Y.}\ \bibnamefont
  {Yao}}, \bibinfo {author} {\bibfnamefont {C.~R.}\ \bibnamefont {Laumann}},
  \bibinfo {author} {\bibfnamefont {S.}~\bibnamefont {Gopalakrishnan}},
  \bibinfo {author} {\bibfnamefont {M.}~\bibnamefont {Knap}}, \bibinfo {author}
  {\bibfnamefont {M.}~\bibnamefont {M\"uller}}, \bibinfo {author}
  {\bibfnamefont {E.~A.}\ \bibnamefont {Demler}}, \ and\ \bibinfo {author}
  {\bibfnamefont {M.~D.}\ \bibnamefont {Lukin}},\ }\href@noop {} {\bibfield
  {journal} {\bibinfo  {journal} {Phys. Rev. Lett.}\ }\textbf {\bibinfo
  {volume} {113}},\ \bibinfo {pages} {243002} (\bibinfo {year}
  {2014})}\BibitemShut {NoStop}%
\bibitem [{\citenamefont {Burin}(2015{\natexlab{a}})}]{BurinPRB2015-1}%
  \BibitemOpen
  \bibfield  {author} {\bibinfo {author} {\bibfnamefont {A.~L.}\ \bibnamefont
  {Burin}},\ }\href@noop {} {\bibfield  {journal} {\bibinfo  {journal} {Phys.
  Rev. B}\ }\textbf {\bibinfo {volume} {91}},\ \bibinfo {pages} {094202}
  (\bibinfo {year} {2015}{\natexlab{a}})}\BibitemShut {NoStop}%
\bibitem [{\citenamefont {Burin}(2015{\natexlab{b}})}]{BurinPRB2015-2}%
  \BibitemOpen
  \bibfield  {author} {\bibinfo {author} {\bibfnamefont {A.~L.}\ \bibnamefont
  {Burin}},\ }\href@noop {} {\bibfield  {journal} {\bibinfo  {journal} {Phys.
  Rev. B}\ }\textbf {\bibinfo {volume} {92}},\ \bibinfo {pages} {104428}
  (\bibinfo {year} {2015}{\natexlab{b}})}\BibitemShut {NoStop}%
\bibitem [{\citenamefont {Gornyi}\ \emph {et~al.}(2017)\citenamefont {Gornyi},
  \citenamefont {Mirlin}, \citenamefont {Polyakov},\ and\ \citenamefont
  {Burin}}]{BurinAnnPhys2017}%
  \BibitemOpen
  \bibfield  {author} {\bibinfo {author} {\bibfnamefont {I.}~\bibnamefont
  {Gornyi}}, \bibinfo {author} {\bibfnamefont {A.}~\bibnamefont {Mirlin}},
  \bibinfo {author} {\bibfnamefont {D.}~\bibnamefont {Polyakov}}, \ and\
  \bibinfo {author} {\bibfnamefont {A.~L.}\ \bibnamefont {Burin}},\ }\href@noop
  {} {\bibfield  {journal} {\bibinfo  {journal} {Annalen der Physik}\ ,\
  \bibinfo {pages} {1600360}} (\bibinfo {year} {2017})}\BibitemShut {NoStop}%
\bibitem [{\citenamefont {Singh}\ \emph {et~al.}(2017)\citenamefont {Singh},
  \citenamefont {Moessner},\ and\ \citenamefont {Roy}}]{Moessner2017}%
  \BibitemOpen
  \bibfield  {author} {\bibinfo {author} {\bibfnamefont {R.}~\bibnamefont
  {Singh}}, \bibinfo {author} {\bibfnamefont {R.}~\bibnamefont {Moessner}}, \
  and\ \bibinfo {author} {\bibfnamefont {D.}~\bibnamefont {Roy}},\ }\href@noop
  {} {\bibfield  {journal} {\bibinfo  {journal} {Phys. Rev. B}\ }\textbf
  {\bibinfo {volume} {95}},\ \bibinfo {pages} {094205} (\bibinfo {year}
  {2017})}\BibitemShut {NoStop}%
\bibitem [{\citenamefont {Halimeh}\ and\ \citenamefont
  {Zauner-Stauber}(2017)}]{halimeh2017dynamical}%
  \BibitemOpen
  \bibfield  {author} {\bibinfo {author} {\bibfnamefont {J.~C.}\ \bibnamefont
  {Halimeh}}\ and\ \bibinfo {author} {\bibfnamefont {V.}~\bibnamefont
  {Zauner-Stauber}},\ }\href@noop {} {\bibfield  {journal} {\bibinfo  {journal}
  {Physical Review B}\ }\textbf {\bibinfo {volume} {96}},\ \bibinfo {pages}
  {134427} (\bibinfo {year} {2017})}\BibitemShut {NoStop}%
\bibitem [{\citenamefont {Homrighausen}\ \emph {et~al.}(2017)\citenamefont
  {Homrighausen}, \citenamefont {Abeling}, \citenamefont {Zauner-Stauber},\
  and\ \citenamefont {Halimeh}}]{homrighausen2017anomalous}%
  \BibitemOpen
  \bibfield  {author} {\bibinfo {author} {\bibfnamefont {I.}~\bibnamefont
  {Homrighausen}}, \bibinfo {author} {\bibfnamefont {N.~O.}\ \bibnamefont
  {Abeling}}, \bibinfo {author} {\bibfnamefont {V.}~\bibnamefont
  {Zauner-Stauber}}, \ and\ \bibinfo {author} {\bibfnamefont {J.~C.}\
  \bibnamefont {Halimeh}},\ }\href@noop {} {\bibfield  {journal} {\bibinfo
  {journal} {Physical Review B}\ }\textbf {\bibinfo {volume} {96}},\ \bibinfo
  {pages} {104436} (\bibinfo {year} {2017})}\BibitemShut {NoStop}%
\bibitem [{\citenamefont {Luitz}\ and\ \citenamefont
  {Lev}(2019)}]{luitz2018emergent}%
  \BibitemOpen
  \bibfield  {author} {\bibinfo {author} {\bibfnamefont {D.~J.}\ \bibnamefont
  {Luitz}}\ and\ \bibinfo {author} {\bibfnamefont {Y.~B.}\ \bibnamefont
  {Lev}},\ }\href@noop {} {\bibfield  {journal} {\bibinfo  {journal} {Phys.
  Rev. A}\ }\textbf {\bibinfo {volume} {99}},\ \bibinfo {pages} {010105}
  (\bibinfo {year} {2019})}\BibitemShut {NoStop}%
\bibitem [{\citenamefont {Botzung}\ \emph {et~al.}(2018)\citenamefont
  {Botzung}, \citenamefont {Vodola}, \citenamefont {Naldesi}, \citenamefont
  {M\"uller}, \citenamefont {Ercolessi},\ and\ \citenamefont
  {Pupillo}}]{Botzung_Muller2018}%
  \BibitemOpen
  \bibfield  {author} {\bibinfo {author} {\bibfnamefont {T.}~\bibnamefont
  {Botzung}}, \bibinfo {author} {\bibfnamefont {D.}~\bibnamefont {Vodola}},
  \bibinfo {author} {\bibfnamefont {P.}~\bibnamefont {Naldesi}}, \bibinfo
  {author} {\bibfnamefont {M.}~\bibnamefont {M\"uller}}, \bibinfo {author}
  {\bibfnamefont {E.}~\bibnamefont {Ercolessi}}, \ and\ \bibinfo {author}
  {\bibfnamefont {G.}~\bibnamefont {Pupillo}},\ }\href@noop {} {\enquote
  {\bibinfo {title} {Algebraic localization from power-law interactions in
  disordered quantum wires},}\ } (\bibinfo {year} {2018}),\ \Eprint
  {http://arxiv.org/abs/1810.09779} {arXiv:1810.09779} \BibitemShut {NoStop}%
\bibitem [{\citenamefont {de~Tomasi}(2019)}]{deTomasi2018}%
  \BibitemOpen
  \bibfield  {author} {\bibinfo {author} {\bibfnamefont {G.}~\bibnamefont
  {de~Tomasi}},\ }\href@noop {} {\bibfield  {journal} {\bibinfo  {journal}
  {Phys. Rev. B}\ }\textbf {\bibinfo {volume} {99}},\ \bibinfo {pages} {054204}
  (\bibinfo {year} {2019})}\BibitemShut {NoStop}%
\bibitem [{\citenamefont {Hashizume}\ \emph {et~al.}(2018)\citenamefont
  {Hashizume}, \citenamefont {McCulloch},\ and\ \citenamefont
  {Halimeh}}]{hashizume2018dynamical}%
  \BibitemOpen
  \bibfield  {author} {\bibinfo {author} {\bibfnamefont {T.}~\bibnamefont
  {Hashizume}}, \bibinfo {author} {\bibfnamefont {I.~P.}\ \bibnamefont
  {McCulloch}}, \ and\ \bibinfo {author} {\bibfnamefont {J.~C.}\ \bibnamefont
  {Halimeh}},\ }\href@noop {} {\enquote {\bibinfo {title} {Dynamical phase
  transitions in the two-dimensional transverse-field ising model},}\ }
  (\bibinfo {year} {2018}),\ \Eprint {http://arxiv.org/abs/1811.09275}
  {arXiv:1811.09275} \BibitemShut {NoStop}%
\bibitem [{\citenamefont {Tikhonov}\ and\ \citenamefont
  {Mirlin}(2018)}]{tikhonov2018many}%
  \BibitemOpen
  \bibfield  {author} {\bibinfo {author} {\bibfnamefont {K.}~\bibnamefont
  {Tikhonov}}\ and\ \bibinfo {author} {\bibfnamefont {A.}~\bibnamefont
  {Mirlin}},\ }\href@noop {} {\bibfield  {journal} {\bibinfo  {journal}
  {Physical Review B}\ }\textbf {\bibinfo {volume} {97}},\ \bibinfo {pages}
  {214205} (\bibinfo {year} {2018})}\BibitemShut {NoStop}%
\bibitem [{\citenamefont {Casetti}\ and\ \citenamefont
  {Gupta}(2014)}]{casetti2014velocity}%
  \BibitemOpen
  \bibfield  {author} {\bibinfo {author} {\bibfnamefont {L.}~\bibnamefont
  {Casetti}}\ and\ \bibinfo {author} {\bibfnamefont {S.}~\bibnamefont
  {Gupta}},\ }\href@noop {} {\bibfield  {journal} {\bibinfo  {journal} {The
  European Physical Journal B}\ }\textbf {\bibinfo {volume} {87}},\ \bibinfo
  {pages} {91} (\bibinfo {year} {2014})}\BibitemShut {NoStop}%
\bibitem [{\citenamefont {Teles}\ \emph {et~al.}(2015)\citenamefont {Teles},
  \citenamefont {Gupta}, \citenamefont {Di~Cintio},\ and\ \citenamefont
  {Casetti}}]{teles2015temperature}%
  \BibitemOpen
  \bibfield  {author} {\bibinfo {author} {\bibfnamefont {T.~N.}\ \bibnamefont
  {Teles}}, \bibinfo {author} {\bibfnamefont {S.}~\bibnamefont {Gupta}},
  \bibinfo {author} {\bibfnamefont {P.}~\bibnamefont {Di~Cintio}}, \ and\
  \bibinfo {author} {\bibfnamefont {L.}~\bibnamefont {Casetti}},\ }\href@noop
  {} {\bibfield  {journal} {\bibinfo  {journal} {Physical Review E}\ }\textbf
  {\bibinfo {volume} {92}},\ \bibinfo {pages} {020101} (\bibinfo {year}
  {2015})}\BibitemShut {NoStop}%
\bibitem [{\citenamefont {Gupta}\ and\ \citenamefont
  {Casetti}(2016)}]{gupta2016surprises}%
  \BibitemOpen
  \bibfield  {author} {\bibinfo {author} {\bibfnamefont {S.}~\bibnamefont
  {Gupta}}\ and\ \bibinfo {author} {\bibfnamefont {L.}~\bibnamefont
  {Casetti}},\ }\href@noop {} {\bibfield  {journal} {\bibinfo  {journal} {New
  Journal of Physics}\ }\textbf {\bibinfo {volume} {18}},\ \bibinfo {pages}
  {103051} (\bibinfo {year} {2016})}\BibitemShut {NoStop}%
\bibitem [{\citenamefont {Campa}\ \emph {et~al.}(2009)\citenamefont {Campa},
  \citenamefont {Dauxois},\ and\ \citenamefont {Ruffo}}]{campa2009statistical}%
  \BibitemOpen
  \bibfield  {author} {\bibinfo {author} {\bibfnamefont {A.}~\bibnamefont
  {Campa}}, \bibinfo {author} {\bibfnamefont {T.}~\bibnamefont {Dauxois}}, \
  and\ \bibinfo {author} {\bibfnamefont {S.}~\bibnamefont {Ruffo}},\
  }\href@noop {} {\bibfield  {journal} {\bibinfo  {journal} {Physics Reports}\
  }\textbf {\bibinfo {volume} {480}},\ \bibinfo {pages} {57} (\bibinfo {year}
  {2009})}\BibitemShut {NoStop}%
\bibitem [{\citenamefont {Bouchet}\ \emph {et~al.}(2010)\citenamefont
  {Bouchet}, \citenamefont {Gupta},\ and\ \citenamefont
  {Mukamel}}]{bouchet2010thermodynamics}%
  \BibitemOpen
  \bibfield  {author} {\bibinfo {author} {\bibfnamefont {F.}~\bibnamefont
  {Bouchet}}, \bibinfo {author} {\bibfnamefont {S.}~\bibnamefont {Gupta}}, \
  and\ \bibinfo {author} {\bibfnamefont {D.}~\bibnamefont {Mukamel}},\
  }\href@noop {} {\bibfield  {journal} {\bibinfo  {journal} {Physica A:
  Statistical Mechanics and its Applications}\ }\textbf {\bibinfo {volume}
  {389}},\ \bibinfo {pages} {4389} (\bibinfo {year} {2010})}\BibitemShut
  {NoStop}%
\bibitem [{\citenamefont {Campa}\ \emph {et~al.}(2014)\citenamefont {Campa},
  \citenamefont {Dauxois}, \citenamefont {Fanelli},\ and\ \citenamefont
  {Ruffo}}]{campa2014physics}%
  \BibitemOpen
  \bibfield  {author} {\bibinfo {author} {\bibfnamefont {A.}~\bibnamefont
  {Campa}}, \bibinfo {author} {\bibfnamefont {T.}~\bibnamefont {Dauxois}},
  \bibinfo {author} {\bibfnamefont {D.}~\bibnamefont {Fanelli}}, \ and\
  \bibinfo {author} {\bibfnamefont {S.}~\bibnamefont {Ruffo}},\ }\href@noop {}
  {\emph {\bibinfo {title} {Physics of long-range interacting systems}}}\
  (\bibinfo  {publisher} {OUP Oxford},\ \bibinfo {year} {2014})\BibitemShut
  {NoStop}%
\bibitem [{\citenamefont {Burin}\ and\ \citenamefont
  {Maksimov}(1989)}]{Burin1989}%
  \BibitemOpen
  \bibfield  {author} {\bibinfo {author} {\bibfnamefont {A.~L.}\ \bibnamefont
  {Burin}}\ and\ \bibinfo {author} {\bibfnamefont {L.~A.}\ \bibnamefont
  {Maksimov}},\ }\href@noop {} {\bibfield  {journal} {\bibinfo  {journal} {JETP
  Lett.}\ }\textbf {\bibinfo {volume} {50}},\ \bibinfo {pages} {338} (\bibinfo
  {year} {1989})}\BibitemShut {NoStop}%
\bibitem [{\citenamefont {Modak}\ \emph {et~al.}(2016)\citenamefont {Modak},
  \citenamefont {Mukerjee}, \citenamefont {Yuzbashyan},\ and\ \citenamefont
  {Shastry}}]{Yuzbashyan_NJP2016}%
  \BibitemOpen
  \bibfield  {author} {\bibinfo {author} {\bibfnamefont {R.}~\bibnamefont
  {Modak}}, \bibinfo {author} {\bibfnamefont {S.}~\bibnamefont {Mukerjee}},
  \bibinfo {author} {\bibfnamefont {E.~A.}\ \bibnamefont {Yuzbashyan}}, \ and\
  \bibinfo {author} {\bibfnamefont {B.~S.}\ \bibnamefont {Shastry}},\
  }\href@noop {} {\bibfield  {journal} {\bibinfo  {journal} {New J. Phys.}\
  }\textbf {\bibinfo {volume} {18}},\ \bibinfo {pages} {033010} (\bibinfo
  {year} {2016})}\BibitemShut {NoStop}%
\bibitem [{\citenamefont {Owusu}\ \emph {et~al.}(2009)\citenamefont {Owusu},
  \citenamefont {Wagh},\ and\ \citenamefont
  {Yuzbashyan}}]{Yuzbashyan_JPhysA2009_Exact_solution}%
  \BibitemOpen
  \bibfield  {author} {\bibinfo {author} {\bibfnamefont {H.~K.}\ \bibnamefont
  {Owusu}}, \bibinfo {author} {\bibfnamefont {K.}~\bibnamefont {Wagh}}, \ and\
  \bibinfo {author} {\bibfnamefont {E.~A.}\ \bibnamefont {Yuzbashyan}},\
  }\href@noop {} {\bibfield  {journal} {\bibinfo  {journal} {J. Phys. A: Math.
  Theor.}\ }\textbf {\bibinfo {volume} {42}},\ \bibinfo {pages} {035206}
  (\bibinfo {year} {2009})}\BibitemShut {NoStop}%
\bibitem [{\citenamefont {Ossipov}(2013)}]{Ossipov2013}%
  \BibitemOpen
  \bibfield  {author} {\bibinfo {author} {\bibfnamefont {A.}~\bibnamefont
  {Ossipov}},\ }\href@noop {} {\bibfield  {journal} {\bibinfo  {journal} {J.
  Phys. A}\ }\textbf {\bibinfo {volume} {46}},\ \bibinfo {pages} {105001}
  (\bibinfo {year} {2013})}\BibitemShut {NoStop}%
\bibitem [{\citenamefont {Celardo}\ \emph {et~al.}(2016)\citenamefont
  {Celardo}, \citenamefont {Kaiser},\ and\ \citenamefont
  {Borgonovi}}]{Borgonovi_2016}%
  \BibitemOpen
  \bibfield  {author} {\bibinfo {author} {\bibfnamefont {G.~L.}\ \bibnamefont
  {Celardo}}, \bibinfo {author} {\bibfnamefont {R.}~\bibnamefont {Kaiser}}, \
  and\ \bibinfo {author} {\bibfnamefont {F.}~\bibnamefont {Borgonovi}},\
  }\href@noop {} {\bibfield  {journal} {\bibinfo  {journal} {Phys. Rev. B}\
  }\textbf {\bibinfo {volume} {94}},\ \bibinfo {pages} {144206} (\bibinfo
  {year} {2016})}\BibitemShut {NoStop}%
\bibitem [{\citenamefont {Deng}\ \emph {et~al.}(2018)\citenamefont {Deng},
  \citenamefont {Kravtsov}, \citenamefont {Shlyapnikov},\ and\ \citenamefont
  {Santos}}]{Kravtsov_Shlyapnikov_PRL2018}%
  \BibitemOpen
  \bibfield  {author} {\bibinfo {author} {\bibfnamefont {X.}~\bibnamefont
  {Deng}}, \bibinfo {author} {\bibfnamefont {V.~E.}\ \bibnamefont {Kravtsov}},
  \bibinfo {author} {\bibfnamefont {G.~V.}\ \bibnamefont {Shlyapnikov}}, \ and\
  \bibinfo {author} {\bibfnamefont {L.}~\bibnamefont {Santos}},\ }\href@noop {}
  {\bibfield  {journal} {\bibinfo  {journal} {Phys. Rev. Lett.}\ }\textbf
  {\bibinfo {volume} {120}},\ \bibinfo {pages} {110602} (\bibinfo {year}
  {2018})}\BibitemShut {NoStop}%
\bibitem [{Per()}]{Periodic_bound_conds_footnote}%
  \BibitemOpen
  \href@noop {} {}\bibinfo {note} {Here and further for simplicity we consider
  one-dimensional systems with periodic boundary conditions meaning that all
  distances are considered modulo $N$, $m-n \equiv m-n (\text{mod }
  N)$}\BibitemShut {NoStop}%
\bibitem [{\citenamefont {Kravtsov}\ \emph {et~al.}(2015)\citenamefont
  {Kravtsov}, \citenamefont {Khaymovich}, \citenamefont {Cuevas},\ and\
  \citenamefont {Amini}}]{Kravtsov_NJP2015}%
  \BibitemOpen
  \bibfield  {author} {\bibinfo {author} {\bibfnamefont {V.~E.}\ \bibnamefont
  {Kravtsov}}, \bibinfo {author} {\bibfnamefont {I.~M.}\ \bibnamefont
  {Khaymovich}}, \bibinfo {author} {\bibfnamefont {E.}~\bibnamefont {Cuevas}},
  \ and\ \bibinfo {author} {\bibfnamefont {M.}~\bibnamefont {Amini}},\
  }\href@noop {} {\bibfield  {journal} {\bibinfo  {journal} {New J. Phys.}\
  }\textbf {\bibinfo {volume} {17}} (\bibinfo {year} {2015})}\BibitemShut
  {NoStop}%
\bibitem [{\citenamefont {Faoro}\ \emph {et~al.}(2018)\citenamefont {Faoro},
  \citenamefont {Feigel'man},\ and\ \citenamefont {Ioffe}}]{faoro2018non}%
  \BibitemOpen
  \bibfield  {author} {\bibinfo {author} {\bibfnamefont {L.}~\bibnamefont
  {Faoro}}, \bibinfo {author} {\bibfnamefont {M.}~\bibnamefont {Feigel'man}}, \
  and\ \bibinfo {author} {\bibfnamefont {L.}~\bibnamefont {Ioffe}},\
  }\href@noop {} {\bibfield  {journal} {\bibinfo  {journal} {arXiv preprint
  arXiv:1812.06016}\ } (\bibinfo {year} {2018})}\BibitemShut {NoStop}%
\bibitem [{\citenamefont {Smelyanskiy}\ \emph {et~al.}(2018)\citenamefont
  {Smelyanskiy}, \citenamefont {Kechedzhi}, \citenamefont {Boixo},
  \citenamefont {Isakov}, \citenamefont {Neven},\ and\ \citenamefont
  {Altshuler}}]{smelyanskiy2018non}%
  \BibitemOpen
  \bibfield  {author} {\bibinfo {author} {\bibfnamefont {V.~N.}\ \bibnamefont
  {Smelyanskiy}}, \bibinfo {author} {\bibfnamefont {K.}~\bibnamefont
  {Kechedzhi}}, \bibinfo {author} {\bibfnamefont {S.}~\bibnamefont {Boixo}},
  \bibinfo {author} {\bibfnamefont {S.~V.}\ \bibnamefont {Isakov}}, \bibinfo
  {author} {\bibfnamefont {H.}~\bibnamefont {Neven}}, \ and\ \bibinfo {author}
  {\bibfnamefont {B.}~\bibnamefont {Altshuler}},\ }\href@noop {} {\bibfield
  {journal} {\bibinfo  {journal} {arXiv preprint arXiv:1802.09542}\ } (\bibinfo
  {year} {2018})}\BibitemShut {NoStop}%
\bibitem [{\citenamefont {Sachdev}\ and\ \citenamefont
  {Ye}(1993)}]{sachdev1993gapless}%
  \BibitemOpen
  \bibfield  {author} {\bibinfo {author} {\bibfnamefont {S.}~\bibnamefont
  {Sachdev}}\ and\ \bibinfo {author} {\bibfnamefont {J.}~\bibnamefont {Ye}},\
  }\href@noop {} {\bibfield  {journal} {\bibinfo  {journal} {Physical review
  letters}\ }\textbf {\bibinfo {volume} {70}},\ \bibinfo {pages} {3339}
  (\bibinfo {year} {1993})}\BibitemShut {NoStop}%
\bibitem [{\citenamefont {Kitaev}(2015)}]{Kitaev-talk}%
  \BibitemOpen
  \bibfield  {author} {\bibinfo {author} {\bibfnamefont {A.}~\bibnamefont
  {Kitaev}},\ }\href {http://online.kitp.ucsb.edu/online/entangled15/kitaev2/}
  {} (\bibinfo {year} {2015}),\ \bibinfo {note} {talks at KITP on April 7th and
  May 27th 2015}\BibitemShut {NoStop}%
\bibitem [{\citenamefont {Micklitz}\ \emph {et~al.}(2019)\citenamefont
  {Micklitz}, \citenamefont {Monteiro},\ and\ \citenamefont
  {Altland}}]{micklitz2019non}%
  \BibitemOpen
  \bibfield  {author} {\bibinfo {author} {\bibfnamefont {T.}~\bibnamefont
  {Micklitz}}, \bibinfo {author} {\bibfnamefont {F.}~\bibnamefont {Monteiro}},
  \ and\ \bibinfo {author} {\bibfnamefont {A.}~\bibnamefont {Altland}},\
  }\href@noop {} {\bibfield  {journal} {\bibinfo  {journal} {arXiv preprint
  arXiv:1901.02389}\ } (\bibinfo {year} {2019})}\BibitemShut {NoStop}%
\bibitem [{\citenamefont {H.~Wang}(2019)}]{Kamenev-talk-1}%
  \BibitemOpen
  \bibfield  {author} {\bibinfo {author} {\bibfnamefont {A.~K.}\ \bibnamefont
  {H.~Wang}},\ }\href {http://meetings.aps.org/Meeting/MAR19/Session/H06.6}
  {\enquote {\bibinfo {title} {Many-body localization in a modified syk
  model},}\ } (\bibinfo {year} {2019})\BibitemShut {NoStop}%
\bibitem [{\citenamefont {Kamenev}(2018)}]{Kamenev-talk-2}%
  \BibitemOpen
  \bibfield  {author} {\bibinfo {author} {\bibfnamefont {A.}~\bibnamefont
  {Kamenev}},\ }\href
  {http://eugenekanzieper.faculty.hit.ac.il/yad8/2018/pages/schedule-abstracts.html}
  {\enquote {\bibinfo {title} {Many-body localization in a modified syk
  model},}\ } (\bibinfo {year} {2018})\BibitemShut {NoStop}%
\bibitem [{\citenamefont {De~Luca}\ \emph
  {et~al.}(2014{\natexlab{a}})\citenamefont {De~Luca}, \citenamefont
  {Altshuler}, \citenamefont {Kravtsov},\ and\ \citenamefont
  {Scardicchio}}]{Luca2014}%
  \BibitemOpen
  \bibfield  {author} {\bibinfo {author} {\bibfnamefont {A.}~\bibnamefont
  {De~Luca}}, \bibinfo {author} {\bibfnamefont {B.~L.}\ \bibnamefont
  {Altshuler}}, \bibinfo {author} {\bibfnamefont {V.~E.}\ \bibnamefont
  {Kravtsov}}, \ and\ \bibinfo {author} {\bibfnamefont {A.}~\bibnamefont
  {Scardicchio}},\ }\href@noop {} {\bibfield  {journal} {\bibinfo  {journal}
  {Phys. Rev. Lett.}\ }\textbf {\bibinfo {volume} {113}},\ \bibinfo {pages}
  {046806} (\bibinfo {year} {2014}{\natexlab{a}})}\BibitemShut {NoStop}%
\bibitem [{\citenamefont {V.E.Kravtsov}\ \emph {et~al.}(2018)\citenamefont
  {V.E.Kravtsov}, \citenamefont {B.L.Altshuler},\ and\ \citenamefont
  {L.B.Ioffe}}]{RRGAnnals}%
  \BibitemOpen
  \bibfield  {author} {\bibinfo {author} {\bibnamefont {V.E.Kravtsov}},
  \bibinfo {author} {\bibnamefont {B.L.Altshuler}}, \ and\ \bibinfo {author}
  {\bibnamefont {L.B.Ioffe}},\ }\href@noop {} {\bibfield  {journal} {\bibinfo
  {journal} {Annals of Physics}\ }\textbf {\bibinfo {volume} {389}},\ \bibinfo
  {pages} {148} (\bibinfo {year} {2018})}\BibitemShut {NoStop}%
\bibitem [{\citenamefont {Pino}\ \emph {et~al.}(2016)\citenamefont {Pino},
  \citenamefont {Ioffe},\ and\ \citenamefont
  {Altshuler}}]{Pino-Ioffe-Altshuler2016}%
  \BibitemOpen
  \bibfield  {author} {\bibinfo {author} {\bibfnamefont {M.}~\bibnamefont
  {Pino}}, \bibinfo {author} {\bibfnamefont {L.~B.}\ \bibnamefont {Ioffe}}, \
  and\ \bibinfo {author} {\bibfnamefont {B.~L.}\ \bibnamefont {Altshuler}},\
  }\href@noop {} {\bibfield  {journal} {\bibinfo  {journal} {PNAS}\ }\textbf
  {\bibinfo {volume} {113}},\ \bibinfo {pages} {536} (\bibinfo {year}
  {2016})}\BibitemShut {NoStop}%
\bibitem [{\citenamefont {Pino}\ \emph {et~al.}(2017)\citenamefont {Pino},
  \citenamefont {Kravtsov}, \citenamefont {Altshuler},\ and\ \citenamefont
  {Ioffe}}]{Pino-Ioffe-Kr2017}%
  \BibitemOpen
  \bibfield  {author} {\bibinfo {author} {\bibfnamefont {M.}~\bibnamefont
  {Pino}}, \bibinfo {author} {\bibfnamefont {V.~E.}\ \bibnamefont {Kravtsov}},
  \bibinfo {author} {\bibfnamefont {B.~L.}\ \bibnamefont {Altshuler}}, \ and\
  \bibinfo {author} {\bibfnamefont {L.~B.}\ \bibnamefont {Ioffe}},\ }\href@noop
  {} {\bibfield  {journal} {\bibinfo  {journal} {Phys. Rev. B}\ }\textbf
  {\bibinfo {volume} {96}},\ \bibinfo {pages} {214205} (\bibinfo {year}
  {2017})}\BibitemShut {NoStop}%
\bibitem [{\citenamefont {Bera}\ \emph {et~al.}(2018)\citenamefont {Bera},
  \citenamefont {De~Tomasi}, \citenamefont {Khaymovich},\ and\ \citenamefont
  {Scardicchio}}]{Bera-Khaym2018}%
  \BibitemOpen
  \bibfield  {author} {\bibinfo {author} {\bibfnamefont {S.}~\bibnamefont
  {Bera}}, \bibinfo {author} {\bibfnamefont {G.}~\bibnamefont {De~Tomasi}},
  \bibinfo {author} {\bibfnamefont {I.~M.}\ \bibnamefont {Khaymovich}}, \ and\
  \bibinfo {author} {\bibfnamefont {A.}~\bibnamefont {Scardicchio}},\
  }\href@noop {} {\bibfield  {journal} {\bibinfo  {journal} {Phys. Rev. B}\
  }\textbf {\bibinfo {volume} {98}},\ \bibinfo {pages} {134205} (\bibinfo
  {year} {2018})}\BibitemShut {NoStop}%
\bibitem [{\citenamefont {Biroli}\ and\ \citenamefont {Tarzia}(2017)}]{Biro17}%
  \BibitemOpen
  \bibfield  {author} {\bibinfo {author} {\bibfnamefont {G.}~\bibnamefont
  {Biroli}}\ and\ \bibinfo {author} {\bibfnamefont {M.}~\bibnamefont
  {Tarzia}},\ }\href {\doibase 10.1103/PhysRevB.96.201114} {\bibfield
  {journal} {\bibinfo  {journal} {Phys. Rev. B}\ }\textbf {\bibinfo {volume}
  {96}},\ \bibinfo {pages} {201114} (\bibinfo {year} {2017})}\BibitemShut
  {NoStop}%
\bibitem [{\citenamefont {Tikhonov}\ and\ \citenamefont
  {Mirlin}(2019)}]{Tikhonov_misc}%
  \BibitemOpen
  \bibfield  {author} {\bibinfo {author} {\bibfnamefont {K.~S.}\ \bibnamefont
  {Tikhonov}}\ and\ \bibinfo {author} {\bibfnamefont {A.~D.}\ \bibnamefont
  {Mirlin}},\ }\href@noop {} {\bibfield  {journal} {\bibinfo  {journal} {Phys.
  Rev. B}\ }\textbf {\bibinfo {volume} {99}},\ \bibinfo {pages} {024202}
  (\bibinfo {year} {2019})}\BibitemShut {NoStop}%
\bibitem [{\citenamefont {Gopalakrishnan}\ \emph {et~al.}(2016)\citenamefont
  {Gopalakrishnan}, \citenamefont {Agarwal}, \citenamefont {Demler},
  \citenamefont {Huse},\ and\ \citenamefont {Knap}}]{Gopa2016}%
  \BibitemOpen
  \bibfield  {author} {\bibinfo {author} {\bibfnamefont {S.}~\bibnamefont
  {Gopalakrishnan}}, \bibinfo {author} {\bibfnamefont {K.}~\bibnamefont
  {Agarwal}}, \bibinfo {author} {\bibfnamefont {E.~A.}\ \bibnamefont {Demler}},
  \bibinfo {author} {\bibfnamefont {D.~A.}\ \bibnamefont {Huse}}, \ and\
  \bibinfo {author} {\bibfnamefont {M.}~\bibnamefont {Knap}},\ }\href@noop {}
  {\bibfield  {journal} {\bibinfo  {journal} {Phys. Rev. B}\ }\textbf {\bibinfo
  {volume} {93}},\ \bibinfo {pages} {134206} (\bibinfo {year}
  {2016})}\BibitemShut {NoStop}%
\bibitem [{\citenamefont {L\"uschen}\ \emph {et~al.}(2017)\citenamefont
  {L\"uschen}, \citenamefont {Bordia}, \citenamefont {Scherg}, \citenamefont
  {Alet}, \citenamefont {Altman}, \citenamefont {Schneider},\ and\
  \citenamefont {Bloch}}]{Bloch2017a}%
  \BibitemOpen
  \bibfield  {author} {\bibinfo {author} {\bibfnamefont {H.~P.}\ \bibnamefont
  {L\"uschen}}, \bibinfo {author} {\bibfnamefont {P.}~\bibnamefont {Bordia}},
  \bibinfo {author} {\bibfnamefont {S.}~\bibnamefont {Scherg}}, \bibinfo
  {author} {\bibfnamefont {F.}~\bibnamefont {Alet}}, \bibinfo {author}
  {\bibfnamefont {E.}~\bibnamefont {Altman}}, \bibinfo {author} {\bibfnamefont
  {U.}~\bibnamefont {Schneider}}, \ and\ \bibinfo {author} {\bibfnamefont
  {I.}~\bibnamefont {Bloch}},\ }\href@noop {} {\bibfield  {journal} {\bibinfo
  {journal} {Phys. Rev. Lett.}\ }\textbf {\bibinfo {volume} {119}},\ \bibinfo
  {pages} {260401} (\bibinfo {year} {2017})}\BibitemShut {NoStop}%
\bibitem [{\citenamefont {Bordia}\ \emph {et~al.}(2017)\citenamefont {Bordia},
  \citenamefont {L\"uschen}, \citenamefont {Scherg}, \citenamefont
  {Gopalakrishnan}, \citenamefont {Knap}, \citenamefont {Schneider},\ and\
  \citenamefont {Bloch}}]{Bloch2017b}%
  \BibitemOpen
  \bibfield  {author} {\bibinfo {author} {\bibfnamefont {P.}~\bibnamefont
  {Bordia}}, \bibinfo {author} {\bibfnamefont {H.}~\bibnamefont {L\"uschen}},
  \bibinfo {author} {\bibfnamefont {S.}~\bibnamefont {Scherg}}, \bibinfo
  {author} {\bibfnamefont {S.}~\bibnamefont {Gopalakrishnan}}, \bibinfo
  {author} {\bibfnamefont {M.}~\bibnamefont {Knap}}, \bibinfo {author}
  {\bibfnamefont {U.}~\bibnamefont {Schneider}}, \ and\ \bibinfo {author}
  {\bibfnamefont {I.}~\bibnamefont {Bloch}},\ }\href@noop {} {\bibfield
  {journal} {\bibinfo  {journal} {Phys. Rev. X}\ }\textbf {\bibinfo {volume}
  {7}},\ \bibinfo {pages} {041047} (\bibinfo {year} {2017})}\BibitemShut
  {NoStop}%
\bibitem [{erg()}]{ergodic_footnote}%
  \BibitemOpen
  \href@noop {} {}\bibinfo {note} {Here and further we distinguish {\it weakly}
  and {\it fully} ergodic phases. By {\it fully ergodic} we mean the states
  which eigenfunction statistics is invariant under basis rotation and which
  level statistics is Wigner-Dyson. This is contrasted by the {\it weak
  ergodicity}~\cite{Luca2014, RRGAnnals} which is defined in a given (e.g.
  coordinate) basis and implies that the wave function support~\cite{RRGAnnals,
  KravtsovSupportSet} takes a finite fraction of all the corresponding 
  coordinate) space. Weak ergodicity does not imply invariance of wave function
  statistics under basis rotation.}\BibitemShut {Stop}%
\bibitem [{Spe()}]{Special_basis_footnote}%
  \BibitemOpen
  \href@noop {} {}\bibinfo {note} {Note that the momentum space basis is
  special for TI-models. However, this principle works for any basis
  $\left|g\ra$ diagonalizing Hamiltonian without on-site disorder $\hat j =
  \sum_g j_g \left|g\ra\la g\right|$, if it is uncorrelated from the
  corresponding eigenvalues $j_g$}\BibitemShut {NoStop}%
\bibitem [{\citenamefont {De~Luca}\ \emph
  {et~al.}(2014{\natexlab{b}})\citenamefont {De~Luca}, \citenamefont
  {Scardicchio}, \citenamefont {Kravtsov},\ and\ \citenamefont
  {Altshuler}}]{KravtsovSupportSet}%
  \BibitemOpen
  \bibfield  {author} {\bibinfo {author} {\bibfnamefont {A.}~\bibnamefont
  {De~Luca}}, \bibinfo {author} {\bibfnamefont {A.}~\bibnamefont
  {Scardicchio}}, \bibinfo {author} {\bibfnamefont {V.~E.}\ \bibnamefont
  {Kravtsov}}, \ and\ \bibinfo {author} {\bibfnamefont {B.~L.}\ \bibnamefont
  {Altshuler}},\ }\href@noop {} {\enquote {\bibinfo {title} {Support set of
  random wave-functions on the bethe lattice},}\ } (\bibinfo {year}
  {2014}{\natexlab{b}}),\ \Eprint {http://arxiv.org/abs/1401.0019}
  {arXiv:1401.0019} \BibitemShut {NoStop}%
\bibitem [{Gau()}]{Gaussian_disorder_footnote}%
  \BibitemOpen
  \href@noop {} {}\bibinfo {note} {Without loss of generality we consider
  $\{\ep_m\}$ to be independent Gaussian random numbers}\BibitemShut {NoStop}%
\bibitem [{\citenamefont {Mott}(1966)}]{Mott1966}%
  \BibitemOpen
  \bibfield  {author} {\bibinfo {author} {\bibfnamefont {N.~F.}\ \bibnamefont
  {Mott}},\ }\href@noop {} {\bibfield  {journal} {\bibinfo  {journal} {Phil.
  Mag.}\ }\textbf {\bibinfo {volume} {13}},\ \bibinfo {pages} {989} (\bibinfo
  {year} {1966})}\BibitemShut {NoStop}%
\bibitem [{\citenamefont {Bogomolny}\ and\ \citenamefont
  {Sieber}(2018{\natexlab{a}})}]{BogomolnyPLRBM2018}%
  \BibitemOpen
  \bibfield  {author} {\bibinfo {author} {\bibfnamefont {E.}~\bibnamefont
  {Bogomolny}}\ and\ \bibinfo {author} {\bibfnamefont {M.}~\bibnamefont
  {Sieber}},\ }\href@noop {} {\bibfield  {journal} {\bibinfo  {journal} {Phys.
  Rev. E}\ }\textbf {\bibinfo {volume} {98}},\ \bibinfo {pages} {042116}
  (\bibinfo {year} {2018}{\natexlab{a}})}\BibitemShut {NoStop}%
\bibitem [{\citenamefont {de~Tomasi}\ \emph {et~al.}(2019)\citenamefont
  {de~Tomasi}, \citenamefont {Amini}, \citenamefont {Bera}, \citenamefont
  {Khaymovich},\ and\ \citenamefont {Kravtsov}}]{RP_R(t)_2018}%
  \BibitemOpen
  \bibfield  {author} {\bibinfo {author} {\bibfnamefont {G.}~\bibnamefont
  {de~Tomasi}}, \bibinfo {author} {\bibfnamefont {M.}~\bibnamefont {Amini}},
  \bibinfo {author} {\bibfnamefont {S.}~\bibnamefont {Bera}}, \bibinfo {author}
  {\bibfnamefont {I.~M.}\ \bibnamefont {Khaymovich}}, \ and\ \bibinfo {author}
  {\bibfnamefont {V.~E.}\ \bibnamefont {Kravtsov}},\ }\href@noop {} {\bibfield
  {journal} {\bibinfo  {journal} {SciPost Phys.}\ }\textbf {\bibinfo {volume}
  {6}},\ \bibinfo {pages} {14} (\bibinfo {year} {2019})}\BibitemShut {NoStop}%
\bibitem [{j_0()}]{j_0_sum_e_p_footnote}%
  \BibitemOpen
  \href@noop {} {}\bibinfo {note} {Note that in TI-models one can remove the
  overall energy shift considering $j_0 = 0$, $\sum_{m=0}^{N-1}\ep_m = 0$, but
  we keep both these parameters free to avoid correlations in
  $\ep_m$}\BibitemShut {NoStop}%
\bibitem [{\citenamefont {Rosenzweig}\ and\ \citenamefont {Porter}(1960)}]{RP}%
  \BibitemOpen
  \bibfield  {author} {\bibinfo {author} {\bibfnamefont {N.}~\bibnamefont
  {Rosenzweig}}\ and\ \bibinfo {author} {\bibfnamefont {C.~E.}\ \bibnamefont
  {Porter}},\ }\href@noop {} {\bibfield  {journal} {\bibinfo  {journal} {Phys.
  Rev. B}\ }\textbf {\bibinfo {volume} {120}},\ \bibinfo {pages} {1698}
  (\bibinfo {year} {1960})}\BibitemShut {NoStop}%
\bibitem [{\citenamefont {Pandey}(1995)}]{Pandey}%
  \BibitemOpen
  \bibfield  {author} {\bibinfo {author} {\bibfnamefont {A.}~\bibnamefont
  {Pandey}},\ }\href@noop {} {\bibfield  {journal} {\bibinfo  {journal} {Chaos
  Solitons Fractals}\ }\textbf {\bibinfo {volume} {5}},\ \bibinfo {pages}
  {1275} (\bibinfo {year} {1995})}\BibitemShut {NoStop}%
\bibitem [{\citenamefont {Br{\'e}zin}\ and\ \citenamefont
  {Hikami}(1996)}]{BrezHik}%
  \BibitemOpen
  \bibfield  {author} {\bibinfo {author} {\bibfnamefont {E.}~\bibnamefont
  {Br{\'e}zin}}\ and\ \bibinfo {author} {\bibfnamefont {S.}~\bibnamefont
  {Hikami}},\ }\href@noop {} {\bibfield  {journal} {\bibinfo  {journal} {Nucl.
  Phys. B}\ }\textbf {\bibinfo {volume} {479}},\ \bibinfo {pages} {697}
  (\bibinfo {year} {1996})}\BibitemShut {NoStop}%
\bibitem [{\citenamefont {Guhr}(1996)}]{Guhr}%
  \BibitemOpen
  \bibfield  {author} {\bibinfo {author} {\bibfnamefont {T.}~\bibnamefont
  {Guhr}},\ }\href@noop {} {\bibfield  {journal} {\bibinfo  {journal} {Ann.
  Phys.}\ }\textbf {\bibinfo {volume} {250}},\ \bibinfo {pages} {145} (\bibinfo
  {year} {1996})}\BibitemShut {NoStop}%
\bibitem [{\citenamefont {Altland}\ \emph {et~al.}(1997)\citenamefont
  {Altland}, \citenamefont {Janssen},\ and\ \citenamefont
  {Shapiro}}]{AltlandShapiro}%
  \BibitemOpen
  \bibfield  {author} {\bibinfo {author} {\bibfnamefont {A.}~\bibnamefont
  {Altland}}, \bibinfo {author} {\bibfnamefont {M.}~\bibnamefont {Janssen}}, \
  and\ \bibinfo {author} {\bibfnamefont {B.}~\bibnamefont {Shapiro}},\
  }\href@noop {} {\bibfield  {journal} {\bibinfo  {journal} {Phys. Rev. E}\
  }\textbf {\bibinfo {volume} {56}},\ \bibinfo {pages} {1471} (\bibinfo {year}
  {1997})}\BibitemShut {NoStop}%
\bibitem [{\citenamefont {Kunz}\ and\ \citenamefont
  {Shapiro}(1998)}]{ShapiroKunz}%
  \BibitemOpen
  \bibfield  {author} {\bibinfo {author} {\bibfnamefont {H.}~\bibnamefont
  {Kunz}}\ and\ \bibinfo {author} {\bibfnamefont {B.}~\bibnamefont {Shapiro}},\
  }\href@noop {} {\bibfield  {journal} {\bibinfo  {journal} {Phys. Rev. E}\
  }\textbf {\bibinfo {volume} {58}},\ \bibinfo {pages} {400} (\bibinfo {year}
  {1998})}\BibitemShut {NoStop}%
\bibitem [{\citenamefont {Shukla}(2000)}]{Shukla2000}%
  \BibitemOpen
  \bibfield  {author} {\bibinfo {author} {\bibfnamefont {P.}~\bibnamefont
  {Shukla}},\ }\href@noop {} {\bibfield  {journal} {\bibinfo  {journal} {Phys.
  Rev. E}\ }\textbf {\bibinfo {volume} {62}},\ \bibinfo {pages} {2098}
  (\bibinfo {year} {2000})}\BibitemShut {NoStop}%
\bibitem [{\citenamefont {Shukla}(2005)}]{Shukla2005}%
  \BibitemOpen
  \bibfield  {author} {\bibinfo {author} {\bibfnamefont {P.}~\bibnamefont
  {Shukla}},\ }\href@noop {} {\bibfield  {journal} {\bibinfo  {journal} {J.
  Phys: Condens. Matter}\ }\textbf {\bibinfo {volume} {17}},\ \bibinfo {pages}
  {1653} (\bibinfo {year} {2005})}\BibitemShut {NoStop}%
\bibitem [{\citenamefont {Facoetti}\ \emph {et~al.}(2016)\citenamefont
  {Facoetti}, \citenamefont {Vivo},\ and\ \citenamefont {Biroli}}]{Biroli_RP}%
  \BibitemOpen
  \bibfield  {author} {\bibinfo {author} {\bibfnamefont {D.}~\bibnamefont
  {Facoetti}}, \bibinfo {author} {\bibfnamefont {P.}~\bibnamefont {Vivo}}, \
  and\ \bibinfo {author} {\bibfnamefont {G.}~\bibnamefont {Biroli}},\
  }\href@noop {} {\bibfield  {journal} {\bibinfo  {journal} {Europhys. Lett.}\
  }\textbf {\bibinfo {volume} {115}},\ \bibinfo {pages} {47003} (\bibinfo
  {year} {2016})}\BibitemShut {NoStop}%
\bibitem [{\citenamefont {Truong}\ and\ \citenamefont
  {Ossipov}(2016)}]{Ossipov_EPL2016_H+V}%
  \BibitemOpen
  \bibfield  {author} {\bibinfo {author} {\bibfnamefont {K.}~\bibnamefont
  {Truong}}\ and\ \bibinfo {author} {\bibfnamefont {A.}~\bibnamefont
  {Ossipov}},\ }\href@noop {} {\bibfield  {journal} {\bibinfo  {journal}
  {Europhys. Lett.}\ }\textbf {\bibinfo {volume} {116}},\ \bibinfo {pages}
  {37002} (\bibinfo {year} {2016})}\BibitemShut {NoStop}%
\bibitem [{\citenamefont {Amini}(2017)}]{Amini2017}%
  \BibitemOpen
  \bibfield  {author} {\bibinfo {author} {\bibfnamefont {M.}~\bibnamefont
  {Amini}},\ }\href@noop {} {\bibfield  {journal} {\bibinfo  {journal}
  {Europhys. Lett.}\ }\textbf {\bibinfo {volume} {117}},\ \bibinfo {pages}
  {30003} (\bibinfo {year} {2017})}\BibitemShut {NoStop}%
\bibitem [{\citenamefont {von Soosten}\ and\ \citenamefont
  {Warzel}(2018)}]{vonSoosten2017phase}%
  \BibitemOpen
  \bibfield  {author} {\bibinfo {author} {\bibfnamefont {P.}~\bibnamefont {von
  Soosten}}\ and\ \bibinfo {author} {\bibfnamefont {S.}~\bibnamefont
  {Warzel}},\ }\href@noop {} {\bibfield  {journal} {\bibinfo  {journal}
  {Electron J. Probab.}\ }\textbf {\bibinfo {volume} {23}},\ \bibinfo {pages}
  {1} (\bibinfo {year} {2018})}\BibitemShut {NoStop}%
\bibitem [{\citenamefont {von Soosten}\ and\ \citenamefont
  {Warzel}(2017)}]{vonSoosten2017non}%
  \BibitemOpen
  \bibfield  {author} {\bibinfo {author} {\bibfnamefont {P.}~\bibnamefont {von
  Soosten}}\ and\ \bibinfo {author} {\bibfnamefont {S.}~\bibnamefont
  {Warzel}},\ }\href@noop {} {} (\bibinfo {year} {2017}),\ \Eprint
  {http://arxiv.org/abs/1709.10313} {arXiv:1709.10313} \BibitemShut {NoStop}%
\bibitem [{\citenamefont {Monthus}(2017)}]{Monthus}%
  \BibitemOpen
  \bibfield  {author} {\bibinfo {author} {\bibfnamefont {C.}~\bibnamefont
  {Monthus}},\ }\href@noop {} {\bibfield  {journal} {\bibinfo  {journal} {J.
  Phys. A: Math. Theor.}\ }\textbf {\bibinfo {volume} {50}},\ \bibinfo {pages}
  {295101} (\bibinfo {year} {2017})}\BibitemShut {NoStop}%
\bibitem [{\citenamefont {Bogomolny}\ and\ \citenamefont
  {Sieber}(2018{\natexlab{b}})}]{BogomolnyRP2018}%
  \BibitemOpen
  \bibfield  {author} {\bibinfo {author} {\bibfnamefont {E.}~\bibnamefont
  {Bogomolny}}\ and\ \bibinfo {author} {\bibfnamefont {M.}~\bibnamefont
  {Sieber}},\ }\href@noop {} {\bibfield  {journal} {\bibinfo  {journal} {Phys.
  Rev. E}\ }\textbf {\bibinfo {volume} {98}},\ \bibinfo {pages} {032139}
  (\bibinfo {year} {2018}{\natexlab{b}})}\BibitemShut {NoStop}%
\bibitem [{\citenamefont {Atas}\ \emph {et~al.}(2013)\citenamefont {Atas},
  \citenamefont {Bogomolny}, \citenamefont {Giraud},\ and\ \citenamefont
  {Roux}}]{Bogomol2013}%
  \BibitemOpen
  \bibfield  {author} {\bibinfo {author} {\bibfnamefont {Y.~Y.}\ \bibnamefont
  {Atas}}, \bibinfo {author} {\bibfnamefont {E.}~\bibnamefont {Bogomolny}},
  \bibinfo {author} {\bibfnamefont {O.}~\bibnamefont {Giraud}}, \ and\ \bibinfo
  {author} {\bibfnamefont {G.}~\bibnamefont {Roux}},\ }\href@noop {} {\bibfield
   {journal} {\bibinfo  {journal} {Phys. Rev. Lett.}\ }\textbf {\bibinfo
  {volume} {110}},\ \bibinfo {pages} {084101} (\bibinfo {year}
  {2013})}\BibitemShut {NoStop}%
\bibitem [{\citenamefont {Mehta}(2004)}]{Mehta2004random}%
  \BibitemOpen
  \bibfield  {author} {\bibinfo {author} {\bibfnamefont {M.~L.}\ \bibnamefont
  {Mehta}},\ }\href {\doibase
  https://www.sciencedirect.com/book/9780124880511/random-matrices} {\emph
  {\bibinfo {title} {Random matrices}}}\ (\bibinfo  {publisher} {Elsevier},\
  \bibinfo {year} {2004})\BibitemShut {NoStop}%
\bibitem [{Note1()}]{Note1}%
  \BibitemOpen
  \bibinfo {note} {This duality is similar to the one in the Aubry-Andr{\'e}
  model~\cite {AA}.}\BibitemShut {Stop}%
\bibitem [{SM()}]{SM}%
  \BibitemOpen
  \href@noop {} {}\bibinfo {note} {See Supplemental Material at [URL will be
  inserted by publisher] for details of calculations}\BibitemShut {NoStop}%
\bibitem [{Note2()}]{Note2}%
  \BibitemOpen
  \bibinfo {note} {Here the width $\Delta _p$ of the energy domain where mean
  level spacing $\delta (N)$ takes a typical value coincides with the total
  hopping bandwidth.}\BibitemShut {Stop}%
\bibitem [{Note3()}]{Note3}%
  \BibitemOpen
  \bibinfo {note} {One can check that even at $\beta <0$ the assumption
  $E_{0}\sim N^{\beta }\gg \Delta _{p}(\beta )\sim N^{(1-\gamma (\beta ))}$
  still holds true.}\BibitemShut {Stop}%
\bibitem [{\citenamefont {Rodriguez}\ \emph {et~al.}(2000)\citenamefont
  {Rodriguez}, \citenamefont {Malyshev},\ and\ \citenamefont
  {Dominguez-Adame}}]{Malyshev2000}%
  \BibitemOpen
  \bibfield  {author} {\bibinfo {author} {\bibfnamefont {A.}~\bibnamefont
  {Rodriguez}}, \bibinfo {author} {\bibfnamefont {V.~A.}\ \bibnamefont
  {Malyshev}}, \ and\ \bibinfo {author} {\bibfnamefont {F.}~\bibnamefont
  {Dominguez-Adame}},\ }\href@noop {} {\bibfield  {journal} {\bibinfo
  {journal} {J. Phys. A: Math. Gen.}\ }\textbf {\bibinfo {volume} {33}},\
  \bibinfo {pages} {L161} (\bibinfo {year} {2000})}\BibitemShut {NoStop}%
\bibitem [{\citenamefont {Rodriguez}\ \emph {et~al.}(2003)\citenamefont
  {Rodriguez}, \citenamefont {Malyshev}, \citenamefont {Sierra}, \citenamefont
  {Martin-Delgado}, \citenamefont {Rodriguez-Laguna},\ and\ \citenamefont
  {Dominguez-Adame}}]{Malyshev_PRL2003}%
  \BibitemOpen
  \bibfield  {author} {\bibinfo {author} {\bibfnamefont {A.}~\bibnamefont
  {Rodriguez}}, \bibinfo {author} {\bibfnamefont {V.~A.}\ \bibnamefont
  {Malyshev}}, \bibinfo {author} {\bibfnamefont {G.}~\bibnamefont {Sierra}},
  \bibinfo {author} {\bibfnamefont {M.~A.}\ \bibnamefont {Martin-Delgado}},
  \bibinfo {author} {\bibfnamefont {J.}~\bibnamefont {Rodriguez-Laguna}}, \
  and\ \bibinfo {author} {\bibfnamefont {F.}~\bibnamefont {Dominguez-Adame}},\
  }\href@noop {} {\bibfield  {journal} {\bibinfo  {journal} {Phys. Rev. Lett.}\
  }\textbf {\bibinfo {volume} {90}},\ \bibinfo {pages} {027404} (\bibinfo
  {year} {2003})}\BibitemShut {NoStop}%
\bibitem [{\citenamefont {Balagurov}\ \emph {et~al.}(2004)\citenamefont
  {Balagurov}, \citenamefont {Malyshev},\ and\ \citenamefont
  {Dominiquez-Adame}}]{Malyshev2004}%
  \BibitemOpen
  \bibfield  {author} {\bibinfo {author} {\bibfnamefont {D.~B.}\ \bibnamefont
  {Balagurov}}, \bibinfo {author} {\bibfnamefont {V.~A.}\ \bibnamefont
  {Malyshev}}, \ and\ \bibinfo {author} {\bibfnamefont {F.}~\bibnamefont
  {Dominiquez-Adame}},\ }\href@noop {} {\bibfield  {journal} {\bibinfo
  {journal} {Phys. Rev. B}\ }\textbf {\bibinfo {volume} {69}},\ \bibinfo
  {pages} {104204} (\bibinfo {year} {2004})}\BibitemShut {NoStop}%
\bibitem [{\citenamefont {de~Moura}\ \emph {et~al.}(2005)\citenamefont
  {de~Moura}, \citenamefont {Malyshev}, \citenamefont {Lyra}, \citenamefont
  {Malyshev},\ and\ \citenamefont {Dominguez-Adame}}]{Malyshev2005}%
  \BibitemOpen
  \bibfield  {author} {\bibinfo {author} {\bibfnamefont {F.~A. B.~F.}\
  \bibnamefont {de~Moura}}, \bibinfo {author} {\bibfnamefont {A.~V.}\
  \bibnamefont {Malyshev}}, \bibinfo {author} {\bibfnamefont {M.~L.}\
  \bibnamefont {Lyra}}, \bibinfo {author} {\bibfnamefont {V.~A.}\ \bibnamefont
  {Malyshev}}, \ and\ \bibinfo {author} {\bibfnamefont {F.}~\bibnamefont
  {Dominguez-Adame}},\ }\href@noop {} {\bibfield  {journal} {\bibinfo
  {journal} {Phys. Rev. B}\ }\textbf {\bibinfo {volume} {71}},\ \bibinfo
  {pages} {174203} (\bibinfo {year} {2005})}\BibitemShut {NoStop}%
\bibitem [{\citenamefont {G{\"a}rttner}\ \emph {et~al.}(2015)\citenamefont
  {G{\"a}rttner}, \citenamefont {Syzranov}, \citenamefont {Rey}, \citenamefont
  {Gurarie},\ and\ \citenamefont {Radzihovsky}}]{Garttner2015disorder}%
  \BibitemOpen
  \bibfield  {author} {\bibinfo {author} {\bibfnamefont {M.}~\bibnamefont
  {G{\"a}rttner}}, \bibinfo {author} {\bibfnamefont {S.~V.}\ \bibnamefont
  {Syzranov}}, \bibinfo {author} {\bibfnamefont {A.~M.}\ \bibnamefont {Rey}},
  \bibinfo {author} {\bibfnamefont {V.}~\bibnamefont {Gurarie}}, \ and\
  \bibinfo {author} {\bibfnamefont {L.}~\bibnamefont {Radzihovsky}},\
  }\href@noop {} {\bibfield  {journal} {\bibinfo  {journal} {Phys. Rev. B}\
  }\textbf {\bibinfo {volume} {92}},\ \bibinfo {pages} {041406} (\bibinfo
  {year} {2015})}\BibitemShut {NoStop}%
\bibitem [{\citenamefont {Cantin}\ \emph {et~al.}(2018)\citenamefont {Cantin},
  \citenamefont {Xu},\ and\ \citenamefont {Krems}}]{Cantin2018}%
  \BibitemOpen
  \bibfield  {author} {\bibinfo {author} {\bibfnamefont {J.~T.}\ \bibnamefont
  {Cantin}}, \bibinfo {author} {\bibfnamefont {T.}~\bibnamefont {Xu}}, \ and\
  \bibinfo {author} {\bibfnamefont {R.~V.}\ \bibnamefont {Krems}},\ }\href@noop
  {} {\bibfield  {journal} {\bibinfo  {journal} {Phys. Rev. B}\ }\textbf
  {\bibinfo {volume} {98}},\ \bibinfo {pages} {014204} (\bibinfo {year}
  {2018})}\BibitemShut {NoStop}%
\bibitem [{\citenamefont {Deng}\ \emph {et~al.}()\citenamefont {Deng},
  \citenamefont {Ray}, \citenamefont {Sinha}, \citenamefont {Shlyapnikov},\
  and\ \citenamefont {Santos}}]{Deng_AA2018}%
  \BibitemOpen
  \bibfield  {author} {\bibinfo {author} {\bibfnamefont {X.}~\bibnamefont
  {Deng}}, \bibinfo {author} {\bibfnamefont {S.}~\bibnamefont {Ray}}, \bibinfo
  {author} {\bibfnamefont {S.}~\bibnamefont {Sinha}}, \bibinfo {author}
  {\bibfnamefont {G.~V.}\ \bibnamefont {Shlyapnikov}}, \ and\ \bibinfo {author}
  {\bibfnamefont {L.}~\bibnamefont {Santos}},\ }\href@noop {} {}\Eprint
  {http://arxiv.org/abs/1808.03585} {arXiv:1808.03585} \BibitemShut {NoStop}%
\bibitem [{\citenamefont {Krishna}\ and\ \citenamefont
  {Bhatt}(2018)}]{Bhatt2018}%
  \BibitemOpen
  \bibfield  {author} {\bibinfo {author} {\bibfnamefont {A.}~\bibnamefont
  {Krishna}}\ and\ \bibinfo {author} {\bibfnamefont {R.~N.}\ \bibnamefont
  {Bhatt}},\ }\href@noop {} {\bibfield  {journal} {\bibinfo  {journal} {Phys.
  Rev. B}\ }\textbf {\bibinfo {volume} {97}},\ \bibinfo {pages} {174205}
  (\bibinfo {year} {2018})}\BibitemShut {NoStop}%
\bibitem [{\citenamefont {Porter}(1965)}]{Porter1965}%
  \BibitemOpen
  \bibfield  {author} {\bibinfo {author} {\bibfnamefont {C.~E.}\ \bibnamefont
  {Porter}},\ }\href@noop {} {\emph {\bibinfo {title} {Statistical Theories of
  Spectra: Fluctuations}}}\ (\bibinfo  {publisher} {Academic, New York},\
  \bibinfo {year} {1965})\BibitemShut {NoStop}%
\bibitem [{Note4()}]{Note4}%
  \BibitemOpen
  \bibinfo {note} {Note that both for TI- and fully uncorrelated variants the
  spectrum in formally delocalized phases, $a<1$ or $\gamma <2$, is unbounded
  in thermodynamic limit $N\to \infty $.}\BibitemShut {Stop}%
\bibitem [{Note5()}]{Note5}%
  \BibitemOpen
  \bibinfo {note} {Here we assume no resonances like $E_0=-2j_0\zeta
  _a$.}\BibitemShut {Stop}%
\end{thebibliography}%

\appendix
\renewcommand{\theequation}{S\arabic{section}.\arabic{equation}}
\setcounter{figure}{0}
\renewcommand{\thefigure}{S\arabic{figure}}
\renewcommand{\thesection}{S\arabic{section}}

\newpage
\begin{widetext}
\section*{Appendix}
This document provides supplemental material 
discussing some technical details of analytic considerations and numerical results.
In particular, the Supplemental Information contains the following notes:
\begin{itemize}
 \item In the main text, we have introduced several translation-invariant (TI) matrix ensembles and made use of scaling criteria in momentum space. In~\ref{App_Sec:DFT_Gaussian} we consider properties of Discrete Fourier Transform (DFT) of Gaussian random variables which lead to the duality for TI-Rosenzweig-Porter (TI-RP) model.
 \item In~\ref{App_Sec:Numerics} we provide some details of numerical calculations and describe the algorithm of extracting the spectrum of fractal dimensions $f(\alpha)$.
 \item The~\ref{App_Sec:Porter-Thomas} is devoted to the comparison of the wave function distributions for RP and TI-RP models in the weakly ergodic delocalized phase, $\gamma<1$, with the fully ergodic random-matrix theory (RMT) prediction.
 \item In the main text we have used scaling arguments for Burin-Maksimov (BM)~\cite{Burin1989} ensemble both in coordinate and momentum spaces, which require calculation of the disorder-free spectrum and level-spacing structure. In~\ref{App_Sec:Fourier} we present its detailed derivation and the estimate for the number of decoupled delocalized states.
 \item Finally, in~\ref{App_Sec:Inverse_Fourier} we present derivation of the effective real-space Hamiltonian for BM and discuss its relations to so-called cooperative shielding~\cite{Borgonovi_2016} and spectrum truncation~\cite{Bhatt2018}. Additionally, we show how the duality $a \leftrightarrow 2-a$ can be violated in this model.
\end{itemize}

\section{Discrete Fourier transform of Gaussian random variables}\label{App_Sec:DFT_Gaussian}
Due to the property
\be
\sum_{m=0}^{N-1} X_m X_m^* = \sum_{p=0}^{N-1} \tilde X_p \tilde X_p^*
\ee
of the discrete Fourier transform (DFT)
\be
\tilde X_p = \tilde X_{-p}^* = \frac{1}{N^{1/2}}\sum_{m=0}^{N-1}X_{m}e^{-2\pi i \frac{p m}N}
\ee
for real (complex) Gaussian statistically independent random variables $X_m = \ep_m/\sqrt{N}$ ($X_m=j_m\sqrt{N}$) with zero mean and fixed variance
\begin{gather}\label{eq:ep_distr}
\la X_{i}\ra = 0, \quad
\la |X_{i}|^2\ra = \sigma^2, \quad
P(X_0,\ldots,X_{N-1}) = \prod_{i=0}^{N-1} \frac{e^{-X_i^2/2\sigma^2}}{\sqrt{2\pi \sigma^2}} \ ,
\end{gather}
the real-valued components of their DFT $\Re \tilde X_p$, $\Im \tilde X_p$, $p=\overline{0,\lceil N/2\rceil}$ are also independent Gaussian random variables
with the same variance $\sigma^2$ for real components $\tilde X_p=\tilde X_p^*$ and
two times smaller one $\sigma^2/2$ for each real-valued components of complex elements $\tilde X_p$.
Note that $\Im \tilde X_0=0$ for all $N$ and $\Im \tilde X_{N/2}=0$ for even $N$ giving in total always $N$ real random numbers.

As a result of this property, the case of the Gaussian unitary ensemble (GUE) provides only the hermitian condition $H_{mn}=H_{nm}^*$ on the matrix elements of the TI-RP Hamiltonian
$H_{mn} = \ep_m\delta_{mn} + j_{m-n} \  $,
keeping the duality $\gamma_p = 2-\gamma$ of the TI-RP ensemble and avoiding approximate degeneracies in its spectrum at small $\gamma$.

However, in the other case of the Gaussian orthogonal ensemble (GOE) the symmetric constriction $H_{mn} = H_{nm}$ on the real-valued matrix elements correlates the TI-hopping integrals $j_{-n}=j_{n}$ and produces the degeneracy $\tilde E_p = \tilde E_{-p}$ in the spectrum
$\tilde E_p =\tilde E_p^* = \sum_{m=0}^{N-1}j_{m}e^{-2\pi i \frac{p m}N}$
of the hopping problem $H_{mn}^{(0)} = j_{m-n}$.
This results in the ratio-statistics going to zero at $\gamma<0$ and lifts the TI-RP symmetry.
Therefore in the main text we focus on GUE case.

\section{Details of numerical calculations}\label{App_Sec:Numerics}
In order to calculate the spectrum of fractal dimensions $f(\alpha)$ in all considered models
we first collect the empirical distribution $P(|\psi_E(m)|^2,N)$ of wavefunction intensities $|\psi_E(m)|^2$ over all $N$ sites $1\leq m\leq N$ averaged over the half of the states in the middle of the spectrum of a finite system and over $N_r = 10^3$ disorder realizations in the form of a histogram.
The extracted finite-size spectrum of fractal dimensions
$f (\alpha, N ) = \ln\lp N^{1-\alpha} P(N^{-\alpha})\ln N\rp/ \ln N$ has been extrapolated to the limit $N\to\infty$ over several system sizes with the linear fit $f (\alpha, N ) = f (\alpha) + c_\alpha/\ln N$, see Fig.~\ref{Fig:f(a)_extrapol}(b).

In order to eliminate the effect of zeros of wave functions $\psi$ which dominate the distribution function $P(|\psi_E(m)|^2)$ at small $|\psi_E(m)|^2\ll N^{-1}$ and extract the distribution function of a smooth envelope of $|\psi_E(m)|^2$ we use the so-called ``rectification'' approach suggested in~\cite{Luca2014} and used in~\cite{Kravtsov_NJP2015}.
Indeed, we represent eigenstates as products $\psi_E(m)=\psi_E^{env}(m)\times \eta$ of GOE random oscillations $\eta$ with
the unit average square and of the envelope $\psi_E^{env}(m)$, which are supposed to be statistically independent from each other.
Then the distribution of $\ln |\psi_E(m)|^2$ is a convolution of the distribution $\ln \psi_E^{env}(m)$
and the known distribution of $\ln \eta$.
Making a numerical de-convolution one obtains the distribution $P_{env}(|\psi_E^{env}(m)|^2)$ in which the effect of zeros of $\eta$ is eliminated.
Note that this method barely affects moderate and large $|\psi_E(m)|^2$ values as the variance of oscillations $\eta$ equals unity.

One can see that $f (\alpha)$ rectified and extrapolated as explained above has a linear in $\alpha$ part which exactly coincides with the analytical predictions dashed lines in Fig.~\ref{Fig:f(a)_extrapol}(a).
The analysis of the spectrum of fractal dimensions $f_p(\alpha_p)$ in the momentum space is completely analogous, see Fig.~\ref{Fig:f(a)_extrapol}(c, d).

 \begin{figure*}[t]
 \includegraphics[width=0.7\textwidth]{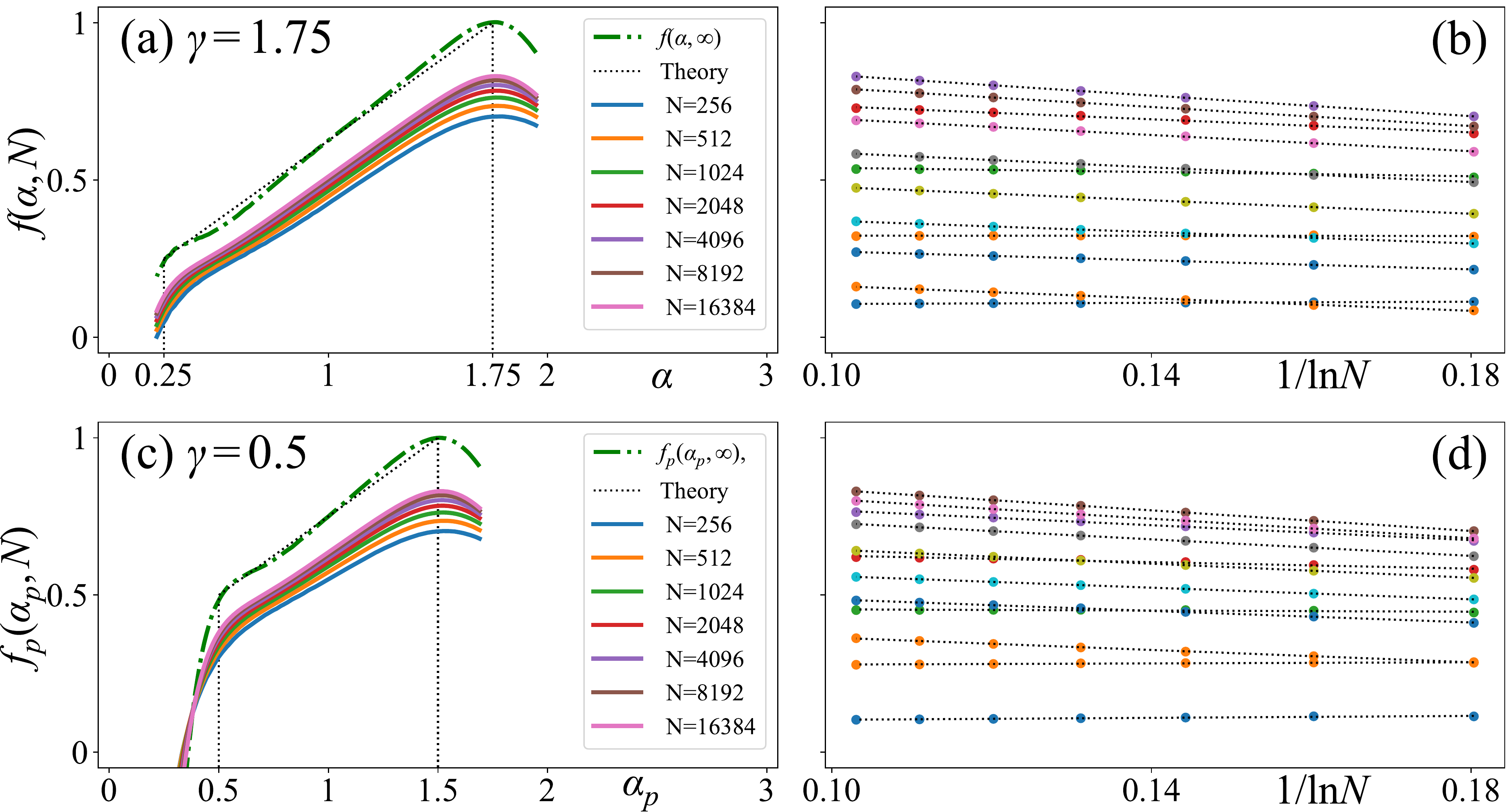}
 \caption{Extrapolation of the spectrum of fractal dimensions for TI-RP ensemble in the system size $N$ in (a, b) coordinate and (c, d) momentum spaces.
 (a, c) The finite size spectrum of fractal dimensions $f(\alpha,N)$ $\lp f_p(\alpha_p,N)\rp$ in the coordinate (momentum) space versus $\alpha$ $\lp\alpha_p\rp$ for different $N$;
 (b, d) $f(\alpha,N)$ $\lp f_p(\alpha_p,N)\rp$ in the coordinate (momentum) space versus $1/\ln N$ for different values of $\alpha$ $\lp \alpha_p\rp$ (symbols) with the linear fitting (dashed lines).
 }
 \label{Fig:f(a)_extrapol}

 \end{figure*}


\section{Comparison of eigenstate distributions in a weakly ergodic phase with RMT predictions}\label{App_Sec:Porter-Thomas}
In this note we consider the eigenstate distributions of RP and TI-RP models in more details, focusing on their deviations from fully-ergodic RMT predictions. \rev{The analysis is similar to the discussion of Fig. 8 in the main text.}

The standard RMT predicts the Gaussian distribution of each real-valued component $\psi_R$ of the eigenvector (see, e.g.,~\cite{Mehta2004random,Porter1965}):
\be
P(\psi_R) = \frac{\exp\lb-\beta N\psi_R^2/2\rb}{\lp2\pi/(\beta N)\rp^{1/2}} \ ,
\ee
with the matrix size $N$ and the ensemble parameter $\beta$ taking the values $\beta=1,2,4$ for GOE, GUE, and GSE (the Gaussian symplectic ensemble), respectively.
Due to the duality of TI-models in the coordinate and momentum spaces in the main text we focus on the GUE case, $\beta=2$,
thus, the RMT distribution of the renormalized wave-function intensity $y=N|\psi|^2$ in this case takes a simple exponential form, the analogue of the Porter-Thomas distribution for GOE:
\be\label{App:P(y)_exp}
P(N|\psi|^2=y) = e^{-y} \ .
\ee

Our motivation of the detailed study of the distribution $P(N|\psi|^2=y)$ is the statement from the main text that
``the sequence of phases in the coordinate space of RP and
TI-RP ensembles and positions of phase transitions are the
same''. To additionally shed some light on this matter we plot the $P(N|\psi|^2=y)$ for several system sizes $N=2^9,\ldots,2^{14}$
for RP and TI-RP ensembles (color solid lines in Figs.~\ref{Fig:RP_P(psi)_Porter-Thomas}) together with the RMT prediction \eqref{App:P(y)_exp} (black dashed lines).

From Fig.~\ref{Fig:RP_P(psi)_Porter-Thomas}(a-d) one can see that for all $\gamma<1$ the RP-ensembles shows exponential distribution at least in the thermodynamic limit $N\to\infty$ (see the flow of the distributions with the system size).
In the critical point $\gamma=1$ of the ergodic transition the distribution does not flow towards~\eqref{App:P(y)_exp} as at that point the spectral statistics is quasi-Poisson with the finite compressibility $\chi$ at large energies~\cite{Kravtsov_NJP2015}.
The TI-RP ensemble, Fig.~\ref{Fig:RP_P(psi)_Porter-Thomas}(e-h), shows the same behavior as its non-TI counterpart for $\gamma>0$ (in the thermodynamic limit), while the point $\gamma=0$
explicitly shows deviations from~\eqref{App:P(y)_exp}.
The deviations at $\gamma=0$ are expected as this point is the Anderson localization transition in the momentum space.
On the other hand, the convergence of the eigenstate statistics to the RMT prediction in TI-RP at $0<\gamma<1$ provides a non-trivial example of
the phase, where the RMT eigenfunction statistics can coexist with a hybrid level statistics.

 \begin{figure*}[t]
 \centering{
 \includegraphics[width=0.75\textwidth]{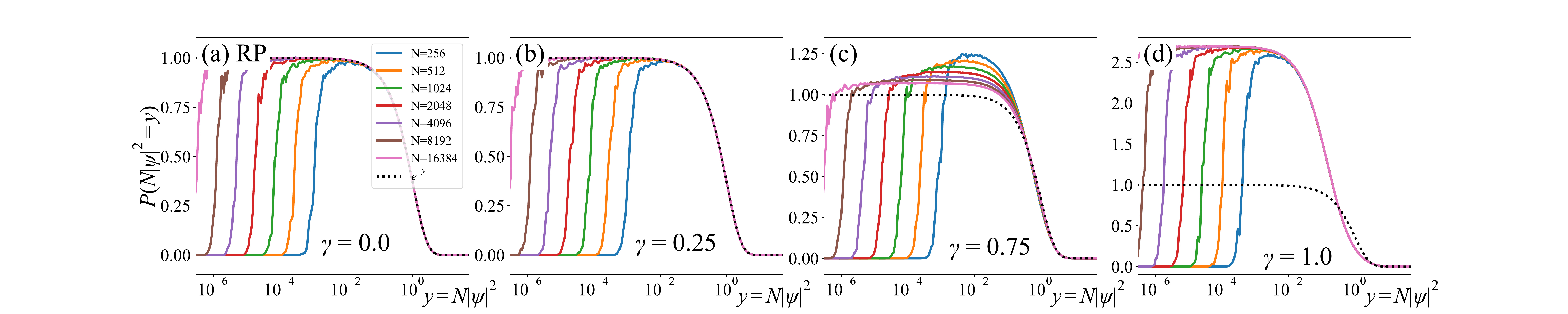}
 \includegraphics[width=0.75\textwidth]{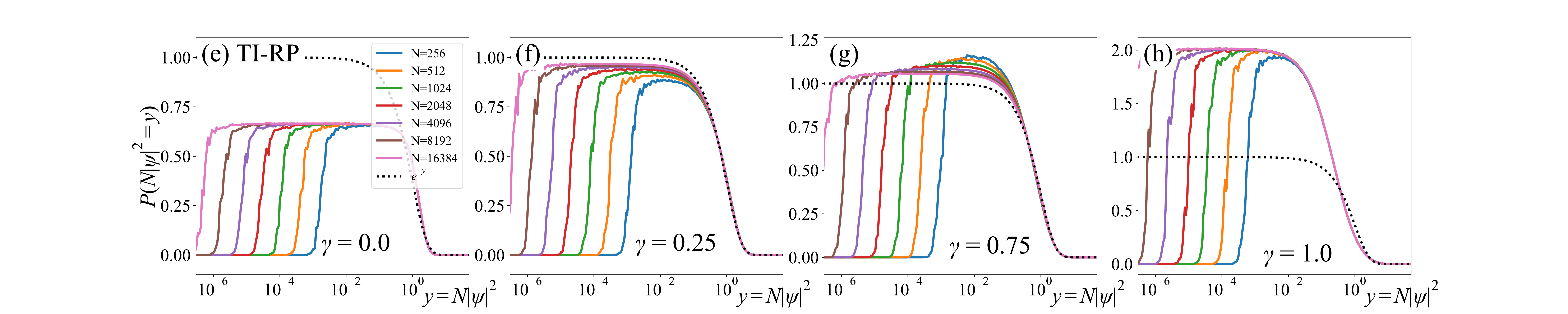}
 }
 \caption{Comparison of eigenstate probability distributions $P(N|\psi|^2=y)$ in the (weakly) ergodic phase of (a-d) RP and (e-h) TI-RP models
 (solid lines for different system sizes $N$) with the RMT prediction $P(y) = e^{-y}$ (black dashed line).
 }
 \label{Fig:RP_P(psi)_Porter-Thomas}

 \end{figure*}

\section{Derivation of the disorder-free spectrum and number of decoupled delocalized states}\label{App_Sec:Fourier}
Here we consider the continuous approximation $N\to\infty$ of the DFT of hopping terms $j_n = (1-\delta_{n,0})/|n|^a$ in BM-model given partially in~\cite{Malyshev2000,Malyshev2005}
\be
\tilde E_p/(2j_0) = \sum_{n\ne0\atop |n|<N/2} \frac{e^{-2\pi i p n/N}}{2|n|^a} =
\Re\sum_{n=1}^{N/2} \frac{e^{-2\pi i p n/N}}{|n|^a} =
\Re\lb(-1)^p e^{2 \pi i p /N} \Phi(e^{2\pi ip /N}, a,
   1 + N/2) + Li_a(e^{2\pi i p /N})\rb
 \ ,
\ee
with Lerch transcendent $\Phi(z,s,b) = \sum_{n=0}^\infty z^n/(n+b)^s$ and polylogarithm $Li_m(z) = \sum_{n=1}^\infty z^n/m^n$ functions.

The expansion of the polylogarithm gives the main result. Indeed,
\be\label{App:E_p_0}
\tilde E_{0}/(2j_0) = \sum_{n=1}^{N/2} \frac{1}{|n|^a} = H_n \simeq \frac{N^{1-a}}{1-a}+\zeta_a + O(N^{-a}), \quad a>0, \; a\ne 1 \ ,
\ee
where $H_n$ is the Harmonic number and $\zeta_a$ is the Riemann zeta function,
\be\label{App:E_p_p_ll_N}
\tilde E_p/(2j_0) \simeq
\zeta_a + A_a\lp\frac{p}{N}\rp^{a-1}   \ ,
\quad 0<|p|\ll N \ ,
\ee
with
\be
A_a = (2\pi)^{a-1}\Gamma_{1-a} \sin \frac{\pi a}{2} \ , a\ne 2m+1, m \in \mathbb{N} \ ,
\ee
\be\label{App:E_p_N2-p_ll_N}
\tilde E_p =
2j_0\Re\sum_{n=1}^{N/2} \frac{(-1)^n e^{2\pi i q n/N}}{|n|^a} \simeq
2j_0\sum_{n=1}^{N/2} \frac{(-1)^n}{|n|^a}\lb 1-\frac{n^2}{2}\lp \frac{2\pi q}{N}\rp^2\rb \simeq 
\tilde E_{\min}+B_a\lp \frac{q}{N}\rp^2 
\ ,
\ee
$q = |N/2-p| \ll N$, and
\be
\tilde E_{\min} = 2 j_0 (2^{1-a}-1)\zeta_a<0 \text{ for } a>-2; \quad B_a = 8 \pi^2 j_0(1-2^{3-a})\zeta_{a-2}\simeq 2\pi^2 j_0 a>0 \ .
\ee


Now we estimate number $N_{dec}$ of decoupled delocalized states both for BM~\cite{Malyshev2000,Malyshev_PRL2003,Malyshev2004,Malyshev2005,Kravtsov_Shlyapnikov_PRL2018} and
the Yuzbashyan-Shastry (YS) ~\cite{Yuzbashyan_JPhysA2009_Exact_solution,Yuzbashyan_NJP2016} models.

In the integrable case of YS-model, $a=0$,
with the scaling of the ratio of the hopping amplitude $j_0$ to the disorder strength $\Delta$: $j_0/\Delta\sim N^{-\gamma/2}$,
the only state, namely the zero-momentum state with the energy $\tilde E_0 \sim N^{1-\gamma/2}$, is decoupled from other $N-1$ degenerate states at $\gamma<2$~\cite{Yuzbashyan_NJP2016,Borgonovi_2016}.

The extensive behavior of the level spacing with $N$, demonstrated in YS model, survives in BM for all $a<1$, but in this case
the number of decoupled states is extensive $N_{dec}\gg 1$.
Indeed, according to Levitov's arguments~\cite{Levitov1989,Levitov1990}
we compare the energy differences ($a\ne 0,1$)
\be\label{App:dE_p}
\delta E_p = |\tilde E_{p+\delta p}-\tilde E_p|\sim 2j_0\left(\frac{N}{2\pi|p|}\right)^{1-a}\left|\left(1+\frac{\delta p}{|p|}\right)^{a-1}-1\right|\sim
2j_0|a-1| \left(\frac{N}{|p|}\right)^{2-a}\frac{\delta p}{N} \ , \quad \delta p\ll |p|
\ee
with the sum of absolute values of hoppings in the same $\delta p$ interval
$\sum_{p'=p}^{p+\delta p} |\tilde{J}_{p'}| \simeq \Delta\frac{\delta p}{\sqrt{N}}$,
and get that for all the states with
\be\label{eq:p*}
|p|^{2-a}<p_*^{2-a}\simeq N^{3/2-a}\frac{2j_0|a-1|}{\Delta}
\ee
they are localized in $p$-space (extended in real space)
and the disorder $\Delta\sim N^0$ cannot delocalize them, meaning that $N_{dec}\sim p_*$ for $a<3/2$ and $N_{dec}=0$ for $a>3/2$.

Similar arguments are given in~\cite{Malyshev2005} for $a>1$.
Note, however, that in the case of $1<a<3/2$, the effect of the decoupled delocalized states is small as all their energies are not increasing with $N$ and there is the critical disorder strength of order of the bare bandwidth
\be\label{eq:polyn_hop_ALT_condition}
\Delta_c \simeq \Delta_p=\tilde E_0 - \tilde E_{\min}
\ee
above which all states $|p|<p_*$ become also localized (see, e.g.,~\cite{Malyshev_PRL2003,Malyshev2005}).
For $j_0/\Delta \sim N^0$ and $a>3/2$ all states are localized for any disorder strength.


\section{Derivation of the effective BM-model}\label{App_Sec:Inverse_Fourier}
One might suggest that in the spirit of~\cite{Borgonovi_2016} for getting the effective theory for deterministic models (YS~\cite{Yuzbashyan_JPhysA2009_Exact_solution,Yuzbashyan_NJP2016} and BM~\cite{Malyshev2000,Malyshev_PRL2003,Malyshev2004,Malyshev2005,Kravtsov_Shlyapnikov_PRL2018})
one simply needs to separate the Hilbert space of the disorder-free hopping model $H_{mn}^0 = j_{m-n}$ in the momentum basis into two subspaces of delocalized $|P|<p_*$ and nearly-degenerated $|P|>p_*$ states, without any coupling between subspaces and consider only the fast oscillation sector $|P|>p_*$.
However, as it was recently pointed out, e.g., in~\cite{Bhatt2018}, this naive procedure leads to the model of the form of the PLRBM ensemble
with the effective power-law decay rate $a_{eff}=1$, equivalent to the critical point of PLRBM, and with strongly correlated random hopping terms.
Despite the fact that this truncating procedure already violates the locator expansion breakdown, the localization phenomenon at $a<1$ and an intriguing duality $a \leftrightarrow 2-a$ are not explained yet, therefore more accurate treatment is required.

Such a more rigorous approach which was introduced in the main text requires calculation of the inverse matrix
$\hat M = \left(\hat j +E_0\right)^{-1}$, with $(-E_0)$ taken to be below the bottom $\tilde E_{\min}$ of the spectrum~(\ref{App:E_p_p_ll_N},~\ref{App:E_p_N2-p_ll_N})~\footnote{Note that both for TI- and fully uncorrelated variants the spectrum in formally delocalized phases, $a<1$ or $\gamma<2$, is unbounded in thermodynamic limit $N\to\infty$.}.
To do so one can use DFT of $j_n$, mentioned in App.~\ref{App_Sec:Fourier}, and the inverse DFT as follows
\be
\frac{M_{m,m+n}}{\rev{E_0}} = \frac{1}{N}\sum_{|p|<N/2} \frac{e^{2\pi i p n/N}}{\tilde E_p +E_0} = \frac{1}{N(2j_0\zeta_a+E_0)+2j_0N^{2-a}/(1-a)}+\frac{2}{N}\Re\sum_{p=1}^{N/2} \frac{e^{2\pi i p n/N}}{\tilde E_p +E_0} \ .
\ee
The latter sum can be split into two ones corresponding to two parts~\eqref{App:E_p_p_ll_N} and~\eqref{App:E_p_N2-p_ll_N} of the spectrum
\be
\frac{2}{N}\Re\sum_{p=1}^{N/2} \frac{e^{2\pi i p n/N}}{\tilde E_p +E_0} =
\frac{2}{N}\Re\sum_{p=1}^{N\alpha} \frac{e^{2\pi i p n/N}}{2j_0\zeta_a +E_0 + 2j_0 A_a(p/N)^{a-1}}+
\frac{2}{N}\Re\sum_{q=0}^{N(1/2-\alpha)} \frac{(-1)^n e^{2\pi i q n/N}}{\tilde E_{\min}+E_0 +B_a (q/N)^2} \ ,
\ee
with the fraction of states taken in the first sum $0<\alpha<1/2$.

For $a>1$ the denominator of the first sum at $p\ll N$ is dominated by a constant $2 j_0 \zeta_a + E_0$~\footnote{Here we assume no resonances like $E_0=-2j_0\zeta_a$.} giving the terms
\be\label{App:S1_a>1}
S_1(N\alpha,a>1) \equiv \frac{2}{N}\Re\sum_{p=1}^{N\alpha} \frac{e^{2\pi i p n/N}}{2j_0\zeta_a +E_0 + 2j_0 A_a(p/N)^{a-1}}\simeq
\frac{2}{2j_0\zeta_a +E_0}\frac{\sin(\pi \alpha n)}{N\sin(\pi n/N)} \ ,
\ee
decaying slower than the original hoppings $j_n\sim |n|^{-a}$ and therefore in this case the transformation to $\hat M$ is not relevant (see, e.g., Appendix~A in~\cite{Malyshev2004}).

In the opposite case of $a<1$, the denominator of $S_1$ is dominated by the polynomial term until some critical index $|p|<p_c$
\be
p_c\simeq (N/2\pi)[E_0/(2j_0 A_a)+\zeta_a/A_a]^{-1/(1-a)}\lesssim N
\ee
giving the result
\be
S_1(N\alpha,a<1) \simeq
\frac{2}{N}\Re\sum_{p=1}^{p_c} \frac{e^{2\pi i p n/N}}{2j_0 A_a(p/N)^{a-1}}+
S_1(N\alpha,a>1)-S_1(p_c,a>1) \ ,
\ee
where last two terms are calculated in terms of~\eqref{App:S1_a>1}.
The first term can be calculated in the continuous limit, $N/p_c\ll n\ll N$, analogously to~\eqref{App:E_p_p_ll_N}
\be\label{App:n_2-a-term}
\frac{2}{N}\Re\sum_{p=1}^{p_c} \frac{e^{2\pi i p n/N}}{2j_0 A_a (p/N)^{a-1}}\simeq
\frac{|n|^{a-2}}{2\pi j_0 A_a}\Re\int_{2\pi n/N}^{2\pi p_c n/N} \frac{e^{ix}dx}{x^{a-1}} \simeq
\frac{A_{a-1}|n|^{-(2-a)}}{2\pi j_0 A_a} =
\frac{|n|^{-(2-a)}}{2\pi j_0 a \tan(\pi a/2)}
\ .
\ee
This term gives the symmetry $a\to 2-a$~\cite{Kravtsov_Shlyapnikov_PRL2018} for the tails of the wavefunctions.
The term analogous to~\eqref{App:E_p_N2-p_ll_N} at $|N/2-n|\ll N$ is not relevant due to larger terms in $S_2$ in that range of $n$.
Note that the amplitude of the term~\eqref{App:n_2-a-term} is just the upper bound as the limits of the integral are not $0$ and $\infty$.

The second sum for $a>0$, which becomes relevant at small distances,
\be\label{App:S_2}
S_2 = \frac{2}{N}\Re\sum_{q=0}^{N(1/2-\alpha)} \frac{(-1)^n e^{2\pi i q n/N}}{\tilde E_{\min}+E_0 +B_a (q/N)^2}
\ee
depends mostly on the sign of the parameter $\tilde E_{\min}+E_0$ as $B_a\simeq 2\pi^2 j_0 a>0$.
For $(1/2-\alpha)^2B_a\ll |\tilde E_{\min}+E_0|$
the denominator is dominated by $\tilde E_{\min}+E_0$ and the sum takes the form similar to~\eqref{App:S1_a>1}
\be
S_2 \simeq \frac{2(-1)^n}{N(\tilde E_{\min}+E_0)}\lb 1+\frac{\sin(\pi(1/2-\alpha)n)}{\sin(\pi n/N)}\rb \ .
\ee

In the opposite case 
%
for $|n|\ll N$ one can use the continuous limit of the sum
\be
S_2 = \frac{2n (-1)^n}{B_a}\Re\int_0^{\pi n(1-2\alpha)}\frac{e^{i x} dx}{x^2+(2\pi n)^2(\tilde E_{\min}+E_0)/B_a}
\ee
and easily calculate it as follows for $\tilde E_{\min}+E_0>0$
\be\label{App:exp-term}
S_2 = (-1)^n\pi\Re \frac{e^{-2\pi|n|\sqrt{(\tilde E_{\min}+E_0)/B_a}}}{\sqrt{B_a(\tilde E_{\min}+E_0)}} \
\ee
and for $\tilde E_{\min}+E_0<0$
\be\label{App:osc-term}
S_2 = C_a(-1)^n\pi\Re \frac{e^{-2\pi i|n|\sqrt{(E-\tilde E_{\min})/B_a}}}{i\sqrt{B_a(E-\tilde E_{\min})}} \ .
\ee
The latter is written up to the constant amplitude $C_a \sim N^0$.

The result in the main text is given by the sum of~\eqref{App:n_2-a-term} and~\eqref{App:exp-term}.


Note that for the models with sign-alternating power-law decaying hoppings the spectrum is also sign-alternating and in the case $a<1$ there is no finite $E_0$ below the spectrum.
Then the result is given by the sum of~\eqref{App:n_2-a-term} and~\eqref{App:osc-term}
and, thus, the system becomes delocalized due to the oscillating term~\eqref{App:osc-term}, but not fully ergodic (see, e.g., the results for TI-PLRBM or~\cite{Cantin2018}).
\end{widetext}

\end{document}